%% file: main.tex
\documentclass[acmsmall,nonacm]{acmart}
\newif\iftechnical
\usepackage{subcaption}
\usepackage{wrapfig}
\usepackage{guyboxes}
\usepackage{thm-restate}
\usepackage[shortcuts]{extdash}
\usepackage{xspace}
\input{minimal}
\input{concurrent-games}

\input{prelude}
\input{know}

\title{Game Semantics: Easy as Pi}
\subtitle{Introducing Programming Game Semantics}

\author{Simon Castellan}
\affiliation{
  \institution{Inria, Univ Rennes, IRISA}            
  \city{Rennes}
  \country{France}
}
\email{simon@phis.me}          
\author{Léo Stefanesco}
\affiliation{
  \institution{Collège de France}            
  \city{Paris}
  \country{France}
}
\email{leo.lveb@gmail.com}          
\author{Nobuko Yoshida}
\affiliation{
  \institution{Imperial College London}            
  \city{London}
  \country{United Kingdom}
}
\email{n.yoshida@ic.ac.uk}          

\setcopyright{none}


\begin{abstract}
  \input{abstract.tex}
\end{abstract}
\begin{document}

\maketitle

\section{Introduction}
\input{introduction}
\section{Overview: Game Semantics as Message-Passing Translation}
\label{sec:overview}
\input{overview}

\section{$\MetalanguageName$: a Metalanguage for Strategies of Game Semantics}
\label{sec:calculus}
\input{calculus}

\section{Translation of $\CML$ into $\MetalanguageName$}
\label{sec:interp ML}
\input{translation}

\section{Causal Game Semantics}
\label{sec:games}
\input{games}


\section{Interpretation of $\MetalanguageName$ into strategies}
\label{sec:interp}
\input{interpretation}

\input{implementation}

\section{Related Work}\label{sec:related}
\input{related-work}
\section{Conclusion}\label{sec:conclusion}
\input{ccl}

\bibliographystyle{ACM-Reference-Format}
\bibliography{biblio}

\appendix

\input{appendix-calculus}
\input{appendix-translation}
\input{appendix-games}
\input{appendix-interpretation}

\end{document}

%% file: concurrent-games.tex
\makeatother
\newcommand{\minconflit}{\mathrel{\hspace{-0.4em}\xymatrix@C=1em { \ar@{~}[r]&}\hspace{-0.4em}}}
\newcommand{\minconflict}{\kl[minimal conflict]{\minconflit}}
\newcommand{\conflict}{\mathrel{\#}}

\makeatletter

\usepackage{tikz}
\usepackage{tikz-cd}

\usetikzlibrary{decorations.pathreplacing,decorations.markings}

\tikzset{module/.style={
    decoration={markings,
      mark= at position 0.45 with {
        \node[] (tempnode) {};
        \draw[-,shorten <=.1em,shorten >=.1em,inner
        sep=1pt] (tempnode.north) -- (tempnode.south);
      }
    },
    postaction={decorate}
  }
}

\newrobustcmd\profto{\!\!\!\!\begin{tikzcd}[ampersand replacement=\&,column
  sep=1.5em,arrow style=tikz]{}
\ar[r,module] \& {}\end{tikzcd}\!\!\!\!}

\DeclareMathOperator{\pol}{\textit{pol}}

\newcommand{\CC}{{\rm C\!\!C}}
\newcommand{\cc}{\ c\!\!\!\!c\,}

\newdir{C}{%
!/4.5pt/@{( }*:(1,-.2)@{}*:(1,+.2)@_{}}

\newdir{|>}{%
!/4.5pt/@{|}*:(1,-.2)@{>}*:(1,+.2)@_{>}}

\newcommand{\ttrue}{\normalfont\textrm{\textnormal{tt}}}
\newcommand{\ffalse}{\normalfont\textrm{\textnormal{ff}}}


%% file: prelude.tex
\usepackage[british]{babel}

\usepackage[nomargin,inline,draft]{fixme}
\fxsetup{theme=color,mode=multiuser}
\FXRegisterAuthor{LS}{aLS}{LS} 
\FXRegisterAuthor{NY}{aNY}{NY} 
\FXRegisterAuthor{SC}{aSC}{SC} 

\usepackage{enumitem}

\newrobustcmd{\demph}[1]{\textbf{#1}}
\newrobustcmd{\Processes}[1]{\mathsf{Proc}(#1)}

\newrobustcmd{\longcov}[1]{{\stackrel{#1}{\mathrel-\joinrel\relbar\joinrel\subset\,}}}

\newrobustcmd{\es}{E}
\newrobustcmd{\Start}{\textsf{q}}
\newrobustcmd\start{\Start}
\newrobustcmd{\esB}{F}
\newrobustcmd{\cess}{\mathcal{\es}}
\newrobustcmd{\cessB}{\mathcal{\esB}}
\newrobustcmd{\causal}{\leq}
\newrobustcmd{\lbl}{\mathrm{lbl}}
\newrobustcmd{\parmap}{\rightharpoonup}
\newrobustcmd{\isofam}[1]{\widetilde{#1}}
\newrobustcmd{\isofambetween}[1]{\cong_{\isofam{#1}}}
\newrobustcmd{\configstep}[1]{\longcov{#1}}
\newrobustcmd{\config}{\mathcal{C}}
\newrobustcmd{\Con}{\kl[consistent sets of events]{Con}}
\newrobustcmd{\causesof}[1]{[#1]}

\newrobustcmd{\game}{\mathcal{A}}
\newrobustcmd{\strat}{\mathcal{S}}

\newrobustcmd{\EmptyProcess}{\textbf 0}
\newrobustcmd{\Naming}[2]{#1 \{ #2 \}}
\newrobustcmd{\Output}[2]{#1\oplus #2}
\newrobustcmd{\Input}[2]{#1\with \left\{#2\right\}_{i\in I}}
\newrobustcmd{\InputII}[3]{#1\with \left\{#3\right\}_{#2}}
\newrobustcmd{\Inputj}[2]{#1\with \left\{#2\right\}_{j\in J}}
\newrobustcmd{\Coder}[2]{#1\#(#2)}
\newrobustcmd{\Der}[2]{#1\wn[#2]}
\newrobustcmd{\Prom}[2]{#1\oc(#2)}
\newrobustcmd{\Contr}[3]{#1\wn\{#2, #3\}}
\newrobustcmd{\CoContr}[3]{#1\oc\{#2, #3\}}
\newrobustcmd{\cut}[4]{(\nu\: #1\, #2 : #3) #4}
\newrobustcmd{\multicut}[2]{(\nu\: #1) \left(#2\right)}
\newrobustcmd{\Typing}[2]{#1 \triangleright #2}
\newrobustcmd{\eqdef}{\stackrel{\text{\tiny{def}}}=}
\newrobustcmd{\ext}{\textsf{ext}}

\newrobustcmd{\imdep}{\kl[immediate causal dependency]{\rightarrowtriangle}}

\newrobustcmd{\Pos}[1]{#1}
\newrobustcmd{\Neg}[1]{#1}
\newrobustcmd{\LinStrat}{\kl[\LinStrat]{\bf LStr}}
\newrobustcmd{\LinDet}{\kl[\LinDetStrat]{\bf LDStr}}
\newrobustcmd{\Strat}{\kl[Strat]{\bf Str}}
\newrobustcmd{\UStrat}{\kl[UStrat]{\bf UStr}}
\newrobustcmd{\ESS}{\kl[ESS]{\bf ESS}}
\newrobustcmd{\ES}{\kl[ES]{\bf ES}}
\newrobustcmd{\fix}{\mathsf {fix}}
\newrobustcmd{\Left}{\mathsf L}
\newrobustcmd{\Right}{\mathsf R}
\newrobustcmd{\Lift}[1]{\kl[lifting maps]{\mathsf {lift}}(#1)}
\newrobustcmd{\LiftS}[1]{\kl[lifting maps with symmetry]{\mathsf {lift}(#1)}}

\newrobustcmd{\E}{\mathcal E}
\newrobustcmd{\F}{\mathcal F}

\newrobustcmd{\Inl}{\mathtt {inl}}
\newrobustcmd{\Inr}{\mathtt {inr}}
\newrobustcmd{\similarMaps}{\mathrel{\kl[similar]{\sim}}}
\newrobustcmd{\symmetric}{\kl[symmetric]{\sim}}

\newrobustcmd{\A}{\mathcal A}
\newrobustcmd{\B}{\mathcal B}

\newrobustcmd{\weakIso}{\mathrel{\kl[weakly isomorphic]{\approxeq}}}
\newrobustcmd{\similarProcesses}{\kl[similarProc]{\equiv}}

\newrobustcmd{\digging}[1]{\kl[digging]{\mu_{#1}}}
\newrobustcmd{\dereliction}[1]{\kl[dereliction]{d_{#1}}}
\newrobustcmd{\contraction}[1]{\kl[contraction]{c_{#1}}}
\newrobustcmd{\codereliction}[1]{\kl[codereliction]{\bar d_{#1}}}
\newrobustcmd{\cocontraction}[1]{\kl[cocontraction]{\bar c_{#1}}}
\newrobustcmd{\codermap}{\bar d}

\newrobustcmd{\sharpG}{\kl[sharpG]{\sharp}}
\newrobustcmd{\ocG}{\kl[ocG]{\oc}}
\newrobustcmd{\ocS}{\kl[ocS]{\oc}}
\newrobustcmd{\wnG}{\kl[wnG]{\wn}}
\newrobustcmd{\weakening}[1]{\kl[weakening]{\oc_{#1}}}
\newrobustcmd{\coweakening}[1]{\kl[coweakening]{\wn_{#1}}}
\newrobustcmd{\fc}[1]{\kl[fc]{\mathsf{fc}}(#1)}
\newrobustcmd{\fn}[1]{\kl[fn]{\mathsf{fn}}(#1)}
\newrobustcmd{\actCtx}[1]{\kl[fc]{\mathsf{fc}}(#1)}
\newrobustcmd{\Names}[1]{\kl[names]{\mathsf{na}}(#1)}
\newrobustcmd{\fcC}[1]{\kl[fcC]{\mathsf{fc}}(#1)}
\newrobustcmd{\Pairing}[1]{\kl[pairing]{\langle}#1\kl[pairing]{\rangle}}
\newrobustcmd{\Forwarder}[3]{\kl[fw]{[}#1 \kl[fw]{\leftrightarrow} #2\kl[fw]{]}_{#3}}
\newrobustcmd{\sumProc}{\mathrel{\kl[sumProc]{+}}}

\newrobustcmd{\ParUnit}{\textbf 1}

\newrobustcmd{\dom}[1]{\kl[dom]{\textsf{dom}}(#1)}
\newrobustcmd{\EmptyES}{\kl[emptyes]{\emptyset}}
\newrobustcmd{\EmptyGame}{\kl[emptygame]{\textbf 1}}
\newrobustcmd{\isoProc}{\kl[isoProc]{\cong}}
\newrobustcmd{\isoStrat}{\kl[isomorphism]{\cong}}
\newrobustcmd{\eqProc}{\kl[eqProc]{ \equiv }}
\newrobustcmd{\depthT}[1]{\kl[depthT]{ \textsf{depth}(#1)}}

\newrobustcmd\linto{\multimap}
\newrobustcmd\abs[2]{\lambda #1.\: #2}
\newrobustcmd\app[2]{#1\, #2}
\DeclarePairedDelimiter{\transl}{{\llparenthesis\,}}{{\,\rrparenthesis}}
\newrobustcmd{\translWk}[1]{\transl {#1}^{\textsf{wk}}}
\newrobustcmd{\translSC}[1]{\transl {#1}^{\textsf{SC}}}
\newrobustcmd{\Split}[3]{\left(#3\right)\!\!\{#1 {\coloneqq} #2\}}
\newrobustcmd{\Wait}[2]{#1(). #2}
\newrobustcmd{\Eval}[3]{\textsf{eval}_{#1,#2,#3}}
\newrobustcmd{\Send}[2]{#1[]. #2}
\newrobustcmd{\Interact}[4]{#3 \:{}_{#1}{\odot_{#2}}\: #4}
\newrobustcmd{\Abs}[3]{#1(#2).\, #3}
\newrobustcmd{\AbsCbv}[4]{\lambda_{#2 \to #1}^{#3}.\,  #4}
\newrobustcmd{\UnLam}[4]{\textsf{unlam}^{#3}_{#1 \to #2}\left(#4\right)}
\newrobustcmd{\App}[4]{\kl[app]{\mathbf{app}}_{#2(#1)} \left( #3, #4 \right)}
\newrobustcmd{\AppCbv}[5]{\mathbf{app}^{\text{cbv}}_{#1:#2 \to #3} \left( #4, #5 \right)}
\newrobustcmd{\AppCbvAnon}[2]{\mathbf{app}^{\text{cbv}} \left( #1, #2 \right)}
\newrobustcmd{\AppT}[6]{\kl[app]{\mathbf{app}}^{#4}_{#1: #2 \to #3} \left( #5, #6 \right)}
\newrobustcmd{\AppAnon}[4]{\kl[app]{\mathbf{app}}_{} \left( #3, #4 \right)}
\newrobustcmd{\AppNL}[4]{\kl[appnl]{\mathbf{app}}_{#2(#1)} \left( #3, #4 \right)}
\newrobustcmd{\ApplyMacro}[4]{\kl[applymacro]{\mathbf{call}}_{#1}^{#2}\left(#3, #4\right)}

\newrobustcmd{\BT}[1]{\textsf{BT}(#1)}

\newrobustcmd{\ValueType}[1]{\textsf{vt}(#1)}
\newrobustcmd{\CompType}[1]{\textsf{ct}(#1)}
\newrobustcmd{\linlamBool}{\mathbb{B}}
\newrobustcmd{\linlamNat}{\mathbb{N}}
\newrobustcmd\subst[2]{[#1 \coloneqq #2]}

\newrobustcmd\redto{\to}

\newrobustcmd{\shareInteract}[4]{#3 \;\:{}_{#1}{\odot_{#2}}\;\: #4}
\newrobustcmd{\shareInteractS}[3]{#2 \stackrel{#1}{\odot} #3}

\newrobustcmd{\DetRedTo}{ \rightarrow _d}
\newrobustcmd\ARel[1]{\:\mathcal R^{#1}\:}
\newrobustcmd\EARel[1]{\:\mathcal{ER}^{#1}\:}

\newrobustcmd\mycut[4]{\nu #1\,#2.\, \left( #4 \right)}

\usepackage{accsupp}
\newcommand*{\llbrace}{%
  \BeginAccSupp{method=hex,unicode,ActualText=2983}%
    \textnormal{\usefont{OMS}{lmr}{m}{n}\char102}%
    \mathchoice{\mkern-4.05mu}{\mkern-4.05mu}{\mkern-4.3mu}{\mkern-4.8mu}%
    \textnormal{\usefont{OMS}{lmr}{m}{n}\char106}%
  \EndAccSupp{}%
}
\newcommand*{\rrbrace}{%
  \BeginAccSupp{method=hex,unicode,ActualText=2984}%
    \textnormal{\usefont{OMS}{lmr}{m}{n}\char106}%
    \mathchoice{\mkern-4.05mu}{\mkern-4.05mu}{\mkern-4.3mu}{\mkern-4.8mu}%
    \textnormal{\usefont{OMS}{lmr}{m}{n}\char103}%
  \EndAccSupp{}%
}

\newrobustcmd\ctxTensor[1]{{\otimes}#1}

\newrobustcmd\posContr[4]{#1 \llbrace #2, #3 \rrbrace. #4 }
\newrobustcmd\Choose[3]{\kappa_{#1,#2}.\, #3}
\newrobustcmd\cbvWait[2]{\uparrow_{#1}.\, #2}
\newrobustcmd\receiveCtx[4]{{(#1{:}#2)} \with \left\{ #4\right\}_{#3}}
\newrobustcmd\Once[2]{\left< #2 \right>_{#1}}

\newrobustcmd{\ContractTree}[1]{\kl[contraction trees]{\mathscr T}_{#1}}

\newrobustcmd\ifte[3]{\mathsf{if}\:#1\:#2\:#3}
\newrobustcmd\share[3]{{#2} \parallel_{#1} {#3}}

\newrobustcmd{\EncodingPairs}[1]{\kl[encopairs]{\langle}#1\kl[encpairs]{\rangle}}
\newrobustcmd{\LiftPlus}[1]{\kl[liftplus]\uparrow^+ #1}
\newrobustcmd{\LiftNeg}[1]{\kl[liftneg]\uparrow^- #1}

\newrobustcmd{\belowStrategies}[1]{\mathrel{\kl[belowstrat]{ \xrightarrow{\;\;#1\;\;} }}}
\newrobustcmd{\belowProc}{\kl[belowProc]{ \xrightarrow{\quad} }}

\newrobustcmd{\Opp}[1]{{#1}^-}
\newrobustcmd{\Pla}[1]{{#1}^+}

\usetikzlibrary{positioning}
\usetikzlibrary{mindmap,trees,shadows,arrows}
\usetikzlibrary{matrix}
\usetikzlibrary{cd}
\usetikzlibrary{positioning}
\usetikzlibrary{shapes}
\usetikzlibrary{calc}

\usetikzlibrary{decorations.pathmorphing, tikzmark}

\tikzset{
  ampersand replacement=\&,
  row sep=0.3cm, column sep=0.3cm,
  pointer/.style={dashed,-},
  cause/.style={-open triangle 60},
  invisible/.style={opacity=0,text opacity=0},
  visible on/.style={alt=#1{}{invisible}},
  alert/.style={alt=#1{color=red}{},anchor=base},
  conflict/.style={snake it,-},
  cov/.style={-(}
  transcausal/.style={wiggle, ->},
  alt/.code args={<#1>#2#3}{%
    \alt<#1>{\pgfkeysalso{#2}}{\pgfkeysalso{#3}} 
  },
}

\tikzset{snake it/.style={decorate, decoration={snake, amplitude=.4mm,segment length=2mm}}}
\tikzset{wiggle/.style={decorate, decoration={
    segment length=4,
    amplitude=.9,post=lineto,
    post length=2pt, zigzag}}}
\tikzstyle{diag}=[matrix of math nodes]

\newrobustcmd{\PCFWM}{\ensuremath{\textsf{IPA}}}
\newrobustcmd{\PCFM}{\ensuremath{\textsf{IPA}}}
\newrobustcmd{\PCF}{\ensuremath{\textsf{PCF}}}
\newrobustcmd{\MALL}{\ensuremath{\textsf{MALL}}}
\newrobustcmd{\LL}{\ensuremath{\textsf{LL}}}
\newrobustcmd{\DiLL}{\ensuremath{\textsf{DiLL}}}
\newrobustcmd{\PIMALL}{\ensuremath{ \pi _{\MALL}}}
\newrobustcmd{\PILL}{\ensuremath{ \pi _{\LL}}}
\newrobustcmd{\PIDiLL}{\ensuremath{ \pi _{\DiLL}}}
\newrobustcmd{\FreeVars}[1]{\textsf{fv}(#1)}
\newrobustcmd{\LocOf}[1]{\textsf{loc}(#1)}
\newrobustcmd{\UnitType}{\bullet}
\newrobustcmd{\NatType}{\textsf{int}}
\newrobustcmd{\BoolType}{\textsf{bool}}
\newrobustcmd{\RefType}{\textsf{ref}}
\newrobustcmd{\LocType}{\textsf{loc}}

\newrobustcmd{\JoinTerm}{\texttt{join}}
\newrobustcmd{\SeqTerm}{\texttt{seq}}
\newrobustcmd{\Alloc}{\texttt{alloc}}
\newrobustcmd{\GetLoc}{\texttt{get}}
\newrobustcmd{\SetLoc}{\texttt{set}}
\newrobustcmd{\SkipTerm}{\texttt{skip}}
\newrobustcmd{\LetIn}[3]{\texttt{let}\, #1 = #2\, \textsf{in}\, #3}
\newrobustcmd{\LetRec}[3]{\texttt{let rec}\, #1 = #2\, \textsf{in}\, #3}
\newrobustcmd{\IfTerm}[3]{\texttt{if}\, {#1}\, {#2}\, {#3}}
\newrobustcmd{\ifSyntax}[3]{\texttt{if}\, {#1}\, {#2}\, {#3}}
\newrobustcmd{\RefAlloc}{\texttt{ref}}

\newrobustcmd{\CML}{\kl[CML]{\textsf{ML}{\text{\scriptsize$\parallel$}}}}
\newrobustcmd{\AllocLbl}[1]{\textsf{Alloc}(#1)}
\newrobustcmd{\GetLbl}[2]{\textsf{Read}(#1, #2)}
\newrobustcmd{\SetLbl}[2]{\textsf{Write}(#1, #2)}
\newrobustcmd{\tuple}[1]{\langle #1 \rangle}
\newrobustcmd{\ProgBisim}{\kl[progbisim]{\approx}}
\newrobustcmd{\CMLbisim}{\kl[cmlbisim]{\approx}_{\CML}}
\newrobustcmd{\CMLObs}{\kl[cmlobs]{\approxeq}_{\CML}}
\newrobustcmd{\MetalanguageName}{\pi_{\textsf{DiLL}}}
\newrobustcmd{\MLbisim}{\kl[cmlbisim]{\approx}_{\MetalanguageName}}
\newrobustcmd{\MLObs}{\kl[mlobs]{\approxeq}_{\MetalanguageName}}
\newrobustcmd{\bisim}{\kl[bisim]{\approx}}

\newrobustcmd{\ClosureOf}[1]{\kl[closureof]{\mathcal C}(#1)}
\newrobustcmd{\ReadMsg}{\textsf{rd}}
\newrobustcmd{\WriteMsg}{\textsf{wr}}
\newrobustcmd{\Done}{\textsf{done}}

\newrobustcmd{\Derout}[2] {#1 \wn \oplus #2}

\newrobustcmd{\LogType}[1]{\textsf{log}_{#1}}
\newrobustcmd{\loggingMemCell}{\textsf{lmem}}
\newrobustcmd{\translClosed}[1]{\transl #1_{\textsf{cl}}}
\newrobustcmd{\EmitLabel}[2]{#1 \oplus #2}

\newrobustcmd{\IPAwb}{\kl[wbIPA]{\approx}}
\newrobustcmd{\Stratwb}{\kl[stratWB]{\approx}}
\newrobustcmd{\Stratwbb}{\kl[stratWBB]{\approx_0}}

\newrobustcmd{\EmptyType}{\kl[emptytype]{\textbf 1}}
\newrobustcmd{\COne}{\kl[emptytype]{\textbf 1}}
\newcommand{\CPar}{\parallel}
\newcommand{\CWith}[2]{\with_{#1} #2}
\newcommand{\CPlus}[2]{\oplus_{#1} #2}
\newcommand {\CBang} {\oc}
\newcommand {\CWN} {\wn}

\newcommand{\PatternVar}{\vec x}
\newcommand{\PatternVari}{\vec y}

\newcommand {\CFork}{\parallel}
\newcommand {\CZero} {\textbf 0}
\newcommand {\CBranching}[3] {#1 \with \{#3\}_{#2}}
\newcommand {\CBranchingBig}[2] {#1 \with \left\{\begin{aligned}#2\end{aligned}\right\}}
\newcommand {\CBranchingOne}[3] {#1 \with #2(#3).\, }
\newcommand {\CBranchingOneS}[2] {#1 \with #2.\, }
\newcommand {\CBranch}[2] {#1(#2).\,}
\newcommand {\CBranchS}[1] {#1.\,}
\newcommand {\CBranchBig}[3] {&#1[#2].\, #3\\}

\newcommand{\CPositiveAction}[3]{#1 \oplus #2[#3]}
\newcommand{\CPositiveActionS}[2]{#1 \oplus #2}
\newcommand {\CSelect}[3] {\CPositiveAction {#1}{#2}{#3}.\, }
\newcommand {\CSelectS}[3] {\CPositiveActionS {#1}{#2}.\, #3}
\newcommand {\CProm} [2] {\oc #1(#2).\, }
\newcommand {\CCoder} [2] {\mathtt{\#}#1(#2).\, }
\newcommand {\CDer} [2] {\wn#1[#2].\, }
\newcommand {\CCut} [3] {(\nu #1 #2) #3}
\newcommand {\fv} [1] {\textsf{fv}(#1)}

\newcommand{\ContextElaboration}[2]{#1 :: #2}

\newcommand{\BoundVars}[1]{\textsf{bv}(#1)}

\newcommand{\Red}{\to}
\newcommand{\HasBarb}[2]{#1 \Downarrow #2}

\newcommand{\GetRequest}{\textsf{get}}
\newcommand{\SetRequest}[1]{\textsf{set}(#1)}
\newcommand{\GetAnswer}[1]{\Return {#1}}
\newcommand{\SetAnswer}{\Return {}}

\newcommand{\Write}[2]{\mathsf{w}\ENCan{#1,#2}}
\newcommand{\Read}[2]{\mathsf{r}\ENCan{#1,#2}}

\DeclarePairedDelimiter\TranslateTerm{\llparenthesis}{\rrparenthesis}

\newcommand{\Constructors}[1]{\kl[Constructor]{\mathfrak C}(#1)}
\newcommand{\ConstructorOfValue}[2]{\kl[ConstructorOfValue]{\mathfrak C}(#1, #2)}

\newcommand{\Request}{\texttt{Req}}
\newcommand{\Call}[1]{\texttt{Call}(#1)}
\newcommand{\Return}[1]{\texttt{Ret}(#1)}

\newcommand{\machine}[4]{{#2}\vdash^{#1}\!\langle #3, #4 \rangle}
\newcommand{\ENCan}[1]{\langle{#1}\rangle}
\newcommand{\machineDefault}[1]{\machine{\Name}{\Delta}{#1}{\mu}}
\newcommand{\machineDefaultt}[1]{\machine{\Name'}{\Delta'}{{#1}}{\mu'}}

\newcommand{\inferruleR}[3][]{
  \infer[\!\!{#1}]{#2}{#3}%
}

\newcommand{\rulename}[1]{\text{\scriptsize[\textsc{#1}]}}

\newcommand{\num}[1]{\underline{#1}}

\newcommand{\mysubsubsection}[1]{\subsubsection*{\bf{#1}}}
\newcommand{\Name}{\vec{y}}

\newcommand{\Program}{{\sf\textit{{Program}}}\xspace}
\newcommand{\Context}{{\sf\textit{{Context}}}\xspace}
\newcommand{\Player}{{\sf\textit{Player}}\xspace}

\makeatletter
\newcommand{\xMapsto}[2][]{\ext@arrow 0599{\Mapstofill@}{#1}{#2}}
\def\Mapstofill@{\arrowfill@{\Mapstochar\Relbar}\Relbar\Rightarrow}
\makeatother

\newcommand{\by}[1]{\stackrel{#1}{\longrightarrow}}
\newcommand{\By}[1]{\stackrel{#1}{\Longrightarrow}}

\newcommand{\Nf}[1]{\textsf{nf}(#1)}
\definecolor{cleargreen}{HTML}{CCFFCC}
\definecolor{fgcleargreen}{HTML}{006600}

\definecolor{clearred}{HTML}{ffe1d1}
\definecolor{fgclearred}{HTML}{66280a}

\definecolor{clearblue}{HTML}{ccccff}
\definecolor{fgclearblue}{HTML}{000066}
\newcommand{\colorabg}[1]{\text{\colorbox{cleargreen}{\textcolor{fgcleargreen}{$#1$}}}}
\newcommand{\colorbbg}[1]{\text{\colorbox{clearred}{\textcolor{fgclearred}{$#1$}}}}

\newcommand{\embedding}[1] {
  \stackrel{#1}{\kl[embedding]{\hookrightarrow}}}


%% file: know.tex
\usepackage[notion,composition,hyperref,quotation,xcolor]{knowledge}

\hypersetup{final}

\knowledgestyle{intro}{}
\definecolor{linkcolor}{rgb}{0,.2,.56}

\IfKnowledgePaperModeTF
   {
    \knowledgestyle{intro notion}{boldface}
    \knowledgestyle{notion}{color}
   }{}
\IfKnowledgeCompositionModeTF
   {
    \knowledgestyle{intro notion}{boldface,color=linkcolor}
    \knowledgestyle{notion}{color=linkcolor}
   }{}
\IfKnowledgeElectronicModeTF
   {
    \knowledgestyle{intro notion}{boldface,color=linkcolor}
    \knowledgestyle{notion}{color=linkcolor}
   }{}

\knowledge {configuration}[configurations]{notion}
\knowledge {game}[games]{notion}
\knowledge {strategy}[strategies|deterministic strategy|Strategy|Strategies]{notion}
\knowledge {isomorphism family}{notion}
\knowledge {event structure with symmetry}[event structures with symmetry]{notion}
\knowledge {event structure}[event structures]{notion}
\knowledge {compatible}{notion}
\knowledge {concurrent}{notion}
\knowledge {app}{notion}
\knowledge {appnl}{notion}
\knowledge {weak bisimulation (strat)}{notion}
\knowledge {weak bisimulation (ipa)}{notion}
\knowledge {applymacro}{notion}
\knowledge {apply}{notion}
\knowledge {extension}[extend|extends|extended]{notion}
\knowledge {essential strategy}[essential strategies]{notion}
\knowledge {program}[programs]{notion}
\knowledge {immediate causal dependency}{notion}
\knowledge {minimal conflict}{notion}
\knowledge {confusion-free}{notion}
\knowledge {cells}[cell]{notion}
\knowledge {race-free}{notion}
\knowledge {dual}{notion}
\knowledge {parallel composition}{notion}
\knowledge {nondeterministic sum}[sum]{notion}
\knowledge {copy index}[copy indices]{notion}
\knowledge {partial map of event structures}[map|partial map|maps|partial maps]{notion}
\knowledge {copycat strategy}[copycat]{notion}
\knowledge {projection}[hiding]{notion}
\knowledge {interaction}{notion}
\knowledge {mapess}[map of event structures with symmetry]{notion}
\knowledge {similar}{notion}
\knowledge {local}{notion}
\knowledge {thin concurrent game}[tcg|thin concurrent games|tcgs|Thin concurrent games]{notion}
\knowledge {$\sim$-game}[$\sim$-games|game with symmetry|games with symmetry|gws]{notion}
\knowledge {localess}[local event structure with symmetry]{notion}
\knowledge {symmetric}{notion}
\knowledge {uniformity witness}{notion}
\knowledge {$\sim$-strategy}[$\sim$-strategies]{notion}

\knowledge {projection}[hiding]{notion}
\knowledge {exponential comonad}{notion}
\knowledge {interaction}{notion}
\knowledge {Composition}[composition]{notion}

\knowledge {weakly isomorphic}[weak isomorphism]{notion}
\knowledge {internal}[neutral|invisible]{notion}
\knowledge {external}[visible]{notion}

\knowledge {map of games}[maps of games]{notion}
\knowledge {map of $\sim$-game}[maps of $\sim$-games]{notion}
\knowledge {lifting maps}[lifting of maps]{notion}
\knowledge {lifting maps with symmetry}{notion}
\knowledge {\LinStrat}{notion}
\knowledge {\Strat}{notion}

\knowledge {forest}{notion}
\knowledge {ESS}{notion}

\knowledge {sharpG}{notion}
\knowledge {ocG}{notion}
\knowledge {wnG}{notion}
\knowledge {dereliction}{notion}
\knowledge {digging}{notion}
\knowledge {contraction}{notion}

\knowledge {weakening}{notion}
\knowledge {pairing}{notion}
\knowledge {copycatS}{notion}
\knowledge {ocS}{notion}
\knowledge {codereliction}{notion}
\knowledge{branching}{notion}
\knowledge{promotion}{notion}
\knowledge {cocontraction}{notion}
\knowledge {fc}{notion}
\knowledge {similarProc}{notion,mathrel}
\knowledge {fw}{notion}
\knowledge {sumProc}{notion}

\knowledge {emptyes}{notion}
\knowledge {dom}{notion}
\knowledge {ES}{notion}
\knowledge {emptygame}{notion}
\knowledge {total map of event structures}[total|total map]{notion}
\knowledge {domain}{notion}
\knowledge {deterministic}{notion}
\knowledge {\LinDetStrat}{notion}
\knowledge {\DetStrat }{notion}
\knowledge {Strat}{notion}
\knowledge {UStrat}{notion}
\knowledge {emptytype}{notion}
\knowledge {isoProc}{notion,mathrel}
\knowledge {isoStrat}{notion,mathrel}
\knowledge {isomorphism}[isomorphic|isoStrat]{notion}
\knowledge {eqProc}{notion,mathrel}

\knowledge {negative contexts}[negative context]{notion}
\knowledge {local injectivity}{notion}
\knowledge {action}[actCtx]{notion}
\knowledge {encpairs}[encopairs]{notion}
\knowledge {depthT}{notion}
\knowledge {cut complexity}{notion}

\knowledge {open process}{notion}
\knowledge {actions}[action]{notion}
\knowledge {positive action}{notion}
\knowledge {negative action}{notion}
\knowledge {naming action}{notion}
\knowledge {liftplus}{notion}
\knowledge {liftneg}{notion}
\knowledge {polarised}{notion}
\knowledge {belowProc}{notion,mathrel}
\knowledge {belowStrat}[belowstrat]{notion,mathrel}
\knowledge {negative game}[negative games]{notion}
\knowledge {internal configuration}{notion}
\knowledge {positive game}[positive games]{notion}
\knowledge {negative strategy}[negative strategies]{notion}

\knowledge {uwe}{notion}
\knowledge {wbIPA}{notion}
\knowledge {conflict}{notion}
\knowledge {stratWB}{notion}
\knowledge {names}{notion}
\knowledge {$\sim$-essential strategy}[$\sim$-essential strategies]{notion}

\knowledge {Definitions}{notion}

\knowledge{CML}{notion}
\knowledge{base type}{notion}
\knowledge{semiclosed}{notion}
\knowledge{constants}{notion}
\knowledge{cmlbisim}{notion,mathrel}
\knowledge{cmlobs}{notion,mathrel}
\knowledge{mlobs}{notion,mathrel}
\knowledge{first-order}{notion}
\knowledge{Constructor}{notion}
\knowledge{stuck}{notion}
\knowledge{ConstructorOfValue}{notion}
\knowledge{rooted}{notion}
\knowledge{interface}{notion}
\knowledge{essential}[essential event|essential events]{notion}
\knowledge{embedding}[embeddings]{notion}
\knowledge{uniform strategy}[uniform strategies]{notion}
\knowledge{receptivity}[Receptivity|receptive]{notion}
\knowledge{Courtesy}[courtesy]{notion}
\knowledge{Secrecy}[secrecy]{notion}
\knowledge{weak bisimilarity}{notion}

%% file: abstract.tex
Game semantics has proven to be a robust method to give compositional
semantics for a variety of higher-order programming languages.
However, due to the complexity  
of most game models, game semantics has remained unapproachable for 
non-experts.

In this paper, we aim at making game semantics more accessible by
viewing it as a syntactic translation into a session typed
$\pi$-calculus, referred to as \emph{metalanguage}, followed by a
semantics interpretation of the metalanguage into a particular game
model. The syntactic translation can be defined for a wide range of
programming languages without knowledge of the particular game model
used.  Simple reasoning on the model (soundness, and adequacy) can be
done at the level of the metalanguage, escaping tedious technical
proofs usually found in game semantics. We call this methodology
\emph{programming game semantics}.

We design a metalanguage ($\MetalanguageName$) inspired from
Differential Linear Logic ($\DiLL$), which is concise but expressive
enough to support features required by concurrent game semantics.  We
then demonstrate our methodology by yielding the first causal,
non-angelic and interactive game model of $\CML$, a higher-order
call-by-value language with shared memory concurrency.  We translate
$\CML$ into $\MetalanguageName$ and show that the translation is
adequate. We give a causal and non-angelic game semantics model using
event structures, which supports a simple semantics interpretation of
$\MetalanguageName$. Combining both of these results, we obtain the first
interactive model of a concurrent language of this expressivity which
is adequate with respect to the standard weak bisimulation, and fully
abstract for the contextual equivalence on second-order terms.

We have implemented a prototype which can explore the generated causal
object from a subset of OCaml.


%% file: introduction.tex
\mysubsubsection{Background} Reasoning about programs requires a
mathematical model of their execution, called semantics. A popular
technique is \emph{operational semantics} which models the concrete
execution on the metal by an abstract machine made of mathematical
symbols and Greek letters. Its popularity is due to its simplicity and
flexibility, modelling a wide range of
programming features. However, operational semantics only gives meaning
to closed programs of ground types. \emph{Open higher-order} programs,
ie.  programs with external functional parameters are left aside. This
creates difficulties, for instance, comparing such programs can only be
done through an untractable quantification over arbitrary contexts.
Denotational semantics tackles this problem by trying to give
\emph{compositional semantics}, usually as a function from its
external parameters to the result. The most common form of
denotational semantics, based on domain theory, copes well with
higher-order functions, but more laboriously with effectful
computations, and almost not at all with concurrency.

At the dawn of the 1990s, a new form of denotational semantics
appeared: \emph{game semantics}
\cite{Hyland00onfull,abramsky2000full}. There, an open program is
modelled by its possible \textbf{interactions} with the
context. Thanks to its interactivity, this methodology has proved to
be very extensible and easily supports a wide range of programming
features (references, control operators, probabilities, concurrency,
quantum, etc.)

Moreover, as observed by \citet{DBLP:journals/entcs/GhicaT12}, game semantics
reconciles denotational and operational semantics together: the interaction
traces of a program can be computed either denotationally (by induction on the
syntax) or operationally (by running an abstract machine).

While game semantics has been recognised as a powerful tool to build 
denotational models, we believe that its most useful and promising
feature is the simplicity with which one can
\emph{describe} the compositional
behaviour of systems in general (eg. with its recent use for verifying
compilers
\cite{DBLP:conf/popl/StewartBCA15,shao-refinement-games} 
or operating
systems \cite{Gu:2018:CCA:3192366.3192381}). So far, its simplicity has not been
apparent, and game semantics is often considered inaccessible to non-experts who
wish to define a denotational model based on games for their favourite language.

\mysubsubsection{The Framework} 
The thesis of this paper is that, the complexity of game semantics
interpretations can be decomposed by introducing a simple
message-passing intermediate language, between the source program and
the model.
Indeed, game semantics bundles two ideas together: the idea of
interpreting a program as a process interacting with its environment,
\emph{and} a semantic interpretation of these processes. 
%
%
Our main contribution in this paper is to make this separation explicit by
factorising the interpretation of a language in game semantics as the following
steps:
\[
\begin{tikzpicture}
  \node (p) [draw, rectangle, text width=2.4cm] { \quad \ {\bf Program} \\ {\small (Source language)} } ;
  \node (pr) [draw, rectangle, right=2.5cm of p, text width=2cm] { \quad {\bf Process
} \\ {\small (Metalanguage)} } ;
  \node (ob) [draw, rectangle, right=2.5cm of pr, text width=2cm] {\quad {\bf Strategy} \\ {\small (Game Model)} } ;
  \draw[->] (p) edge node[above] {syntactic} node[below] {translation} (pr);
  \draw[->] (pr) edge node[above] {semantics} node[below] { interpretation } (ob);
\end{tikzpicture}
\]

\noindent
This factorisation offers several advantages. First, it allows to
decouple the interpretation of the source language from the details of
the model. One can use the same translation to obtain different models
(eg. traces, event structures, LTSs, ...), and, conversely, one can
interpret a language becomes a matter of writing a syntactic
translation, an easier task than doing the interpretation from a
language to a game model directly, and which becomes model-agnostic.

\mysubsubsection{A Language for Strategies} What would a good
intermediate language for strategies be? Strategies represent the
interaction of the program with the environment in the form of
messages that are exchanged between them.
For that reason, a message-passing language such as the $\pi$-calculus
\cite{DBLP:journals/iandc/MilnerPW92a} is an ideal candidate. 
Dating back to encodings of the call-by-name and call-by-value
$\lambda$-calculi by \citet{DBLP:journals/mscs/Milner92}, the
$\pi$-calculus has been used to encode a wide range of programming
languages, including functional, concurrent, and distributed
languages. Its connection with game semantics has been studied ever
since the introduction of game semantics
\cite{DBLP:conf/fpca/HylandO95}.
Some game semantics interpretations rely on the use of the categorical
semantics of Linear Logic \cite{mellies2008categorical}.  Recently,
\citet{DBLP:conf/concur/CairesP10} have discovered a Curry-Howard
correspondence between the $\pi$-calculus and Linear Logic through
\emph{session types}.  Specifically, this paper uses an extension of
the calculus in \cite{DBLP:journals/jfp/Wadler14} to Differential
Linear Logic ($\DiLL$) \citep{DiLL_tutorial} where $\otimes$ and
$\parr$ are identified to be able to interpret languages with
deadlocks. Session types are a natural fit here since their connection
with game semantics have been recently discovered~\cite{POPL19} (in a
prototypical setting). Our metalanguage is going beyond other session
type-based calculi to be able to express game semantics
interpretations.

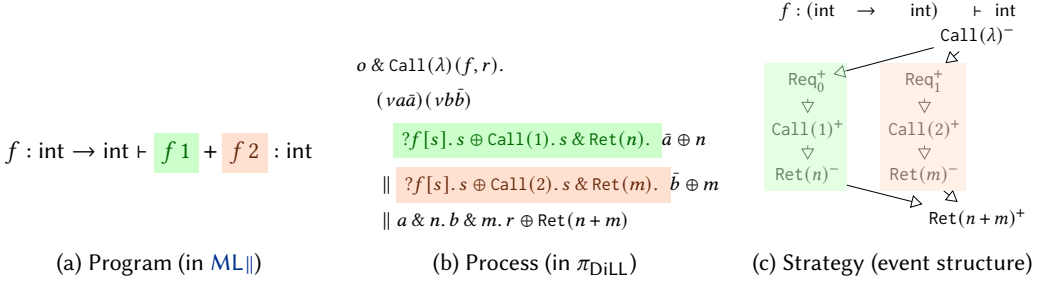
\begin{figure}\small
  \subcaptionbox{\label{fig:over_IPA}Program (in $\CML$)} {
    $f:  \NatType   \rightarrow  \NatType  \vdash  \colorabg{f\, 1}  + \colorbbg{f\, 2} : \NatType$
    \bigskip
    \bigskip
    \bigskip
  }
  \quad
  \subcaptionbox{\label{fig:over_IPA}Process (in $\MetalanguageName$)} {
    \scriptsize
    $
    \begin{aligned}
      &\CBranchingOne o {\Call{ \lambda }} {f, r} \\
      &\quad\CCut {a \bar a} {} \CCut {b \bar b} {}\\
      &\quad\quad\colorabg{\CDer f s \CSelectS s {\Call 1} \CBranchingOneS s {\Return{n}}} \CPositiveActionS {\bar a} {n} \\
      &\quad\parallel\colorbbg{\CDer f s \CSelectS s {\Call 2} \CBranchingOneS s {\Return{m}}} \CPositiveActionS {\bar b} {m} \\
      &\quad\parallel\CBranchingOneS a {n} \CBranchingOneS b {m} \CPositiveActionS r {\Return {n+m}}
    \end{aligned}$
  }\quad
  \subcaptionbox{\label{fig:over_IPA}Strategy (event structure)} {
    \scalebox{0.8} {
      \begin{tikzpicture}
        \matrix[diag, column sep=0,row sep=0.25cm] {
          f:(\NatType \&  \rightarrow  \& \NatType) \&   \vdash   \& \NatType \\[-2mm]
          \& \& \& |(aa)|  \\[1mm]
          |(r1)| \Request_0^+ \& \& |(r2)| \Request_1^+ \\
          |(c1)| \Call 1^+ \& \& |(c2)| \Call 2^+\\
          |(re1)| \Return n^- \& \& |(re2)| \Return m^-\\[1mm]
          \& \& \& |(rr)| \\[-2mm]
        };
        \node (a) at (aa) { $\Call{\lambda}^-$} ;
        \node (r) at (rr) { $\Return{n+m}^+$} ;
        \path[cause]
        (a) edge (r1) (a) edge (r2)
        (r1) edge (c1) (c1) edge (re1) (re1) edge (r)
        (r2) edge (c2) (c2) edge (re2) (re2) edge (r)
        ;
        \fill [fill=cleargreen, opacity=0.5]
        ($(r1.north west)+(-0.3,0)$) rectangle (re1.south east);
        \fill [fill=clearred, opacity=0.5]
        ($(r2.north west)+(-0.3,0)$) rectangle (re2.south east);
      \end{tikzpicture}
    }
  }
  \qquad
  \vspace{-2mm}
  \caption{An overview of the methodology on an example}
  \label{fig:overview}
  \vspace{-5mm}
\end{figure}
\mysubsubsection{Contributions and Outline of the Paper} 
This paper proposes a uniform framework where we view game semantics
interpretations as syntactic translations to a process calculus,
followed by a semantic interpretation of this calculus into a game
model.

We demonstrate this methodology using one translation and one semantics
interpretation: we translate $\CML$, which is a mini ML extended with
shared memory concurrency into the metalanguage, $\MetalanguageName$
based on Differential Linear Logic.\footnote{It differs from
  $\pi\mathtt{DILL}$ (dual intuitionistic linear logic) in
  \cite{DBLP:conf/concur/CairesP10}.} Our semantic interpretation of
$\MetalanguageName$ is based on an extension of concurrent games
\cite{lics11}, that deals with internal actions and
replication. Combining the syntactic translation and semantics
interpretation gives us the first causal and interactive
interpretation of a concurrent call-by-value language, which is
adequate -- and fully-abstract for second-order interfaces -- for a
strong notion of contextual equivalence. This bridges the gap with
real-world programming languages, as our implementation of the model
allows to explore the causal behaviour of concurrent programs written
in (a subset of) OCaml.

The methodology is illustrated in Figure~\ref{fig:overview} on a
program calling an external function $f$ twice in parallel, as function
parameters are executed in parallel in this language. This program
is translated to the $\pi$-calculus which is then interpreted as an
event structure describing the causal relationships between the
different actions of the program (see~\S~\ref{sec:interp ML} and
\S~\ref{sec:games}).

Our paper is structured as follows: \textbf{\S~\ref{sec:overview}}
presents an overview of the paper and introduces the source language
$\CML$; \textbf{\S~\ref{sec:calculus}} presents our metalanguage
$\MetalanguageName$ and its equational theory;
\textbf{\S~\ref{sec:interp ML}} presents the translation of $\CML$
into $\MetalanguageName$; \textbf{\S~\ref{sec:games}} presents the
causal model of the metalanguage; \textbf{\S~\ref{sec:interp}}
presents the interpretation of $\MetalanguageName$ into the model;
\textbf{\S~\ref{sec:implementation}} outlines our prototype
implementation of a causal interpretation of a subset of OCaml.
\textbf{\S~\ref{sec:related}} provides related work
and \textbf{\S~\ref{sec:conclusion}} concludes with future work. 
work. Proofs can be found in \emph{\textbf{Appendix}}, and we shall
submit our prototype implementation to \emph{\textbf{Artifact
    Evaluation}}.

We make use of the \texttt{knowledge} package.  ""Definitions"" of
mathematical concepts appear blue boldface, and their "uses@Definitions"
occur in blue and are linked to the original definition.




%% file: overview.tex
In this section, we informally introduce our methodology. In
particular, we explain basic ideas behind game semantics
interpretations, and their syntactic counterpart in our
metalanguage. This section paves the way to the formal definition of
$\MetalanguageName$ in \S~\ref{sec:calculus}. In \S~\ref{sec:CML}, we
first introduce our source language, a concurrent call-by-value
language, $\CML$. We then illustrate different aspects of the language
and the techniques we use to represent them: \S~\ref{sec:first-order}
focuses on first-order terms; \S~\ref{sec:higher-order} on
higher-order behaviour; and finally \S~\ref{sec:memory} discusses
shared memory.

\subsection{$\intro*\CML$: Mini-ML with shared-memory concurrency}\label{sec:CML} We use a concurrent call-by-value language with integer
references, called $\CML$:
\[
\begin{array}{lrcl}
\text{type} &  \sigma ,  \tau  & ::= & \UnitType \mid \NatType \mid \BoolType \mid \RefType \mid  \sigma   \rightarrow   \tau  \quad\quad 
  \text{value} \quad V  \ ::=  \ x \mid \lambda x.\, M \mid \num n \mid \ttrue \mid \ffalse \mid ()\\[1mm]  
\text{term} &  M, N & ::= & V \mid  M\, N \mid Y \mid \bot 
         \mid \IfTerm M N {N'} \mid \mathtt{plus} \mid \mathtt{equal} 
         \mid  \mathtt{get} \mid \mathtt{set} \mid \RefAlloc 
\end{array}
\]
A ""base type"" is either the unit type written $\UnitType$, $\NatType$ or $\BoolType$.  $Y$
is the fixpoint operator.  We use standard syntactic sugar:
$\LetIn x M N$ for $( \lambda x.\, N)\, M$; $\LetIn {f\, x} M N$ for
$\LetIn f { \lambda x.\, M} N$. Note that our language does not
include products or sum types for presentation purposes. It is
straightforward to extend our approach to those, and our
implementation supports them.

Instead of having specific constructs (eg. $M :=
N$), we chose to have higher-order constants which makes the
metatheory simpler. Hence, we write $M + N$ for $\mathtt{plus}\, M\,
N$; $M = N$ for $\mathtt{equal}\, M\, N$; $\oc M$ for $\texttt{get}\,
M$; $M := N$ for $\mathtt{set}\, M\,
N$. We have not included any sequential or parallel composition
operators in the language as it is definable in the language: $M;
N$ is a shorthand for $( \lambda x.\, N)\, M$ when
$x$ not occurring free in $N$ and $M \parallel
N$ is a shorthand for $( \lambda xy.\, ())\, M\,
N$: this behaves as expected due to our semantics of application,
which evaluates arguments in parallel.

Typing of this language is standard, given by a judgement
$ \Delta \vdash M : \sigma $ where
$ \Delta = x_1: \sigma _1, \ldots , x_n : \sigma _n$ and is shown in
Figure~\ref{fig:cml typing}. The terms \texttt{equal}, \texttt{plus},
\texttt{get}, \texttt{set}, and $\RefAlloc$ are called
""constants"". An ""interface"" is a pair of a context $ \Delta $ and a type
$ \sigma $, written $ \Delta \vdash \sigma $.

\input{figures/cml_typing.tex}

\mysubsubsection{Operational Semantics} As it is customary for such
languages mixing higher-order and shared state, we define the
operational semantics in two steps: we first define a confluent
reduction relation $ \rightarrow $ on all terms, which only reduces
functional redexes, leaving reference operations unchanged. The rules
for this reduction relation are standard and given in Figure
\ref{fig:cml:pure op}, where we use the following evaluation contexts:
$E ::= [\ ] \mid E\, M \mid M\, E\mid \IfTerm E M N$. This choice of
contexts implies in particular that in $M\, N$ the evaluation of $M$ and $N$ are
done \emph{in parallel}: This allows for operators that evaluate
their arguments in parallel; e.g.  $M + N$, which unfolds to
$\mathtt{plus}\, M\, N$, evaluates $M$ and $N$ in
parallel. Sequentiality is only ensured between the body of a function
and its argument: in $(\lambda x.\, M)\, N$, $N$ is always evaluated
first. Since $M \parallel N$ is simply defined as $(\lambda xy. ())\, M\, N$,
this also guarantees that $M$ and $N$ are evaluated in parallel.

In a second step, we look at terms of the form
$ \Delta \vdash M : \sigma $ where (1) $ \sigma $ is not a functional
type; and $ \Delta (a) = \RefType$ for all $a \in \dom \Delta$. Such terms
are called ""semiclosed"", and are going to be executed by a machine
whose states are tuples $\machine \Name \Delta M \mu $ where
$ \Delta \vdash M : \sigma $ is "semiclosed", $\Name$ is a subset of
$\dom \Delta$, the \emph{public} locations; and
$\mu : \dom \Delta \rightarrow \mathbb{N}$ is the memory state,
mapping locations to values. The type $ \sigma $ is the type of the machine
$\machineDefault M$.

We chose to give the operational semantics of such machine states
under the form of a labelled transition system. The visible actions
will correspond to memory actions described by
$ \Sigma _{\CML} ::= \Read x k \mid \Write x k$ where $x$ is a free
reference variable and $k \in \mathbb{N}$ is a value: $\Read x k$
means the program has read value $k$ from $x$, and $\Write x k$ that
it has written $k$ to $x$. The transitions of the LTS are labelled
over the set $ \Sigma _{\CML} \cup \{ \tau \}$, which is ranged over
by $ \alpha$, $\beta$. The silent action $ \tau $ is often omitted.
The LTS of $\CML$ is given in Figure~\ref{fig:cml:lts}.  Note that
$\Name$ stays unchanged through out the reduction.

\begin{lemma}[Subject Reduction]
  {\rm (1)} If $ \Delta   \vdash  M :  \sigma$
  and $M \Red N$, then
  $ \Delta   \vdash  N :  \sigma$; and {\rm (2)} 
  if $ \Delta   \vdash  M :   \sigma $ is "semiclosed"
  and $\machineDefault M  \by{\alpha}
  \machine {\Name}{\Delta'}{M'}{\mu'}$ then $ \Delta '  \vdash  M :   \sigma $.
\end{lemma}
\input{figures/cml_op.tex}

\mysubsubsection{Weak bisimulation and observation equivalence.}  We define the standard weak
bisimulation relation between two machines as equivalence on base
terms. By the LTS rules in Figure~\ref{fig:opsem CML}, weak
transitions are defined as expected: we write $\By{}$ for the
reflexive, transitive closure of $\by{\tau}$, $\By{~\alpha~}$ for
transitions $\By{}\by{~\alpha~}\By{}$, and $\By{\hat{~\alpha~}}$ for
$\By{\alpha}$ if $\alpha\not = \tau$ and $\By{}$ otherwise.

\begin{definition}[Weak bisimulation of $\CML$ machines]\label{def:wb_cml}
  A relation $\mathcal{R}$ over machines of type $\UnitType$ is a \emph{weak bisimulation} if
  \begin{enumerate}
   \item
     for all
     $\machine{\Name} {\Delta_1} {M_1} {\mu_1} \ \mathcal{R}
     \ \machine {\Name}{\Delta_2}{M_2}{\mu_2}$,
    whenever
    $\machine{\Name} {\Delta_1} {M_1} {\mu_1} \by{\alpha} 
    \machine{\Name} {\Delta_1'} {M_1'} {\mu_1'}$, there exists  
    $\machine{\Name}{\Delta_2}{M_2}{\mu_2} \By{\hat{\alpha}}
    \machine{\Name}{\Delta_2'}{M_2'}{\mu_2'}$ such that 
    $\machine{\Name} {\Delta_1'} {M_1'}{\mu_1'} \ \mathcal{R}
     \ \machine {\Name}{\Delta_2'}{M_2'}{\mu_2'}$.
    \item The symmetric case of 1. 
  \end{enumerate}
  The largest such bisimulation is called weak bisimilarity, denoted
  by $\intro*\CMLbisim$.
\end{definition}
We define the \emph{observational equivalence} on terms by quantifying
over machine contexts.

\begin{definition}[Observational equivalence of $\CML$]\label{def:cml:obs}
  Given $ \Delta \vdash M, N : \sigma$, $M$ and $N$ are
  observationally equivalent, written $M \intro*{\CMLObs} N$ when for
  all machine contexts $C$ of $\UnitType$ type (that is a machine of
  $\UnitType$ type with a hole in the term component),
  $C[M] \CMLbisim C[N]$.
\end{definition}

\subsection{First-order programs}\label{sec:first-order}
\label{sec:interactive-semantics}
We start our investigation of the interactive semantics of $\CML$ with
the first-order case. An interface $ \Delta \vdash \sigma $ is
""first-order"" when $ \sigma $ is a base type and $ \Delta $ only
contains base types. In call-by-value, such terms have a limited
interaction with their environment, as variables stand for values:
waiting for the values of its parameters from the environment,
computing the result, and finally possibly returning it to the
environment (or diverging).

For each type $ \sigma $, we define a set $\intro*{\Constructors { \sigma }}$ of messages
describing values of type $ \sigma $. For base types, $\Constructors{ \sigma }$ is
exactly the set of values of $ \sigma $:
$$\Constructors{\NatType} =  \mathbb{N}  \qquad \Constructors{\UnitType} = \{ () \} \qquad \Constructors {\BoolType} = \{\ttrue, \ffalse\}.$$
For non-base types, as we will see in the next section, this
correspondence will not hold. This extends to context by letting
$\Constructors \Delta$ be the set of tuples
$\prod_{a \in \dom { \Delta }} \Constructors{ \Delta (a)}$. So, the
protocol of a first-order interface $ \Delta \vdash \sigma $ is very
simple: $\Program$ expects a tuple of constructors for its parameters
(an element of $\Constructors \Delta $) and then may return its
result, a constructor of $ \sigma $.

Unsurprisingly, our language of session types must then include type
constructors representing communication from $\Context$ to $\Program$
and from $\Program$ to $\Context$. Since there are only two
participants, we choose to describe the protocol from the point of
view of $\Program$: the first kind of messages will be seen as
\emph{inputs} (and be assigned a negative polarity) while the second
kind will be seen as \emph{outputs} (and assigned a positive
polarity).

We use here a dialect of session types inspired by Linear Logic, and
this data flow is captured by the \emph{additive connectives}:
$\with$ (input) and $\oplus$ (output). Concretely, the translation of
a "first-order" interface $ \Delta   \vdash   \sigma $ gives the following protocol:
$\TranslateTerm { \Delta   \vdash   \sigma } = \with_{i\in \Constructors{ \Delta }} \Call i^- \cdot
\oplus_{o  \in  \Constructors{ \sigma }} \Return o^+$: $\Program$ must input a
message $\Call i$ containing the parameter values and may output a
message $\Return o$ with its return value.

The interpretation of a first-order type is simply the process that
waits for the input parameter, computes the value and returns it. This
process is parametrised by a name~$o$ representing the communication
channel with $\Context$. For instance, the negation is:
$$\TranslateTerm{x:\BoolType  \vdash \mathtt{neg} \eqdef \IfTerm x \ffalse \ttrue : \BoolType}_o
= \CBranching o {}
{\CBranch {\Call \ttrue} r \CPositiveActionS r {\Return \ffalse};\ \ \CBranch {\Call \ffalse} r \CPositiveActionS r {\Return \ttrue}}
.$$

It starts by an input on $o$ and there are two branches since we can
receive two possible values. Along with the value received, there is
also a new name $(r)$ which is used for the continuation of the
protocol: in this case, returning the result. (We will see soon that
in this particular case it is possible to reuse the name $o$, as is
customary in session types, but in general it is not possible.)

\begin{figure}
    \small
  \centering
  \begin{tikzpicture}[baseline]
    \matrix[diag] {
      \& \& |(x)|  \\[-2mm]
      \& |(a)| \Call{\ttrue}^- \& \& |(b)| \Call{\ffalse}^- \\
      |(a1)| \Return{\ttrue}^+ \& |(a2)| \Return{\ffalse}^+ \& \& |(b1)| \Return{\ttrue}^+ \& |(b2)| \Return{\ffalse}^+ \\
    };
    \path [cause]
    (a) edge (a1) (a) edge (a2)
    (b) edge (b1) (b) edge (b2);
    \path
    (a) edge[conflict] (b)
    (a1) edge[conflict] (a2)
    (b1) edge[conflict] (b2);
    \node at (x) { $\TranslateTerm{x:\BoolType  \vdash  \BoolType}$};
  \end{tikzpicture}
  \quad  \quad  \quad  \quad  \quad
  \begin{tikzpicture}[baseline]
    \matrix[diag] {
      \&  |(x)|  \\[-2mm]
      |(a)| \Call{\ttrue}^- \& \& |(b)| \Call{\ffalse}^- \\
      |(a2)| \Return{\ffalse}^+ \& \& |(b1)| \Return{\ttrue}^+ \\
    };
    \path [cause]
    (a) edge (a2)
    (b) edge (b1);
    \path
    (a) edge[conflict] (b);
    \node at (x) { $\TranslateTerm{\mathtt{neg}}$};
  \end{tikzpicture}
  \vspace{-3mm}
  \caption{Semantics interpretation of the negation: game (left) and strategy (right)}
    \vspace{-5mm}
  \label{fig:ex:negation}
\end{figure}
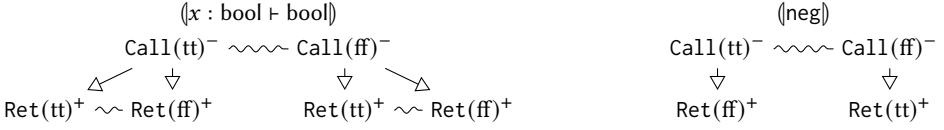

In the semantic world, the session type
$\TranslateTerm { \Delta \vdash \sigma }$ becomes a \textbf{game}
(where moves are messages), and the process
$\TranslateTerm { \Delta \vdash M : \sigma }$ becomes a
\textbf{strategy}, which in our setting explicits the causal
relationships between the different actions of the program (in $\CML$:
memory operations, calls and returns). Both games and strategies are described by the
same mathematical object: an \textbf{event structure}, which is a
partial order (written $ \rightarrowtriangle $) along with a conflict
relation ($\minconflict$). In the example of the negation, we have
drawn both the game and the strategy in Figure
\ref{fig:ex:negation}. Conflict represents incompatibility within an
execution: two events in conflict will occur in different
executions. In games, this represents the fact that the protocol forbids to
play these two moves in the same execution, corresponding to choices
($\with$ and $\oplus$) in the syntax of session types: $\Context$
has to choose a set of parameters, and $\Player$ a return value. Note
that in this simple case, the strategy is simply a subset of the game.

\subsection{Higher-order programs}
\label{sec:higher-order}
This might seem a lot of complexity to capture simple input/output
behaviours. However, it becomes justified when looking at
higher-order programs where the control jumps between $\Program$ and
$\Context$ due to calling external functions, i.e.. functions of the
interface. Let us start with a simple (second-order) example
$f: \UnitType \rightarrow \UnitType \vdash M := f (); f () :
\UnitType$. As before, $\Context$ starts by sending a constructor for
each parameter. But what should be a constructor for the type
$\UnitType \rightarrow \UnitType$? In game semantics, we explicitly observe
function calls and returns across the "interface". To model
this, $\Context$ sends a token, which we write simply $\lambda$, instead
of sending a function or some code. However, now the session between
$\Program$ and $\Context$ is split in two independent sub-sessions:
one, as before, for $\Program$ to return the result, and a second one
to make calls to $f$. To represent this splitting of a session into
several sub-sessions, we use the notation $ \parallel $, which
represents the multiplicative connectives of Linear Logic
$( \otimes , \parr)$ identified into one.\footnote{The distinction in
  Linear Logic is crucial to ensure deadlock-freedom and hence
  cut-elimination. Since programs of $\CML$ do have deadlocks, we
  explicitly want to identify the multiplicative connective.} The
interface thus becomes:
$$\TranslateTerm {f: \UnitType \rightarrow \UnitType  \vdash  \UnitType} =
\Call{\lambda} ^- \cdot (\Call {}^+ \cdot \Return {}^- \parallel \Return
{}^+).$$ where we use $ \Call{\lambda} ^- \cdot S$ instead of the unary choice
$\with_{e  \in  \{ \ast \}}  \Call{\lambda} ^- \cdot S$ and similarly for $\Call{}^+ \cdot S$.

In this protocol, $\Program$ receives the $ \lambda $ token
from $\Context$, and then can either call $f$ or return. If $\Program$
calls $f$, then it might return $()$. However, there is some
information that is not captured here. $\Program$ should be allowed to
call $f$ several times, but may only return only once: certain
sessions can be repeated while some others may not. To encode this
information in the session type, we use the standard exponentials of
Linear Logic: by default every message is linear (can only occur
once). Then, we have two dual modalities that allow to specify
duplication: $\wn S$ means $\Program$ can spawn as many instances
of $S$ as they like; while $\oc S$ means $\Context$ can spawn as
many instances of $S$ as they like, hence $\Program$ must behave as a
server accepting requests from $\Context$.

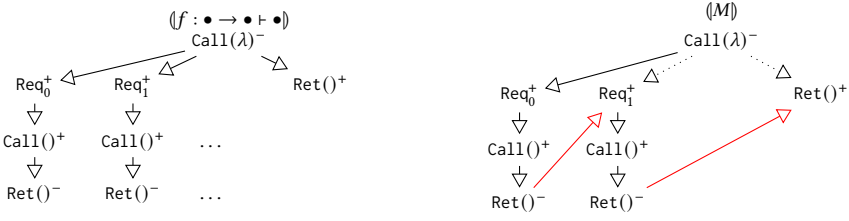
\begin{figure}
 \scriptsize
  \centering
  \begin{tikzpicture}[baseline]
    \matrix[diag] {
      \& \& |(x)| \\[-2mm]
      \& \& |(y)| \\
      |(re0)| \Request^+_0 \& |(re1)| \Request^+_1 \& \& |(r)| \Return {}^+ \\
      |(c0)| \Call {}^+ \& |(c1)| \Call {}^+ \&  {\ldots\qquad} \\
      |(r1)| \Return {}^- \& |(r2)| \Return {}^- \&  {\ldots\qquad}  \\
    };
    \node at (x) { $\TranslateTerm{f : \UnitType  \rightarrow  \UnitType  \vdash  \UnitType}$  };
    \node (lambda) at (y) { $\Call{\lambda}^-$ };
    \path[cause]
    (lambda) edge (re0) (lambda) edge (re1) (lambda) edge (r)
    (re0) edge (c0) (re1) edge (c1)
    (c0) edge (r1) (c1) edge (r2);
    \hfill
  \end{tikzpicture}
  \qquad\qquad\qquad
  \begin{tikzpicture}[baseline]
    \matrix[diag] {
      \& \& |(x)| \\[-2mm]
      \& \& |(lambda)| \Call{\lambda}^- \\
      |(a0)| \Request^+_0 \& |(c0)| \Request ^+_1 \& \& |(e)| \Return {}^+ \\
      |(a)| \Call {}^+ \& |(c)| \Call {}^+ \& \& \\
      |(b)| \Return {}^- \& |(d)| \Return {}^- \\
    };
    \node at (x) { $\TranslateTerm{M}$ };
    \path[cause]
    (lambda) edge (a0)
    (a0) edge (a)
    (a) edge (b)
    (b) edge[red] (c0)
    (c0) edge (c)
    (c) edge (d)
    (d) edge [red] (e);
    \path[cause, dotted] (lambda) edge (e);
    \path[cause, dotted] (lambda) edge (c0);
  \end{tikzpicture}
  \vspace{-3mm}
  \caption{Interpretation of higher-order terms}
  \vspace{-5mm}
  \label{fig:ex:ho}
\end{figure}

Our example requires $\wn(\cdot)$ since $\Program$ decides how many times $f$
is called. We get:
$$
\begin{aligned}
\TranslateTerm {f: \UnitType \rightarrow \UnitType  \vdash  \UnitType} &=
\Call{\lambda} ^- \cdot (\wn (\Call {}^+ \cdot \Return {}^-) \parallel \Return
{}^+). \\
\TranslateTerm M_o &=
\CBranchingOne {o} {\Call {\lambda}} {f, r}
\CDer f s \CSelect s {\Call {}} { r_1 }
\CBranchingOneS { r_1 } {\Return {}}\\
&\qquad\qquad\qquad\qquad\ \ \ \CDer f s \CSelect s {\Call {}} { r_2 }
\CBranchingOneS { r_2 } {\Return{}}
\CPositiveActionS r  {\Return {}}
\end{aligned}
$$
The above example features two new constructs. First, due to
$ \parallel $, it is possible to introduce several names (in the input
on $o$), as there might be several sub-sessions now: for instance, $f$
denotes here the session related to $f$ while $r$ that related to
returning the result. Second, we have the construction $?f[s]$ which
opens a new session (of type $S$) on $f$ called $s$ when $f$ has type
$\wn S$. In this case, sessions on name $f$ are used to make a call to
parameter $f$. Note that we can reuse the name $s$ across sessions as,
being a linear name, it disappears after the
$\CPositiveActionS s {\Call {}}$. The interpretation of $M$ and its
interface are given in Figure \ref{fig:ex:ho}. The game is now more
complex: we see clearly the possible independent sub-sessions as
being incomparable for the causal order. The $\wn S$ construction is
putting infinitely many copies of $S$, each prefixed by a positive
move $\Request_i$ equipped with a copy index $i \in \mathbb{N}$. When
the strategy plays, it chooses indices for the positive moves:
this choice is irrelevant (two different choices will yield equivalent
strategies). In this case, the strategy is still a subset of the game,
but there are now extra causal links indicating the sequentiality
specific to the term (arrows in red).

By duality, the $\oc$ construction is used when \emph{defining}
functions. To construct a process implementing a session type $\oc S$,
we use a construct called \textbf{promotion}, written $\CProm a x P$, which represents a server spawning
up $P$ each time a request is made, with $x$ being the request. For
instance, we have:
$$
\begin{aligned}
  \TranslateTerm { \vdash  \UnitType  \rightarrow  \UnitType} &= \Call {}^-
  \cdot ( \lambda ^+ \cdot \oc (\Call {}^- \cdot \Return {}^+)) \\
  \TranslateTerm { \lambda x. ()}_o &= \CBranchingOneS o {\Call{}}
  \CSelectS o { \lambda }
  \CProm o c
  \CBranchingOneS c {\Call {}} \CPositiveActionS c {\Return{}}.
\end{aligned}$$
A function starts by telling \Context that it does normalise to a
$\lambda$-abstraction, by sending $ \lambda $, and then sets up a
server to handle calls to the function by \Context. In this term, we
have not specified a list of names for some actions: this means we
simply reuse the subject of the action, for instance
$\CPositiveActionS o \lambda$ stands for
$\CPositiveAction o \lambda o$.

\subsection{Shared memory concurrency}\label{sec:memory}
To conclude our illustration of the phenomena at play in the
game semantics of a language like $\CML$, we now outline the case of
references. Our viewpoint is that each reference has an owner -- the
one that declared it with $\RefAlloc$, and when an open program has a
free variable of type $\RefType$, it is not the owner of the
reference, but $\Context$ is. Operations on non-owned references are
simply forwarded to the owner: if a term $M$ has a
free reference $r$, then what it receives from $\Context$ is simply a
name at which it can send the memory operations. In our simple
setting, there are only two operations: \texttt{get} and
\texttt{set}. For instance,
$r : \RefType \vdash r := !r + 1 : \UnitType$ will be interpreted as a
process that does two requests on the object $r$: a \texttt{get}, and
then a \texttt{set}. Dually, when interpreting the construct
$\RefAlloc(\cdot)$, $\Program$ is the owner, and $\Context$ is the
user (since the reference is returned to $\Context$): this means that
$\Program$ must create a server that can adequately handle these
requests.

However, such a server cannot be implemented using the promotion
described in \S~\ref{sec:higher-order}. Indeed, such servers are
\emph{stateless}, meaning that the answer to a request does not depend
on previous or parallel requests. In particular, two independent
requests can (and are) treated independently. This is not the case for
a server implementing a reference, which has to treat requests in a
sequential order. Two independent request must thus be
sequentialised. To represent this phenomenon, we turn our attention to
an extension of Linear Logic, called \emph{Differential Linear Logic}
\cite{DiLL_tutorial}, which extends Linear Logic with such servers. We
use here a slight variant of the standard \emph{codereliction}, which
allows to set up a one-time server. Such ``servers'', written
$\CCoder a x P$ in our metalanguage introduced in the next section,
wait for \emph{one} request $x$, and execute $P$ on it. $P$ has still
access to $a$ so it can start a new such server if it wants to, after
having processed the request $x$. In the meanwhile, the server is
down, and requests are left pending. In particular two concurrent
requests are racing to be satisfied by this server. This phenomenon
induces a natural sequentialisation of the requests on $a$.
The implementation of this reference server is given in
\S~\ref{subsec:translation:terms}.


%% file: figures/cml_typing.tex
\begin{figure}
$
  \small
\begin{array}{c}
  \inferruleR[\rulename{var}]{ \Delta , x:  \sigma   \vdash  x :  \sigma } { }
  \quad 
  \inferruleR[\rulename{abs}]{\Delta  \vdash   \lambda x.\, M :  \sigma   \rightarrow   \tau } { \Delta , x:  \sigma   \vdash  M :  \tau  }
  \
  \inferruleR[\rulename{app}]{
    \Delta  \vdash M\, N :  \tau } {
\Delta   \vdash  M :  \sigma   \rightarrow   \tau\quad 
\Delta   \vdash  N :  \sigma
  } \
  \inferruleR[\rulename{bot}]{ \Delta   \vdash  \bot :  \sigma }{ }
  \\[1mm]
\inferruleR[\rulename{fix}]{ \Delta   \vdash  Y : (( \sigma   \rightarrow   \tau )  \rightarrow  ( \sigma   \rightarrow   \tau ))  \rightarrow   \sigma   \rightarrow   \tau }{ }
\
  \inferruleR[\rulename{bool}]{ \Delta   \vdash  \ttrue, \ffalse : \BoolType}{ }
\
\inferruleR[\rulename{if}]{ \Delta   \vdash  \IfTerm M {N_1} {N_2} :  \sigma } { \Delta   \vdash  M : \BoolType\quad  \Delta   \vdash  N_1 :  \sigma  \quad  \Delta   \vdash  N_2 :  \sigma }\\[1mm]
\inferruleR[\rulename{num}]{ \Delta   \vdash  \underline n : \NatType}{ }\
\inferruleR[\rulename{plus}]{ \Delta   \vdash  \texttt{plus} : \NatType  \rightarrow  \NatType  \rightarrow  \NatType}{ }\quad 
\inferruleR[\rulename{unit}]{ \Delta   \vdash  () : \UnitType}{ }\\[1mm]
\inferruleR[\rulename{ref}]{ \Delta   \vdash  \RefAlloc : \NatType  \rightarrow  \RefType}{}\
\inferruleR[\rulename{get}]{ \Delta   \vdash  \texttt{get} : \RefType  \rightarrow  \NatType}{}\
\inferruleR[\rulename{set}]{ \Delta   \vdash  \texttt{set} : \RefType  \rightarrow\NatType  \rightarrow  \UnitType}{}\
\end{array}
$
\vspace{-3mm}
\caption{Typing system of $\CML$}
\vspace{-5mm}
\label{fig:cml typing}
\end{figure}

%% file: figures/cml_op.tex
\begin{figure}
  \small
  \begin{subfigure}{\textwidth}
\[    
\begin{array}{c}
  \inferruleR[\rulename{$\beta$}]{( \lambda x. \, M)\, V \rightarrow M[V/x]} {V\text{ a value}}
  \quad 
    \inferruleR[\rulename{sum}]{\underline n + \underline m  \rightarrow  \underline{n+m}} {}\quad 
    \inferruleR[\rulename{et}]{\underline n = \underline n  \rightarrow  \ttrue} {}\quad 
    \inferruleR[\rulename{ef}]{\underline n = \underline m  \rightarrow  \ffalse} {n\neq m}\\[2mm]
    \inferruleR[\rulename{ift}]{\IfTerm \ttrue M N  \rightarrow  M}{}\quad
    \inferruleR[\rulename{iff}]{\IfTerm \ffalse M N  \rightarrow  N}{}\quad
    \inferruleR[\rulename{Y}]{Y\, V\, V'  \rightarrow  V (Y\, V)\, V'}{V, V'\text{ are values}}\quad
    \inferruleR[\rulename{context}]{E[M]  \rightarrow  E[N]}{M  \rightarrow  N}
\end{array}
\]
    \caption{Deterministic reduction for functional fragment of $\CML$}
    \label{fig:cml:pure op}
  \end{subfigure}
  \begin{subfigure}{\textwidth}
    \[
    \begin{array}{c}
      \inferruleR[\rulename{base}]
      {\machineDefault {M} \xrightarrow{ \tau } \machineDefault {N}}
      {M \rightarrow N}
\quad \inferruleR[\rulename{write}]
      {\machine \Name { \Delta }{x := \num n} \mu  \xrightarrow{ \alpha }\machine \Name{ \Delta } {()} {\mu [x := n]}}
      {x \in \dom \Delta \quad x \in \Name  \Rightarrow   \alpha  = \Write x n \quad x \not\in \Name  \Rightarrow   \alpha  =  \tau }
\\[2mm]
            \inferruleR[\rulename{cxt}]
      {\machineDefault {E[M]} \xrightarrow{ \alpha } \machineDefaultt {E[N]}}
      {\machineDefault M \xrightarrow{ \alpha } \machineDefaultt N}
\quad 
      \inferruleR[\rulename{read}]
      {\machine \Name { \Delta }{!x} \mu  \xrightarrow{ \alpha }\machine \Name{ \Delta } {\underline{\mu(x)}} {\mu}}
      {x \in \dom \Delta \quad x \in \Name  \Rightarrow   \alpha  = \Read x {\mu(x)} \quad x \not\in \Name  \Rightarrow   \alpha  =  \tau }
      \\[2mm]
      \inferruleR[\rulename{alloc}]
      {\machine \Name \Delta {\RefAlloc(\underline n)} \mu \xrightarrow{  \tau  } \machine \Name { \Delta ,r:\RefType} {r} {\mu[r:=n]}}
        {r \not\in \dom {\Delta}}
    \end{array}
    \]
    \caption{Labelled Transition System for $\CML$ semiclosed terms}
    \label{fig:cml:lts}
  \end{subfigure}
   \vspace{-2mm}
  \caption{Operational semantics of $\CML$}
    \vspace{-3mm}
  \label{fig:opsem CML}
\end{figure}

%% file: calculus.tex
In this section, we present a formal treatment of the intuitions
presented in \S~\ref{sec:overview}, and formalise it into a calculus
that we call $\MetalanguageName$, as its typing rules are close to
\emph{Differential Linear Logic} (\DiLL). In \S~\ref{sec:calculus
  types} we present the types of $\MetalanguageName$; in
\S~\ref{sec:calculus syntax} we present the syntax of the calculus for
session types and processes along with the typing discipline. In
\S~\ref{sec:calc:sem} we define the operational semantics of
$\MetalanguageName$; and in \S~\ref{sec:interp ML} we present the
interpretation of $\CML$ into $\MetalanguageName$.

\subsection{Types of $\MetalanguageName$}\label{sec:calculus types}
As explained in \S~\ref{sec:overview}, our types closely
mirror the syntax of Linear Logic formulas:
$$T \Coloneqq \CWith {i \in I} {\ell_i^- \cdot T_i} \mid \CPlus {i \in I} {\ell_i^+ \cdot T_i} \mid (T \CPar  \ldots  \CPar T) \mid \wn S \mid \oc S.$$
These types describe protocols from the point of view of
$\Program$. The first two are the \emph{additive} connectives and
represent data input and data output respectively:
$\CWith {i \in I} {\ell_i^- \cdot T_i}$ means that $\Program$ should
expect a message consisting in one of the $\ell_i$ from $\Context$ and
then behave as the corresponding $T_i$; and dually
$\CPlus {i \in I} {\ell_i^+ \cdot T_i}$ indicates that $\Program$ should
output to $\Context$ one of the $\ell_i$ and continue as the
corresponding $T_i$. In both instances, $I$ is an arbitrary countable
set of choices; and $\ell_i$ represents the \textbf{payload} of the
message, to be thought of simple datatypes such as integers or
booleans choosen in a set $\mathbb L$. When $I$ is a singleton we
write $\ell^- \cdot S$ and $\ell^+ \cdot S$ directly.

The construct $T_1 \parallel \ldots \parallel T_n$ is the parallel
composition and represents both $\parr$ and $ \otimes $: in this
setting with deadlocks, we do not need to distinguish between the two
multiplicatives connectives of Linear Logic and can identify them (as
in compact-closed models of Linear Logic). From a protocol standpoint,
this means that the current session splits into $n$ independent
subsessions. In Session Types, these two connectives are called input
and output respectively and represents delegation, or
channel-passing. However, in this setting closer to the internal
$\pi$-calculus \cite{DBLP:journals/tcs/Sangiorgi96a}
where free names are never sent, the right intuition
is that channels are not passed around, but messages occurring in one
of the subsession is just prefixed with the index of the sub-session
(thus encoding the so-called \emph{multiplicative addresses}), and the
names are just abstraction around these addresses. A parallel
composition can be empty, leading to the type $\intro* \COne$, which is the
equivalent of the (identified) multiplicative identities of Linear
Logic.

The dual types $\oc S$ and $\wn S$ are the exponentials and
represent replicable sessions: $\oc S$ means that
$\Program$ is expected to receive arbitrarily many queries from
$\Context$ to start a session of type $S$ (ie. $\Program$ is the
server, and $\Context$ the client); while $\wn S$ means that
$\Program$ can contact $\Context$ to start as many sessions of type
$S$ as it wants (ie. $\Program$ is a client and $\Context$ the
is a server).

Types come equipped with a notion of \emph{duality} stemming from the
usual De Morgan laws: $\with$ and $\oplus$ are dual to each other, as
well as $\wn$ and $\oc$ while $ \parallel $ and $\EmptyType$ are
self-dual. Session types whose toplevel constructors are not
$ \parallel $ are called ""rooted"" as they have a unique minimal
action. Only such types will be bound to channels in
$\MetalanguageName$.

\subsection{Processes of $\MetalanguageName$}\label{sec:calculus syntax}
The syntax of $\MetalanguageName$ follows 
the notation of \cite{DBLP:journals/jfp/Wadler14}.
$$
\begin{aligned}
  P, Q ::= \, &\CZero \mid P \CFork Q \mid \CCut a b P \mid
\CBranching c {i \in I} {\CBranch{\ell_i}{\vec x_i} P_i} \mid \CSelect c
  k {\vec x} P
  \mid\, \CProm c \PatternVar P
  \mid \CCoder c \PatternVar P \mid
  \CDer c \PatternVar P
\end{aligned}$$ $\CZero$ denotes a nil process, $P \CFork Q$ is a
\emph{parallel} composition, $\CCut a b P$ is a \emph{restriction}
where binding $a$ and $b$ in $P$.
$\CBranching c {i \in i} {\CBranch i {\vec x_i} P_i}$ is a
""branching"" with labels $\ell_i \in \mathbb{L}$ indexed by $i \in I$
where $I$ can be infinite; its dual, $\CSelect c k {\vec x} P$, is a
\emph{selection} to label $k$.  ""promotion"" $\CProm c \PatternVar P$
(stateless server), ""codereliction"" $\CCoder c \PatternVar P$
(stateful server), and ""dereliction"" $\CDer c \PatternVar P$ (their
client) have been explained in \S~\ref{sec:overview}. In all
constructs, $\vec{x}$, $\vec{x}_i$ are lists of bound names 
as in the internal $\pi$-calculus \cite{DBLP:journals/tcs/Sangiorgi96a}.

We consider the finite or \textbf{infinite} processes generated by
this grammar. Concretely, we consider the set of processes to be the
ideal completion of the finite processes ordered by the subtree
ordering. This is standard in some literature of the $\pi$-calculus
(eg.~\cite{CP09}) and necessary to give a syntactic counterpart to
infinite strategies which are used in the interpretation of most
languages, for recursion and memory.  Definition of such infinite
processes will mostly be by computing the least fixpoint of continuous functions. Such fixpoints are written $\fix f$.

\mysubsubsection{Typing rules.} We now describe the typing rules for the
calculus. Typing contexts are of the form
$ \Gamma  := a_1: A_1, \ldots , a_n: A_n$ where each $A_i$ is a "rooted"
type. This is key since we want every action of our $ \pi $-calculus to
correspond to a meaningful computational event.

While typing judgements are flat lists, session types have more
structure since $(A \parallel B) \parallel C$ is only
\emph{equivalent} to $A \parallel (B \parallel C)$ (ie. isomorphic as games
as will be seen in \S~\ref{sec:interpinterp}), but not equal. This prompts us to
define a way to flatten a non-rooted session type into a context. For
that, we define a partial function taking a list of names and a session type,
and possibly returning a context:
$$
\ContextElaboration {x} T \;\coloneqq\; x: T \quad \text{(when $T$ is rooted)} \qquad\quad
\ContextElaboration {\vec x, \vec y} {(T  \parallel  U)} \;\coloneqq\; \ContextElaboration {\vec x} T, \ContextElaboration {\vec y} U.
$$
and undefined otherwise. Note that for the second equation, the
decomposition in $\vec x, \vec y$ of any list $\vec z$ is unique,
given type $T \parallel U$ of $z$ where $\vec x$ and $\vec y$ are assigned by
$T$ and $U$, respectively. We write $\cdot$ for the empty list of
names. Note that $\ContextElaboration {\cdot} 1$. Typing judgements
are simply of the form $\Typing P {\Gamma}$. The rules are described
in Figure \ref{fig:mall typing}. (As for the syntax of untyped terms,
we consider the infinite derivation tree generated by these rules.)

The rule \rulename{nil} allows us to weaken any name preserving
typability; note that
$\MetalanguageName$ is \emph{affine} and not linear: this mirrors a similar
phenomenon in game semantics. The rule \rulename{par} forces the split
of the linear part of the context in a parallel composition. The rule
\rulename{res} introduces two connected channels with dual types.

The rule \rulename{rep} is the introduction rule for $\oc$, and
corresponds to setting up a server on $a$ which spins up a new copy of
$P$ whenever a request is made. As usual in Linear Logic, since $P$
might be duplicated, the session types in $ \Gamma$ must be of the
form $\wn S$, which we indicate with the notation $\wn \Gamma$. The
dual rule \rulename{req} sends a request to a server. Note that $a$
is still usable in the continuation.

The
rule \rulename{sel} is \emph{selection}, an output of a
payload to the other party on a channel. In the continuation, the name
$a$ disappears as it has been consumed, and the new names $\vec x$
appear: they are used to deconstruct the continuation type $T_k$. We
have a non-trivial list as soon as $T_k$ is not rooted. The dual rule
\rulename{br} represents \emph{branching}: every possible message
that can be received must be handled.

The rule \rulename{nd} arises from Differential Linear Logic and
corresponds to setting up a server that accepts exactly one request,
and then disappears. Unlike rule \rulename{rep}, the continuation
might use $a$ again, and set up a new server later on. Two concurrent
requests made on these one-time servers will race to be accepted,
creating nondeterminism. This rule merges codereliction and
cocontraction of \DiLL{} in one single rule. We found easier to give
non-angelic semantics to this presentation as semantics of \DiLL{} is
usually angelic. However, the usual codereliction and cocontraction are
admissible from this rule.

When writing down processes, we often omit the list of successors
$\vec x$. This may mean two things: if the move is maximal, it has no
successors and we omit the empty list $ \epsilon$; if the move has a
unique successor, we use the standard convention of session types
which reuses the same session name, eg. we write
$\CPositiveActionS a k$ for $\CPositiveAction a k a$.  This is only
valid if $a$ does not have type $\wn S$ since it can then be
reused. In general we also reuse linear names along a session.
\input{figures/meta_typing.tex}

\mysubsubsection{The forwarding agent} We introduce the
\emph{forwarding agent}
$\Typing {\intro*{\Forwarder {\vec a} {\vec b} S}} {\ContextElaboration {\vec
    a} S, \ContextElaboration {\vec b} {S^\perp}}$ for any session
type $S$ and $\ContextElaboration {\vec a} S$ and
$\ContextElaboration {\vec b} {S^\perp}$ defined. The forwarding agent
represents copycat in game semantics. It
is defined by induction on~$S$, by following the $\eta $-expansion
laws in Linear Logic:
$$
\small
\begin{array}{rclrcl}
\Forwarder a b {\CWith {i \in I} { \ell _i \cdot S_i}} &=&
\CBranching a {i \in I} {\CBranch { \ell _i} {\vec x} \CSelect b { \ell_i} {\vec y} \Forwarder {\vec x} {\vec y} {S_i}}
&
  \Forwarder a b {\oc S} &= & 
   \CProm a {\vec x} \CDer a {\vec y} {\Forwarder {\vec x} {\vec y} {S}}\\
\Forwarder {\vec a} {\vec b} {S_1  \parallel   \ldots   \parallel  S_n} &=&
  \parallel _{1  \leq  i  \leq  n} \Forwarder {\vec a_i} {\vec b_i} {S_i}
&      
  \Forwarder a b {S^\perp} &= & \Forwarder b a S
\end{array}
$$
\mysubsubsection{Composition of processes} A key operation in game
semantics is \textbf{composition}. It allows defining the
interpretation of a programming language by introducing small bricks
that
can be composed together. This composition can be expressed in this setting as is well-known:\\
  \centerline{
$\Typing{\shareInteractS {\vec a} P Q :=
  \CCut {\vec a} {\vec {\bar a}} (P \parallel Q)}{ \Gamma _1, \Gamma _2, \wn \Gamma } \quad \mbox{where}\quad 
\Typing P { \Gamma _1, \wn \Gamma , \vec a: \vec S}
\mbox{ and }
\Typing Q { \Gamma _2, \wn \Gamma , \vec{\bar a}:\vec S^\perp}$
}\\
This mimimcks the traditional definition using parallel interaction
with hiding. We use here the notation $\overline{a}$ but not with its
standard $\pi$-calculus meaning: here $a$ and $\overline{a}$ are simply two
independent names which only become related via the restriction. 

\subsection{Operational semantics of the metalanguage}\label{sec:calc:sem}
We define the operational semantics of
$\MetalanguageName$. As for $\CML$, it will be done in two steps: (1)
a congruence that will contain the usual structural congruence of the
$ \pi $-calculus, the permutation rules due to courtesy in game
semantics, and finally the deterministic communication, and (2) an
operational semantics which resolves the races induced by the
codereliction/dereliction pairs.

\mysubsubsection{Congruence}\label{sec:calc:permutations} We start with
the congruence on our calculus, which takes into account the
\emph{asynchronous permutations}, needed for the game semantics. These
permutations are necessary to obtain a category, which is essential to
interpret functional languages. These asynchronous permutations
include all permutations between actions on distinct channels of the
same polarity, among others. To formalise them, we use the notion of
\textbf{prefix}.  A prefix $\mathfrak a$ is a pair of an arity and a
context with as many holes as the arity, as defined by:
$$
\begin{aligned}
  \mathfrak a ::=\, &\CSelect a k \PatternVar [] \mid \CDer a \PatternVar [] 
  \mid\, \CBranching a {i \in I} {\CBranch{\ell_i}{\PatternVar_i} {[]_i}} \mid
  \CProm a \PatternVar [] \mid \CCoder a \PatternVar []
\end{aligned}$$
The arity of all prefixes is one, except the branching
prefix whose arity is the indexing set~$I$. Given a prefix $\mathfrak a$
with arity $I$, and a family of process $(P_i)_{i \in I}$, we write
$\mathfrak a[P_i]_{i \in I}$ for the substitution. A prefix is
positive (written $\mathfrak a^+$) when it is a selection or dereliction;
and negative otherwise (written $\mathfrak a^-$).

We also adopt the convention that a prefix $\mathfrak a^p$ is a prefix
of polarity $p$ on channel $a$.  We write $\BoundVars {\mathfrak a}$
for the set of bound variables occurring in the pattern in $\mathfrak
a$. Two prefixes $\mathfrak a$ and $\mathfrak b$ are orthogonal,
written $\mathfrak a \bot \mathfrak b$ when
$(\{a\} \cup \BoundVars {\mathfrak a}) \cap (\{b\} \cup \BoundVars
{\mathfrak b}) = \emptyset$. The structural congruence on finite
processes is defined as the smallest congruence on well-typed terms
(over the same context) satisfying the rules described in
Figure~\ref{fig:calc:cong}.  The first box defines the standard
structural congruence rules. The second box explains how prefixes
commute with each other and other construct. The premise of
\rulename{swap} ensures the typability of both sides of processes.
The last rule \rulename{id} is very important and states that the
asynchronous forwarder behaves as an identity. This induces most
asynchronous permutations. The fourth box is about deterministic
communication. We consider them as a part of the congruence rules as
we perform them under context -- since they are deterministic, it does
not matter when or where they are performed.

\input{figures/meta_str}

These rules are extended to infinite processes in the obvious
continuous way: $P \equiv Q$ when for all finite $P_0  \leq  P$, then there
exists $Q_0  \leq  Q$ with $P_0 \equiv Q_0$ and vice-versa.

\begin{example}
  Remember the process $P$ of Figure~\ref{fig:overview} representing
  two parallel calls to the same external function. This term is equivalent to the following:
  $$P  \equiv  \CBranchingOne o {\Call \lambda} {f, r}
  \CDer f s \CSelectS s {\Call 1} \CDer f {s'} \CSelectS {s'} {\Call
    2} \CBranchingOneS s {\Return n} \CBranchingOneS {s'} {\Return m}
  \CPositiveActionS {r} {\Return {n+m}}.$$ This illustrates that, due
  to those permutations, processes that seem sequential (ie. without
  $\parallel$) can actually be concurrent due to these asynchronous
  permutations: only syntactic dependence from input to output is
  preserved.
\end{example}
\mysubsubsection{Reduction rules of $\MetalanguageName$}
\label{subsec:meta_reduction} We define the operational semantics of
$\MetalanguageName$, which now only reduces redexes involving a
dereliction and a codereliction. This redex
is nondeterministic because the reduction consumes the codereliction:
hence two concurrent derelictions race for the codereliction. The
rules of the LTS for finite processes are defined in Figure~\ref{fig:ml_op},
where $E ::= [] \mid E \CFork P \mid \CCut x y E$. As
before, we extend the LTS to infinite processes by letting
$P \to Q$ when $Q$ is the limit of all $Q_0  \leq  Q$ such that
there exists $P_0  \leq  P$ with $P_0 \to Q_0$.
For the proof of Lemma~\ref{lem:SR}, see Appendix~\ref{sec:calculus:proofs}.

\begin{restatable}[Subject Reduction]{lemma}{SR}
\label{lem:SR}
If $\Typing P \Gamma$ and $P\to Q$,
then $\Typing Q \Gamma$.  
\end{restatable}

\mysubsubsection{Observational theory}\label{sec:calc:obs} From this
operational semantics, we deduce a notion of behaviour equivalence
through the standard concept of \emph{reduction-closed barbed
  congruence} \cite{MiSa92,HondaKYoshida95}. First, let us define the notion of barbs: a process $P$
has a barb on $a$, written $\HasBarb P a$ if
$P \equiv \CCut {\vec x}{\vec y} {(\mathfrak a^+[P] \CFork Q)}$.

\begin{definition}
  A reduction-closed barbed congruence is an equivalence relation on
  typed processes $\mathcal R$ containing $ \equiv $ such that:
  \begin{itemize}
  \item If $(P, Q) \in \mathcal R$, then $\HasBarb P a$ iff $\HasBarb Q a$
  \item If $(P, Q) \in \mathcal R$ and $P \Red P'$ then there exists
    $(P', Q') \in \mathcal R$ such that $Q \Red^* Q'$.
  \item If $(P, Q) \in \mathcal R$ then for all context $C$,
    $(C[P], C[Q]) \in \mathcal R$.
  \end{itemize}
\end{definition}
We write $\intro*{\MLObs}$ for the largest reduction-closed barbed congurence.


%% file: figures/meta_typing.tex
\begin{figure}[t]
$
  \small
  \begin{array}{c}
    \begin{array}{llllll}
      \rulename{nil} &  \rulename{par} & \rulename{res} & \rulename{rep} &
      \rulename{req}\\
  \inferruleR
      {
        \Typing \CZero { \Gamma }} {\ }
      & 
      \inferruleR
      {\Typing  {P \CFork Q}{ \wn  \Gamma , \Gamma_1,  \Gamma _2}}
      {\Typing {P}{ \wn  \Gamma , \Gamma _1} \quad \Typing  {Q} { \wn  \Gamma , \Gamma _2}}&
      \inferruleR
      {\Typing  {\CCut a b P}{ \Gamma }}
      {\Typing  {P}{ \Gamma , a: A, b: A^\perp}}
&
               \inferruleR
      {\Typing  {\CProm a \PatternVar P}{ \wn \Gamma , a: \CBang T}}
      {\Typing  P{ \wn \Gamma , \ContextElaboration {\PatternVar} T}}
      &
      
            \inferruleR
                       {\Typing  {\CDer a \PatternVar P}{ \Gamma , a: \CWN T}}{\Typing {P}{\Gamma, a: \CWN T, \ContextElaboration {\PatternVar} T}}
    \end{array}
    \\[5mm]
    \begin{array}{llllll}
      \rulename{br}& \rulename{sel} & \rulename{nd}\\
            \inferruleR
      {\Typing {\CBranching a {i \in I} {\CBranch {\ell_i} {\PatternVar_i} {P_i}}}{ \Gamma, a: \CWith {i \in I} {\ell_i^- \cdot T_i} }}
      {\forall i \in I, \Typing  {P_i}{ \Gamma, \ContextElaboration {\PatternVar_i} T_i}}
      & \inferruleR
      {\Typing {\CSelect a {\ell_k} \PatternVar P}{ \Gamma, a: \CPlus {i \in I} {\ell_i ^ + \cdot T_i} }}
      {k\in I \quad \Typing  {P}{\Gamma, \ContextElaboration \PatternVar T_k}}
&      \inferruleR
      {\Typing  {\CCoder a \PatternVar P}{\Gamma , a: \CBang T}}
      {\Typing  P{ \Gamma , a: \CBang T, \ContextElaboration {\PatternVar} T}}
    \end{array}
    \end{array}
$
  \vspace{-2mm}
  \caption{\label{fig:mall typing}Typing rules of the metalanguage}
   \vspace{-5mm}
\end{figure}

%% file: figures/meta_str.tex
\begin{figure}
  \small
  \boxit[$\parallel$ and restriction] {
\vspace{2mm}
\[
    \begin{array}{c}
P \CFork \CZero  \equiv  P\quad 
P \CFork Q  \equiv  Q \CFork P\quad 
(P \CFork Q) \CFork R  \equiv  P \CFork (Q \CFork R)\quad
\CCut a b {\CZero}  \equiv  \CZero\quad 
\CCut a b {P}  \equiv  \CCut b a P\\[3mm]
\infer {\CCut a b {\CCut c d P}  \equiv  \CCut c d {\CCut a b P}}{\{a, b\} \cap \{c, d\} = \emptyset}  \quad 
\infer {\CCut a b {(P \CFork Q)}  \equiv  (\CCut a b P) \CFork Q}{\fv Q \cap \{a, b\} = \emptyset}
    \end{array}
    \]
    }
  \boxit[Permutations of prefixes] {
\vspace{2mm}    
    \begin{mathpar}
      \inferruleR[\rulename{nil}]
      {\CCut a b {\mathfrak a^-[P_i]_{i \in I}}  \equiv  \CZero}{} \and
      \inferruleR[\rulename{res}]{\CCut a b {\mathfrak c[P_i]_{i \in I}}  \equiv  \mathfrak c[\CCut a b {P_i}]_{i \in i}} {c \not\in \{a, b\}} \and
      \inferruleR[\rulename{swap}]{\CCut a b {(\mathfrak a^-[P_i]_{i \in I} \CFork \mathfrak c^-[Q_j]_{j \in J})}  \equiv  \mathfrak c^-[\CCut a b {(\mathfrak a^-[P_i]_{i \in I} \CFork Q_j)}]_{j \in J}}{c \not\in \{a, b\}}
      \quad 
\inferruleR[\rulename{id}]{\shareInteractS a P {\Forwarder {\bar a} b} {S} \equiv P[b/a]}{\Typing {P} { \Gamma , a: S} }
   \end{mathpar}
  }
  \boxit[Deterministic communication] {
    \vspace{1mm}
    \begin{mathpar}
      \inferruleR[\rulename{com}]
      {\CCut a b {(\CSelect a k \PatternVar P \CFork
          {\CBranching b {i \in I} {\CBranch i {\PatternVari_i} {Q_i}}})}
        \equiv
        \CCut {\PatternVar} {\PatternVari_k}
        {(P \CFork Q_k)}} {k \in I}
      \and
      \inferruleR[\rulename{rep}]
      {\CCut a b {(\CDer a \PatternVar P \CFork \CProm b \PatternVari Q)}
        \equiv
        \CCut a b {(\CProm b {\PatternVari} Q \CFork \CCut {\PatternVar} {\PatternVari} {(P \parallel Q)})}}
      {\ }\and
      \inferruleR[\rulename{share}]
      {\CCut a b {(P \CFork Q \CFork \CProm b \PatternVar R)}
        \equiv
        \CCut a b {(P \CFork \CProm b \PatternVar R)}
        \CFork
        \CCut a b {(Q \CFork \CProm b \PatternVar R)}}
      {}
    \end{mathpar}
  }
  \boxit[Reduction rules]{
 \vspace{1mm}
\begin{mathpar}
  \inferruleR[\rulename{race}]{\CCut a b {(\CDer a \PatternVar P \CFork \CCoder b \PatternVari Q \CFork R)} \rightarrow \CCut a b {(R \CFork \CCut {\PatternVar} {\PatternVari} {(P \parallel Q)})}}{\ }\\
    \inferruleR[\rulename{cxt}]{E[P]  \rightarrow  E[Q]} {P \rightarrow Q}
    \quad \quad  
    \inferruleR[\rulename{str}]{P  \rightarrow  Q} {P'\equiv P  \rightarrow 
      Q\equiv Q'}
\end{mathpar}
}
\vspace{-3mm}
\caption{\label{fig:calc:cong} \label{fig:ml_op} The structural congruence rules and reductions for $\MetalanguageName$}
\vspace{-5mm}
\end{figure}

%% file: translation.tex
This section defines the translation of $\CML$ into $\MetalanguageName$, and proves it correct.  

\subsection{Translation of types}
\label{subsec:translation:types}
We start by translating types of $\CML$ into session types.  As
introduced in \S~\ref{sec:first-order}, labels used for communication
in translated types are based on the constructors of types:
$$
\begin{array}{c}
\text{constructor:} \quad  
c ::= n  \in   \mathbb{N}  \mid () \mid \ttrue \mid \ffalse \mid  \lambda  \mid \RefAlloc\qquad\qquad
\text{label:} \quad
\ell ::= \Call c \mid \Return c
\end{array}$$
We often abbreviate $\Call {()}$ and $\Return {()}$ as
$\Call{}$ and $\Return{}$. A type $ \sigma $ of $\CML$ will be
interpreted as a positive session type of the form
$\oplus_{c \in \Constructors{ \sigma }} \Return {c} \cdot
\TranslateTerm \sigma _c$. Positive types are adequate to model call-by-value reduction as the initial positive move is used to tell \Context that the term does
converge to a value. The session type $\TranslateTerm{\sigma}_c$ represents what happens
\emph{after} the initial constructor has been sent. For base types,
the constructor describes already the value, hence there is nothing to be
done (ie. $ \TranslateTerm{\sigma} _c = \COne$). For non base types, this is more subtle. Given two positive
session types we define their arrow as follows:
$$
\begin{aligned}
  (\oplus_{i  \in  I} \Return {\ell_i} \cdot S_i)  \rightarrow  T &:= \with_{i \in I} \Call {\ell_i} \cdot (S_i^\perp  \parallel  T).\\
\end{aligned}$$

The arrow type between positive types starts with a call from context
with the return values of $S$, and then continues along $S_i^\perp$
and $T$. In particular $T$ only starts after the call message has been
received. Note that this construction does not preserve
positivity. With these elements in hand, we can now define
$\TranslateTerm \sigma $ and $\TranslateTerm \sigma _c$ by mutual
induction:
$$
\begin{array}{rclrcl}
  \TranslateTerm { \sigma } &= &
  \oplus_{c \in \Constructors{ \sigma}} \Return c \cdot \TranslateTerm{ \sigma }_c
  &
    \TranslateTerm {  \sigma   \rightarrow   \tau }_{ \lambda } &= & \oc ( \llbracket  \sigma  \rrbracket  \rightarrow   \llbracket  \tau  \rrbracket) \\
    \TranslateTerm { \sigma }_v & =& \COne\quad (\sigma\text{ base type})& 
  \TranslateTerm {\RefType}_{\mathtt{ref}} &=& \oc (\GetRequest \cdot \TranslateTerm \NatType \with \with_{n  \in   \mathbb{N} } \SetRequest n \cdot \TranslateTerm \UnitType)
\end{array}$$
Reference types are interpreted as objects that can receive two
methods calls: \texttt{get} and $\texttt{set}(k)$ and that return the
appropriate type. The interpretation of a context $ \Delta $ is 
$\TranslateTerm \Delta := \oplus_{c \in \Constructors\Delta} \Return c \cdot
\TranslateTerm \Delta _c$ where
$\TranslateTerm \Delta _c :=
\parallel _{a \in \dom \Delta}
\TranslateTerm { \Delta (a)}_{c(a)}$:
for a context, all the
constructors of all the parameters arrive bundled together. The
interpretaton of an interface is
$\TranslateTerm{ \Delta \vdash \sigma } := \TranslateTerm \Delta
\rightarrow \TranslateTerm \sigma $ (a negative type).

\subsection{Translation of terms}
\label{subsec:translation:terms}
We want to translate a term $ \Delta \vdash M : \sigma $ into
a process
$\Typing{\TranslateTerm M_o} {o: \TranslateTerm{ \Delta \vdash \sigma
  }}$. Doing it directly by induction would lead an inefficient
translation filled with cuts that can be eliminated, as
$\TranslateTerm M_o$ will always be of the shape
$\CBranching o {c \in \Constructors{ \Delta }} {\CBranch c {\vec x}
  S_c}$ (due to the shape of $\TranslateTerm { \Delta   \vdash   \sigma }$). So the primitive
object for the translation of $ \Delta \vdash M : \sigma $ is what
comes after receiving $c \in \Constructors { \Delta }$, ie. a family
of processes
$(\Typing{\TranslateTerm { \Delta \vdash M : \sigma }_{c,
    o}}{\TranslateTerm { \Delta \vdash \sigma }_{c,o}^{\text{ctx}}})$
where $c \in \Constructors \Delta$, $o$ is a name not present in
$ \Delta $, and:
$$\TranslateTerm {a_1:  \sigma _1,  \ldots , a_n:  \sigma _n  \vdash   \sigma }_{c, o} = a_1: \TranslateTerm{ \sigma _1}_{c(a_1)}^\perp,  \ldots , a_1: \TranslateTerm{ \sigma _n}_{c(a_n)}^\perp, o: \TranslateTerm  \sigma .$$
This is well-defined because for any type $ \sigma $ and compatible
$c$, both $\TranslateTerm \sigma $ and $\TranslateTerm \sigma _c$ are
"rooted" session types. From this, we can recover the usual strategy
as follows:
$$\TranslateTerm {a_1:  \sigma _1,  \ldots , a_n:  \sigma _n  \vdash  M :  \sigma }_o := \CBranching o {c \in \Constructors{ \Delta }} {\CBranch {\Call c} {a_1,  \ldots , a_n, o} {\TranslateTerm { \Delta   \vdash  M :  \sigma }_{c, o}}}.$$

 
The interpretation is mostly straightforward and follows the usual
game semantics intuition. To define the behaviour of
$\RefAlloc(\cdot)$, we need to implement a reference server, ie. a
server that waits for requests (get or set) and answers appropriately,
while maintaining as internal state the current value of the
reference. This is done using our codereliction construct
$\CCoder a x$ The server accepts requests one by one, and as such is
infinite.  We write $\Processes \Gamma $ for the $ \omega $-CPO of
well-typed processes on $ \Gamma $. In particular, the least element
of $\Processes \Gamma$ is $\CZero$.  We then define the server via a
least fixpoint of a function on the $ \omega $-CPO
$\Processes {a: \TranslateTerm \RefType_{\mathtt{ref}})}^ \mathbb{N} $
(ordered pointwise) as follows:
$$\mathsf{RefServer}_a ::= \fix \left( \lambda S.\,  \lambda n.\, \CCoder a x
\CBranchingBig x
{\CBranchBig {\GetRequest} r {\CSelectS r {\Return n} S(n)}
  \CBranchBig {\SetRequest k} r {\CSelectS r {\Return {}} S(k)}}\right)$$

The interpretation of typing rules for finite $\CML$ terms
(ie. without fixpoint) is given in Figure \ref{fig:CMLinterp}.
Recursion is simply dealt with by unfolding it. We define the $n$-th
approximant of the fixpoint operator by induction as: 
$Y_0 =  \lambda  \varphi .\,  \lambda x.\,  \bot$ and 
$Y_{n+1} =  \lambda  \varphi .\,  \lambda x.\,  \varphi  ( \lambda y.\, Y_{n}\,  \varphi \, y)\, x$. 
It is straightforward to see that
$\TranslateTerm {Y_n}  \leq  \TranslateTerm {Y_{n+1}}$ in
$\Processes {\TranslateTerm { \:\vdash  (( \sigma   \rightarrow   \tau )  \rightarrow  ( \sigma   \rightarrow   \tau ))  \rightarrow  ( \sigma   \rightarrow   \tau )}}$.
Then, for any $\CML$ term $M$, the $n$-th approximant of $M$, written
$M_n$ is defined as $M$ where $Y$ is replaced by $Y_n$. The chain
$(\TranslateTerm {M_n})_{n  \in   \mathbb{N} }$ is an increasing chain, hence has a
least fixpoint which is the desired translation $\TranslateTerm M$.

\input{figures/translation}

\subsection{Translation of machines}\label{sec:interpcorrect}
\label{subsec:translation:machines}
We now define the translation of machines. As we have seen in
\S~\ref{sec:CML}, the operational semantics of machines observes memory
operation via labelled transitions. To model this in the world of
processes, we show how to emulate these labelled transitions.

\mysubsubsection{Account of visible actions.} Given an alphabet
$ \Sigma $, we construct a session type
$\TranslateTerm \Sigma = \wn {\oplus_{ \alpha \in \Sigma } \alpha
  \cdot \mathtt{ok}^-}$. We translate machines to processes with a
free name $\mathfrak l$ of type $\TranslateTerm \Sigma $. Such
processes support the emission of actions in $ \Sigma $ as follows:
$ \alpha ^{\mathfrak l}.\, P := \CDer {\mathfrak l} x \CSelectS x
\alpha \CBranchingOneS x {\mathtt{ok}} P$. This allows us to define a
variant of the reference server that emits a visible action at each
operation performed:

$$\mathsf{RefServer}^{\mathfrak l}_a ::= \fix \left( \lambda S.\,  \lambda n.\, \CCoder a x
  \CBranchingBig x {\CBranchBig {\GetRequest} r {\Read \ell n^{\mathfrak l}.\,
      \CSelectS r {\Return n} S(n)} \CBranchBig {\SetRequest k} r
    {\Write \ell k^{\mathfrak l}.\, \CSelectS r {\Return {}} S(k)}}\right)$$ Unlike
the previous server, this one logs (emit a visible action) everytime
an action is performed. Given a state $\machine{\vec{y}}{\Delta}{M}{\mu}$ we define the process
$$\TranslateTerm{\machine{\vec{y}}{\Delta}{M}{\mu}}_{\mathfrak l} =
\CCut o {\bar o} ( \nu \vec a\vec {\bar a})
  \left(\TranslateTerm
  {M}_{c_\Delta, o} \parallel ( \parallel _{a \in
    \vec{y}}\mathsf{RefServer}^{\mathfrak l}_{\bar a}) \parallel ( \parallel _{a \in \dom
    \Delta \setminus \vec{y}}\mathsf{RefServer}_{\bar a})\right)
  $$
with $\vec{a} \in \dom \Delta$. 
We only use the logging reference server for public names: every
operation gives rise to a visible action. Note that the interpretation
of a machine is a closed process, thus there is no need for an output
channel as for the terms, as the return value is discarded. Moreover,
in this case, $\Constructors \Delta$ is a singleton, hence the
parameter $c_ \Delta $ is simply the tuple
$(\RefType, \ldots , \RefType)$.

\mysubsubsection{Weak bisimulation and barbed reduction-closed congruence of $\MetalanguageName$} On processes well-typed in the
context $\mathfrak l : \TranslateTerm  \Sigma $, we can define a LTS as
follows:
$$P \xrightarrow{ \tau } Q\, \text{whenever}\, P  \rightarrow  Q\qquad P\xrightarrow{ \alpha } Q\,\text{whenever}\, P  \equiv   \alpha ^{\mathfrak l}.\, Q.$$
Using this LTS, we can define a notion of weak bisimulation on such
processes.
\begin{definition}[Weak bisimulation of $\MetalanguageName$]\label{def:wb_ml}
  An equivalence relation $\mathcal{R}$ over processes typed in the context
  $\mathfrak l : \TranslateTerm  \Sigma $, is a \emph{weak bisimulation} if
  for all $P\, \mathcal{R}\, Q$ with $\Typing{P,Q}{\emptyset}$,
  $P \by{\alpha} P'$, there exists $Q \By{\hat{\alpha}} Q'$ with
  $P'\, \mathcal R\, Q'$.  The largest such bisimulation is called
  ""weak bisimilarity"", denoted by $\MLbisim$.
\end{definition}

\subsection{Correctness of the translation}
\label{subsec:translation:correctness}
\mysubsubsection{Soundness of call-by-value} We first show that our
semantics is well-behaved with respect to the parallel call-by-value
strategy.  First, we observe that values get mapped to specific
processes -- those which emit a constructor right away, without
interrogating the context. Formally, given a value
$ \Delta \vdash V : \sigma $ and given $c \in \Constructors \Delta $,
we define $\intro*{\ConstructorOfValue c V} \in \Constructors \sigma $ by
induction on $V$:
$$\ConstructorOfValue c x = c(x) \qquad \ConstructorOfValue c {\underline n} = n \qquad
\qquad\ConstructorOfValue c {b} = b\quad (b \in \{\ttrue, \ffalse, ()\})\qquad
\ConstructorOfValue c { \lambda x.\, M} =  \lambda $$

\begin{lemma}\label{lemma:interp of values}
  Assume $ \Delta   \vdash  V :  \sigma $. Then
  $\TranslateTerm { \Delta   \vdash  V :  \sigma }_{c, o} \equiv \CSelectS o
  {\Return {\ConstructorOfValue c V}} P$ for some process $P$.
\end{lemma}
\begin{lemma}\label{lemma:substitution}
For $ \Delta   \vdash  V :  \sigma $ and $ \Delta , x:  \sigma   \vdash  M : \tau$, $\TranslateTerm{M[V/x]}_{c, o} \equiv \shareInteractS x {\TranslateTerm M_{c[x:=\ConstructorOfValue c V], o}} {\CProm {\bar x} r \TranslateTerm V_{c, r}} $. 
\end{lemma}
\begin{lemma}\label{lemma:soundness}
  Consider $ \Delta   \vdash  M, N :  \sigma $. If $M  \rightarrow  N$, then for all $c, o$, 
  $\TranslateTerm M_{c, o} \equiv \TranslateTerm N_{c, o}$.
\end{lemma}

\mysubsubsection{Full-abstraction on machines} We now turn our
attention to the LTS on machines. We first start with a 
simulation lemma.
\begin{lemma}
  \label{lemma:cml sim}
If 
  $\machineDefault M \xrightarrow{ \alpha } \machineDefaultt N$, then
  $\TranslateTerm {\machineDefault M}_{\mathfrak l} \By{ \alpha } \TranslateTerm
  {\machineDefaultt N}_{\mathfrak l}$.
\end{lemma}
We now show an adequacy lemma which shows the converse. However,
formulating this result is subtle due to slight differences in
the LTS. In $\CML$, operations on visible locations are performed
directly as visible transitions:
$\machine {x: \RefType} {x} {x := 1 \parallel x := 2} {x \mapsto 0}$
can do either $\Write x 1$ or $\Write x 2$ as initial
transitions. In $\MetalanguageName$, any transition is done in
two stesps: first the program connects to the memory server, which
implies that a "dereliction" meets a "codereliction"; and only then
the action is logged (cf. the definition of
$\mathtt{RefServer}^\ell$). This means that the translation of this
machine can do two $ \tau $-transitions, each followed by a write
action. Hence, the two LTSs are not directly weakly bisimilar, but we
can still show the following result:

\begin{restatable}[Adequacy]{lemma}{Adequacy}\label{lemma:adequacy lemma}
  If $\TranslateTerm{\machineDefault M}_{\mathfrak l} \xrightarrow{ \alpha } P$,
  then $ \alpha = \tau $, and there are two cases:
  \begin{itemize}
  \item Either $\machineDefault M \xrightarrow{ \tau } \machineDefaultt N$
    with $\TranslateTerm {\machineDefaultt N}_{\mathfrak l} \equiv P$,
  \item Or $P  \equiv   \alpha^{\mathfrak l} .\, Q$ and
    $\machineDefault M \xrightarrow{ \alpha } \machineDefaultt N$ with
    $\TranslateTerm {\machineDefaultt N}_{\mathfrak l} \equiv Q$.
  \end{itemize}
\end{restatable}
From this, we can show that our translation preserves and reflects
weak bisimulation:
\begin{restatable}[Soundness and completeness]{theorem}{TranslationCorrect}
  \label{the:trans:correct}
  For any typed 
  $\machine {\vec y} { \Delta } M \mu$ and
  $\machine {\vec y} { \Delta '} N {\mu'}$, the following are
  equivalent:
  \emph{(1)} $\machine {\vec y} { \Delta } M {\mu} \CMLbisim \machine
  {\vec y} { \Delta } N {\mu'}$, \emph{(2)}
  $\TranslateTerm{\machine {\vec y} { \Delta } M \mu}_{\mathfrak l}
  \MLbisim \TranslateTerm {\machine {\vec y} { \Delta } N
    {\mu'}}_{\mathfrak l}$ and \emph{(3)}
  $\TranslateTerm{\machine {\vec y} { \Delta } M \mu}_{\mathfrak l}
  \MLObs \TranslateTerm {\machine {\vec y} { \Delta } N
    {\mu'}}_{\mathfrak l}$
\end{restatable}

From Theorem
\ref{the:trans:correct}, it follows immediately that:
\begin{theorem}[Soundness]
  \label{thm:sound}
  For all $ \Delta   \vdash  M, N :  \sigma $, if
  $\TranslateTerm N_o \MLObs \TranslateTerm N_o$ then $M \CMLObs N$.
\end{theorem}
The converse, known as \emph{full abstraction} in the denotational
semantics community does not hold in general for very well-understood
reasons.  See the Appendix \S~\ref{subsec:discussion} for a
counterexample. However, we still have full abstraction at
second-order interfaces, cf. \S~\ref{sec:cml semantics}.


%% file: figures/translation.tex
\begin{figure}
 \centering
  \small
  $$\begin{aligned}
    \TranslateTerm { \Delta , a: \sigma   \vdash  a:  \sigma }_{c, o} &= \CSelect {o} {\Return{c(a)}} {x} \Forwarder {a} {x} {\TranslateTerm{ \sigma }_{c(a)}} \quad\quad\quad
    \TranslateTerm { \Delta  \vdash  \underline n : \NatType}_{c, o} = \CPositiveActionS {o} {\Return n} \\
    \TranslateTerm { \Delta  \vdash  b : \BoolType}_{c, o} & = \CPositiveActionS {o} {\Return b}
    \hspace{3.7cm}
    \TranslateTerm { \Delta  \vdash  () : \UnitType}_{c, o} = \CPositiveActionS {o} {\Return {}} \\
    \TranslateTerm { \Delta   \vdash   \lambda a.\, M :  \sigma   \rightarrow   \tau }_{c, o} &= \AbsCbv { o } a \sigma {(\TranslateTerm M_{c[a:=c'], o)})_{c'  \in  \Constructors\sigma}} \\
    \TranslateTerm { \Delta   \vdash  M\, N :  \sigma }_{c, o} &=
    \shareInteractS {x, y} {\left(\TranslateTerm M_{c, \bar x} \CFork {\TranslateTerm N_{c, \bar y}}\right)}
    {\\&\CBranchingOneS {y} {\Return {\lambda}} 
      \CBranching {z} {c} {\CBranch {\Return {c}} {\vec w'}
      \CDer {y} {y_0}
      \CSelect {y_0} {\Call c} {\vec w, o'}
     (\Forwarder {\vec w} {\vec {w'}} {} \CFork \Forwarder o {o'} {})}}
    \\
  \end{aligned}$$    
$$\begin{aligned}
\TranslateTerm{ \vdash  \mathtt{plus} : \NatType  \rightarrow  \NatType  \rightarrow  \NatType}_{(), o} &=
\AbsCbv{o}{ \cdot } {\NatType} {(\AbsCbv{o}{ \cdot }{\NatType}{(\CPositiveActionS o {\Return {i+j}})_{j \in  \mathbb{N} }})_{i \in  \mathbb{N} }} \\
\TranslateTerm{ \vdash  \mathtt{equal} : \NatType  \rightarrow  \NatType  \rightarrow  \BoolType}_{(), o} &=
\AbsCbv{o}{ \cdot } {\NatType} {(\AbsCbv{o}{ \cdot }{\NatType}{(\CPositiveActionS o {\Return {i=j}})_{j \in  \mathbb{N} }})_{i \in  \mathbb{N} }} \\
\TranslateTerm{\vdash \mathtt{get} : \RefType \rightarrow \NatType}_{(), o} &=
\AbsCbv{o}{ r } {\RefType} {
  (\CDer r {x}
  \CSelectS x {\GetRequest}
  \CBranching r {n \in  \mathbb{N} } {\CBranchS {\GetAnswer n} \CPositiveActionS o {\Return n}})}\\
\TranslateTerm{\vdash \mathtt{set} : \RefType \rightarrow \NatType \rightarrow \UnitType}_{(), o} &=
\AbsCbv{o}{ r } {\RefType} {
(\AbsCbv{o}{  \cdot  } {\NatType} {(
  \CDer r {x}
  \CSelectS x {\SetRequest i}
  \CBranchingOneS r {\SetAnswer}
  \CPositiveActionS o {\Return {}}
  )_{i \in  \mathbb{N} }})}\\
\TranslateTerm{\vdash \mathtt{ref} : \NatType \rightarrow \RefType}_{c, o} &=
\AbsCbv{o}{  \cdot  } {\NatType} {\left(\CSelect o {\mathsf{ref}} a
\mathsf{RefServer}_a(n)\right)}_{n  \in   \mathbb{N} }\\
  \end{aligned}$$
  \[
  \begin{array}{rcl}
\textbf{Shortcuts:} \quad\quad 
\AbsCbv o {\vec x}{ \sigma } {(P_c)_{c \in \Constructors \sigma}} & := & 
\Typing{\CSelectS o  {\Return{\lambda}}  \CProm o r \CBranching r {c \in \Constructors \sigma} {\CBranch c {\vec x} P_c}} {o: \TranslateTerm { \sigma   \rightarrow   \tau }, \wn  \Delta }
\end{array}
\]
\vspace{-3mm}  
\caption{Interpretation of $\CML$ into the metalanguage}
\vspace{-5mm}  
  \label{fig:CMLinterp}
\end{figure}

%% file: games.tex
In this section, we describe a causal semantics interpretation of
$\MetalanguageName$. This model interprets session types and processes
as event structures, a causal model of concurrent and nondeterministic
computation. The event structures denoted by session types are called
\textbf{games}, while the event structures denoted by processes are
called \textbf{strategies}. The contribution of this section is to
extend the non-angelic model of \citep{fossacs18} with the tools for
replication of \cite{cg2}. This results in the first game semantics model
providing adequate modelling of nondeterministic and nonlinear
languages up to weak bisimulation.

In \S~\ref{sec:es}, we recall Winskel's event structures and introduce
the notations we will be using. In \S~\ref{sec:es:games}, we define the
notion of games of \cite{lics11}, as well as the interpretation of
type formers on it. In \S~\ref{sec:strategies}, we present strategies
based on event structures of \cite{fossacs18}. In
\S~\ref{sec:comp}, we introduce composition of strategies as
well as the categorical structure which is key to interpret
$\MetalanguageName$. Finally, in \S~\ref{sec:replication}, we show
problems related to replication and introduce \emph{\textbf{symmetry}}
to solve them.

\subsection{Event Structures}\label{sec:es}
Our model is based on event structures \cite{eventstructures},
introduced to give a semantic model of causal concurrency (or true
concurrency). From this viewpoint, a system is a set of events along
with a partial order describing the \textbf{causal relationship}
between events; and a binary relation describing \textbf{conflict} (or
incompatibility) between events, ie. those events that cannot occur
together in the same run of the system. As a result, event structures
model the system \textbf{globally} rather than describe independent
executions.
We use here \emph{prime event structures with binary conflict}.

\begin{definition}
An ""event structure"" is a triple $(E,  \leq _E, \#)$ where $(E,  \leq _E)$ is
a partial order, and $\# \subseteq E^2$ is a binary irreflexive symmetric
relation satisfying: (1) 
$[e] \eqdef \{ e'  \in  E \mid e'  \leq  e\}$ is finite, and 
(2) if $e \# e'$ and $e'  \leq  e''$ then $e \# e''$.
\end{definition}
Two events $e,e'$ are in ""conflict"" when $e \# e'$ and are
""compatible"" otherwise.  Two compatible events which are not ordered
are called ""concurrent"". We write $e \imdep e'$ (""immediate causal
dependency"") when $e <_E e'$ with no events in between, and
$e \minconflict e'$ (""minimal conflict"") when $(e, e')$ is the only
conflicting pair in $[e] \cup [e']$. From $\imdep$ and $\minconflict$,
we can recover $ \leq $ and $\#$ via axiom (2). Depictions of "event
structures" will use $\imdep$ and $\minconflict$.

A notion of partial execution can be recoverd on event structures
through the notion of configuration.  Given an "event structure" $E$,
a ""configuration"" of $E$ is a subset $x \subseteq E$ down-closed for
$ \leq _E$ and conflict-free. We write $\config (E)$ for the set of
finite configurations of $E$. For $x \in \config (E)$, an
""extension"" of $x$ is an event $e \in E \setminus x$ such that
$x \cup \{e\} \in \config (E)$, which we write $x \longcov e$.

An event structure $E$ is ""confusion-free"" when (1)
$e \minconflict e'$ implies $[e] \setminus \{e\} = [e'] \setminus \{e'\}$ and
(2) the relation $(\minconflict_E\, \cup\, =_E)$ is an equivalence relation. Its
equivalence classes are called ""cells"". Confusion-free event structure have
\textbf{local} nondeterminism and cells represent choices between different
branches of the program. Finally, given a set $V \subseteq E$, the
""projection"" of $E$ to $V$ is the event structure $E \downarrow V \eqdef (V,
\leq _E \cap V^2, \# \cap V^2)$.

\subsubsection*{Constructions on Event Structures} Given a family of event
structures $(E_{i \in I})$ we define their ""parallel composition"" $ \parallel
_{i \in I} E_i$ as follows. Its events are pairs $(i \in I, e \in
E_i)$. Causality and conflict are obtained by lifting those from the $E_i$:\\
{\centerline{
  $(i, e)  \leq _{ \parallel _{E_i}} (j, e') \eqdef (i = j \wedge e  \leq _{A_i}
  e') \qquad(i, e) \#_{ \parallel _{E_i}} (j, e') \eqdef (i = j \wedge e \#_{A_i}
  e')$
}}\\
Similarly, the ""nondeterministic sum"" of the $E_i$, written
$\sum_{i \in I} E_i$ has the same components as
$ \parallel _{i \in I} E_i$ except for conflict:\\
{\centerline
{$(i, e) \#_{\sum_{i \in I} E_i} (j, e') \eqdef (i = j  \Rightarrow  e \#_{E_i} e').$}}\\[1mm]
The empty event structure, written $\intro*{\EmptyES}$ is the unit for
both parallel composition and sum.

\subsection{Games based on Event Structures}\label{sec:es:games}
We use event structures to represent the semantic information of
session types. This is a natural fit as session types feature both
\textbf{causality}, for some messages must come before some other, and
\textbf{conflict}, for choices are offered to both participant.
In this paper, we use a simple notion of games.

\begin{definition}
  A ""game"" is an event structure $A$ along with a labelling
  $\pol : A  \rightarrow  \{\Neg -, \Pos +\}$ such that:
  \begin{enumerate}
  \item $A$ is "confusion-free" and ""race-free"": if $a \minconflict_A a'$, then $\pol(a) = \pol(a')$
  \item $(A,  \leq _A)$ is a ""forest"": elements of $[a]$ are totally ordered
    by $ \leq _A$ for any $a  \in  A$.
  \end{enumerate}
\end{definition}
\noindent 
The first axiom comes from the local and polarised aspects of
choices in session types of $\MetalanguageName$. Parallel
composition of event structures extends to games. Given a "game" $A$
and $\ell \not\in A$, we write $\ell^p \cdot A$, where
$p \in \{\Neg -, \Pos +\}$ is a polarity, for the game $A$ prefixed by
$\ell^p$ defined as follows. Its events are $A \cup \{\ell\}$;
causality is $ \leq _A \cup \{\ell\} \times (A \cup \{\ell\})$; and
conflict is $\conflict_A$. In general sums of games are not
games as they may fail confusion-freeness or race-freeness, but the
prefixed sum of games is well-defined: if $A_i$ is a family of games,
then $\sum_{i \in I} \ell_i^p \cdot A_i$ is also a game (because the
polarity $p$ does not depend on on $i$). Given a "game" $A$, its
""dual"" $A^\perp$ is the game obtained by reversing
polarities in $A$. 

Moves of a game are meant to be played once, therefore to represent
exponentials we need to make copies of a game to allow a move to
be played several times. We define the unpolarised exponential
$\intro*{\sharpG A}$ as the infinite parallel composition
$ \parallel _{i \in \mathbb{N} } A$. Events of $\sharpG A$ are pairs
$(i, a)$ of a ""copy index"" $i \in \mathbb{N} $ and an element
$a \in A$. We sometimes write $a_i$ (especially in diagrams).

Examples of games related to the interpretation of $\CML$ are found in
Figures~\ref{fig:ex:negation} and \ref{fig:ex:ho}.

\subsection{Strategies as Event Structures}\label{sec:strategies}
We recall the strategies of \cite{fossacs18}. As we have seen in
\S~\ref{sec:overview}, the game expresses all possible behaviours
allowed by the type, while the strategy selects those behaviours
exhibited by the program. As a result, in a deterministic setting, a
strategy on a game $A$ is actually a subset of $A$. In general though,
the strategy might make nonidempotent choices, so a strategy will be an event
structure $S$, along with a labelling function $S  \rightarrow  A$. For the
strategy to respect the rules of the game (eg. if $a < b$, then always
play $a$ before $b$), we need the function to be a \emph{map of event
  structures}, ensuring that every behaviour of $S$
is included in that of $A$.

\begin{definition}
  A ""partial map of event structures"" is a partial function
  $f : E \rightharpoonup F$ such that for $x  \in  \config (E)$,
  $f\, x  \in  \config (F)$ and which is ""locally injective@local injectivity"":
  $f$ restricted to any configuration of $E$ is injective. We write
  $\intro*{\dom f}$ for the domain of $f$, and  $f$ is said to be ""total@total
  map"" when $\dom f = E$. A map represents a simulation of $E$ by $F$. Event
  structures and total maps form a category
  $\intro*{\ES}$.
\end{definition}
\begin{definition}
  A ""strategy"" on a "game" $A$ is a pair $(S,  \sigma )$ of an event
  structure $S$ and a partial map of event structures $ \sigma : S
  \rightharpoonup  A$ with the following properties:
  \begin{description}
  \item [""Receptivity""] For $x \in \config (S)$ and $a$ a negative
    "extension" of $ \sigma \, x$, there is a unique $x \longcov s$ with
    $ \sigma \, s = a$.
  \item[""Courtesy""] For any $s \imdep _S \, s'$ such that $ \sigma \, s$
    and $ \sigma \, s'$ are not related by $ \leq _A$ (possibly because one
    of them is undefined), the event $s$ is not mapped to a positive move
    of $A$, and $s'$ not mapped to a negative move of $A$.
  \item[""Secrecy""] If $s \minconflict s'$, then $s$ and $s'$ are not
    mapped to positive moves of $A$.
  \end{description}
\end{definition}
Partiality allows $S$ to have internal events (akin to $ \tau $
transitions) not corresponding to any moves of the game. An event
$s \in S$ is ""internal"" or "neutral" when $ \sigma \, s$ is not
defined; ""visible"" or "external" when it is.  "Receptivity" ensures
that a strategy cannot prevent Opponent from playing. "Courtesy"
restricts the shape of immediate causal links in the strategy to only be
from negative or neutral events to positive or neutral. Those links
are asynchronous and are essential to ensure that we have a
categorical identity. Finally, "secrecy" (together with "receptivity")
restricts the shape of minimal conflict: either between negative moves
or between internal moves. The former are external choices (Opponent
chooses) while the latter are internal choices (Player chooses).

We tend to write simply $ \sigma , \tau , \ldots $ for strategies and
assume the underlying event structure is called $S, T, ...$ A strategy
from a game $A$ to a game $B$ is a strategy on the composite game
$A^\perp \parallel B$. We write $ \sigma : A \profto B$ to distinguish
them from maps. Strategies are displayed as labelled
\begin{wrapfigure}[7]{r}{0.33\linewidth}
  \vspace{-1.4em}
  \centering
  \begin{tikzpicture}
    \matrix[diag] {
      |(a)|  \ast _{\Write x 1} \& |(b)|  \ast _{\Write x 2}\\
      |(c)|  \ast _{\Write x 2} \& |(d)|  \ast _{\Write x 1}\\
      |(e)| \Pla{\Return 1} \& |(f)| \Pla {\Return 1}\\
    };
    \path[cause] (a) edge (c) (c) edge (e) (b) edge (d) (d) edge (f);
    \path (a) edge[conflict] (b);
  \end{tikzpicture}
  \vspace{-1.0em}
  \caption{Nondeterministic strategy}
  \label{fig:ex:nd}
\end{wrapfigure}
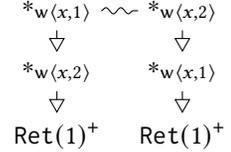 event structures, as in
Figures~\ref{fig:ex:negation} and \ref{fig:ex:ho}, which are examples
of total strategies, ie. for which there are no internal
moves. Because of the conflict axiom, such strategies must be also
deterministic (no internal choices, only external choices).  On the
right, the interpretation as a strategy of the nondeterministic
program $(x := 1 \parallel x := 2); 1$ is depicted. The nondeterminism
arises from a race between two writes on the same location. In this
example, the map $ \sigma $ is not injective (it plays twice
$\Return 1$), because there might be two reasons for returning this
value, depending on how the race is resolved. In the diagram, we use
$\ast$ for neutral events, annotated with a label describing the
operation \textemdash but this information is not part of the
mathematical object.  \mysubsubsection{Copycat strategy}For a "game"
$A$, we form the ""copycat strategy"" on $A$,
$\cc_A : \CC_A \rightarrow A^\perp \parallel A$, as follows. The
events of $\CC_A$ are exactly those of $A^\perp \parallel A$, and
$ \leq _{\CC_A}$ is obtained from the transitive closure of
$ \leq _{A^\perp \parallel A} \cup \{ (a, \bar a) \mid a \in A^\perp
\parallel A\text{ is negative}\}$ where we write
$\bar a \in A^\perp \parallel A$ for the corresponding event to $a$ on
the other side. Then $a \#_{\CC_A} a'$ holds when
$[a]_{\CC_A} \cup [a']_{\CC_A} \not\in \config (A^\perp \parallel A)$. The
labelling function $\cc_A$ is simply the identity. A strategy is
""negative@negative strategy"" when its minimal events are all
negative. Copycat for example is a negative strategy.

The empty strategy $\intro*{\weakening A}$ on a game $A$ is defined as
the inclusion $A_0 \subseteq A$ where $A_0$ contains events $a \in A$
such that $[a]$ only contains negative events (ie. it is the minimal
receptive strategy).

\mysubsubsection{Isomorphism of strategies} Because $S$ is arbitrary,
equality of strategies is not meaningful. Rather, we consider strategies
up to isomorphism. Two strategies $ \sigma : S \rightharpoonup A$ and
$ \tau : T \rightharpoonup A$ are ""isomorphic"" (written
$ \sigma \cong \tau $) when there exists an isomorphism of event
structures $ \varphi : S \cong T$ (that is a map $S \to T$ with an
inverse map) such that $ \tau \circ \varphi = \sigma $.

\subsection{Composition of strategies and categorical structure}\label{sec:comp}
We now explore an important algebraic structure on strategies:
composition. Composition is the semantic counterpart to the operation
$\stackrel {a}{\odot}$ which connects two strategies and make them
interact on a common part. Concretely, given $ \sigma  : A \profto B$ and
$ \tau  : B\profto C$, we wish to define $ \tau   \odot   \sigma  : A \profto C$ obtained by
making $ \sigma $ and $ \tau $ interact on $B$ and hiding the resulting
interaction. This composition allows us to define a category of games
and strategies. This is useful for two main reasons:
\begin{itemize}
\item It allows us to give a simple meaning to the restriction
  $\CCut a b P$ by simply composing $P$ with an asynchronous forwarder
  that connects $a$ and $b$.
\item It allows us to define the interpretation of constructs of the
  metalanguage by composing with simple blocks rather than some ad-hoc
  construction. For instance, to interpret $\CSelectS a \ell P$,
  instead of somehow prefixing $ \llbracket P \rrbracket$ by an event labelled with
  $\CPositiveActionS a \ell$, we prefer to compose $ \llbracket P \rrbracket$ by a strategy
  $ \iota _{\CPositiveActionS a \ell}$ which represents an injection for a
  weak coproduct structure. This allows us to show the correctness of
  the translation by simple calculations on these little blocks rather
  than with an arbitrary context $P$.
\end{itemize}
\subsubsection*{Composition of Strategies}\label{sec:compStrat}
Composition is defined in two steps: interaction, and then
hiding. Consider a strategy $ \sigma : A \profto B$ from $A$ to $B$,
and a strategy $ \tau : B \profto C$ from $B$ to $C$. We first start
by forming their interaction, which is an event structure $T  \circledast  S$
along with a map $ \tau  \circledast  \sigma  : T  \circledast  S  \rightharpoonup  A^\perp  \parallel  B  \parallel  C$ (during interaction,
polarities on $B$ are meaningless). This event structure represents
the common behaviours of $S$ and $T$ that can be reached through a
common order. Intuitively, on the causal aspect, this amounts to make
the union of the partial orders of $S$ and $T$ and remove cycles. This
operation is somewhat involved to define concretely, but enjoys the following
universal property.\footnote{In the case without internal events, this is a
  pullback, but in this setting, the pullback exists but does not
  capture the right operational intuitions.}

\begin{lemma}[\cite{fossacs18}]
  There exists a unique tuple $(T  \circledast  S,  \tau   \circledast   \sigma ,  \Pi _1 : T  \circledast  S  \rightharpoonup  S,  \Pi _2 : T  \circledast  S  \rightharpoonup  T)$ such that
  \begin{enumerate}
  \item If $ \Pi _1\, e$ and $ \Pi _2\, e$ are defined then $ \sigma ( \Pi _1\, e)$ and
    $ \tau ( \Pi _2\, e)$ are both defined and equal (in $B$).
  \item Moreover, for any other tuple
    $(X,  \chi  : X  \rightharpoonup  A  \parallel  B  \parallel  C,  \Pi '_1 : X  \rightharpoonup  S,  \Pi '_2 : X  \rightharpoonup  T)$ satisfying
    the first property, there exists a unique map $ \varphi : X  \rightharpoonup  T  \circledast  S$ such that
    $ \Pi _1 \circ  \varphi  =  \Pi '_1$ and $ \Pi _2 \circ  \varphi  =  \Pi '_2$.
  \end{enumerate}
\end{lemma}
\noindent This ensures that $T \circledast S$ is well-defined up to
isomorphism. The second step is to recover a strategy on
$A^\perp \parallel C$ by hiding those events of $T \circledast S$
projected to $B$. Because of "secrecy", events in $B$ cannot
partake in a minimal conflict, so they can be hidden without losing
information up to bisimulation.

Letting $V$ be the set of events $e$ of $T  \circledast  S$ such that if
$ \tau   \circledast   \sigma (e)$ is defined, it is not in $B$, we define $T \odot S$ to be
$T \circledast S \downarrow V$. The map $ \tau \circledast \sigma $
restricts to a map
$ \tau \odot \sigma : T \odot S \rightharpoonup A^\perp \parallel C$
which is the ""composition"" of $ \sigma $ and $ \tau $.  It is a
strategy.

\begin{lemma}[\cite{fossacs18}]
  "Composition" of strategies is associative, and "copycat" is an
  identity on "strategies", both up to isomorphism.
\end{lemma}
\noindent For copycat to be an identity, the conditions of "receptivity" and
"courtesy" are necessary. This is because composing a
strategy with copycat amounts to prefixing it with an asynchronous
buffer; and if the strategy is too ``synchronous'', this buffer may
alter its observable behaviour. 

\subsubsection*{Categorical structure}\label{sec:lindet cat} We define $\intro*{\Strat}$, the category
whose objects are finite "games" and morphisms from $A$ to $B$ are
"strategies" $ \sigma : A \profto B$, considered up to
isomorphism. First, duality of games extends to an isomorphism of
category $\Strat \cong \Strat^{\text{op}}$: any strategy
$ \sigma : A \profto B$ can be regarded as a strategy
$ \sigma ^\perp : B^\perp \profto A^\perp$. Moreover, parallel
composition of games forms a monoidal structure on
$\Strat$. Therefore, $\Strat$ is compact-closed:
\begin{lemma}[\cite{fossacs18}]
  $(\Strat,  \parallel , \EmptyGame, (-)^\perp)$ is a compact-closed category.
\end{lemma}
\noindent Though $\Strat$ does not have (co)products, it has weak
(co)products. Given a family of games
$(A_k)_{k \in K}$ and of labels $(\ell_k)_{k \in K}$, we define
$\with_{k \in K} \ell_k^- \cdot A_k \eqdef \sum_{k \in K} \ell_k^-
\cdot A_k$. There is a natural strategy
$ \pi _k : \with_{k  \in  K} \ell_k^- \cdot A_k \profto A_k$ which plays
$\ell_k$ on the right (a positive move as it is on the left), and then
plays copycat between the left $A_k$ and the right $A_k$.
\begin{restatable}{lemma}{productLinDet}\label{lemma:product}
  For every family of strategies $( \sigma _k : A \profto B_k)_{k  \in  K}$
  there exists a strategy
  $ \intro*{\Pairing { \sigma _k}_{k  \in  K}} : A \profto \with_{k  \in  K} \ell_k^- \cdot B_k$ such that
  $ \pi _k  \odot   \Pairing {\sigma_k}_{k  \in  K}  \cong  \sigma _k$
\end{restatable}
By duality, we get weak coproducts:
$\oplus_{k \in K} \ell_k^+ \cdot A_k \eqdef \sum_{k \in K} \ell_k^+
\cdot A_k$, and define
$$ \iota _k \eqdef  \pi _k^\perp : A_k \profto \oplus_{k \in K} \ell_k^+ \cdot A_k.$$
These are used to interpret the additive fragment of the metalanguage.

\subsection{Replication and symmetry}\label{sec:replication}
We have weak co-products, and compact-closedness so we can interpret a
large part of $\MetalanguageName$. For exponentials, we would be
tempted to use $\sharpG {(-)}$ construction to interpret
$\oc(\cdot)$ and $\wn(\cdot)$. However, for the interpretation be to
sound, we would require special structural morphisms. It is well-known
from the Linear Logic categorical semantics
\citep{mellies2008categorical} that we would need $\sharpG {(-)}$ to
have a (co)monadic structure to be able to interpret correctly the
promotion, and dereliction rule.

There are natural candidates for a monad structure on $\sharpG {(-)}$:
$$
   \eta _E: (a \in E) \mapsto (0, a) \in \sharpG E \qquad \qquad \mu _E: \big((i, (j, a)) \in \sharpG {\sharpG E}\big)  \mapsto \big((\EncodingPairs {i, j}, a) \in \sharpG E\big) 
$$
These can be lifted to strategies, however one can see already in
$\ES$ those do not satisfy the monadic laws: indeed
$\mu_E \circ \eta_E : \sharpG {A}  \rightarrow  \sharpG{A}$ sends $(i, a)$ to
$( \langle i, 0 \rangle , a)$ which may not be the same as $(i, a)$.

\mysubsubsection{Event Structures with Symmetry} The solution to this
problem is to relax the notion of equality to be \emph{up to copy
  indices}. This intuition is formalised through \textbf{event
  structures with symmetry}:
\begin{definition}
  An ""isomorphism family"" on an event structure $E$ is a family
  $\tilde E$ of order\-/isomorphisms $ \varphi : x  \cong  y$ with $x, y  \in   \config (E)$ such
  that
  \begin{enumerate}
  \item $\tilde E$ contains all identities, and is stable under inverse and composition
  \item If $x \subseteq x'$ and $ \varphi  : x'  \cong  y'  \in  \tilde E$, then $ \varphi \vert_x : x  \cong   \varphi \, x \in \tilde E$
  \item If $x \subseteq x'$ and $ \varphi  : x  \cong  y \in \tilde E$, then there
    exists a (non-necessarily unique) $y' \in \config(E)$ such that $ \varphi $
    extends to $ \varphi ' : x'  \cong  y' \in \tilde E$.
  \end{enumerate}
  An ""event structure with symmetry"" is a pair $(E, \tilde E)$ of an
  event structure $E$ and an "isomorphism family" $\tilde E$ on $E$.
We write $\mathcal E, \mathcal F, ...$ for "event structures with
symmetry".  
\end{definition}
Most constructions on event structures (such as "parallel
composition", "sum", prefixing) can be extended seamlessly to "event
structures with symmetry" \cite{winskel2007event}. 
The main purpose of these objects is to support a weaker equivalence
relation on maps. Consider two partial maps
$f, g : E \rightharpoonup F$ between event structures and an
isomorphism family $\tilde F$ on $F$. These maps are \textbf{similar} wrt
$\tilde F$, written $f \intro*{\similarMaps}_{\tilde F} \, g$ when for
all $x \in \mathscr{C} (E)$, the mapping $f\, e \mapsto g\, e$ defines
a bijection $f\, x \cong g\, x$ which is in $\tilde F$. When the
family is clear from context, we write
$f \similarMaps g$.

Now, we can extend the $\sharpG{(-)}$ construction to work at the
level of symmetry. Given an event structure with symmetry $\E$, define
an isomorphism family $\sharpG \tilde E$ on $\sharpG E$ containing all
the $ \theta : x \cong y$ for which there exists
$ \pi : \mathbb{N} \rightarrow \mathbb{N} $ such that
$(1)\, \text{for $(i, e)  \in  x$, $ \theta \, (i, e)$ is of the form $( \pi \, i, e')$} \text{and} (2)\, \{ (e, e') \mid \theta (i, e) = ( \pi \,
i, e') \} \in \tilde E$. We define the "event structure with
symmetry" $\sharpG \E$ as $(\sharpG E, \sharpG \tilde E)$. The induced
equivalence formally captures to be equal ``up to copy indices''.

A ""map of event structures with symmetry""
$f : \mathcal E \rightharpoonup \mathcal F$ is a "map"
$f : E \rightharpoonup F$ such that for all $ \theta \in \tilde E$,
$f\, \theta \eqdef \{ (f\, e, f\, e') \mid (e, e') \in \theta \}$ is
in $\tilde F$. Event structures with symmetry and total maps form a
category $\intro*{\ESS}$. \citet{winskel2007event} proved that
$\sharpG {(-)} : \ESS \rightarrow \ESS$ is a monad up to
$\similarMaps$.

\mysubsubsection{Thin concurrent games with symmetry} Simply equipping
games with one symmetry works \cite{lics14} but leads to a setting
uncomfortable for semantics. We follow here \cite{cg2} and add three
symmetries on games: one global symmetry, and one subsymmetry for each
player. (Details of this section appear in the unpublished thesis
\cite{these}).

\begin{definition}
  A ""thin concurrent game"" (tcg) is a tuple
  $\A = (A, \tilde A, \tilde A_+, \tilde A_-)$ where $A$ is a game,
  and $\tilde A, \tilde A_+, \tilde A_-$ are isomorphism families on
  $A$ subject to axioms listed in \cite{cg2}; in particular
  $\tilde A_-$ and $\tilde A_+$ are sub-isomorphism families of $\tilde A$.
\end{definition}
The intuition is that $\tilde A_-$ (resp. $\tilde A_+$) contains only
isomorphisms that affect negative (resp. positive) events. This idea
of polarised decomposition of the symmetry was first introduced by
\citet{mellies2003asynchronous} in a simpler setting. Game operations
(parallel composition, dual, prefixed sums) extend to tcgs.

However, in general $\sharpG {\mathcal A}$ is not a tcg even if
$\mathcal A$ is. It only works for polarised games: a game is
""negative@negative game"" (resp. ""positive@positive game""), when
all its minimal events are negative (resp. positive). A game is ""polarised"" when it is either negative or positive.
\begin{lemma}[\cite{cg2}]
  If $\A$ is a "polarised" tcg, then $\sharpG \A$ can be made into a tcg.
\end{lemma}
A negative symmetry in $\sharpG \A$ leaves \emph{positive} copy
indices unchanged, and conversely for positive symmetries. Since the
interpretation of session types is not polarised, we need to
explicitly \emph{lift} them before using $\sharpG (-)$: indeed a lifted
game is always polarised as it has a unique minimal event. Depending
on the polarity of the minimal event, we get the two exponentials. We
define:
$$\intro*{\ocG \A} = \sharpG {(\Request^-\cdot \A)} \qquad \intro*{\wnG \A} = \sharpG {(\Request^+\cdot \A)}.$$

\mysubsubsection{Uniform Strategies} Symmetry on games induces
naturally an equivalence relation on strategies, relaxing
"isomorphism" by asking that the triangle commutes up to symmetry on
the game. This equivalence, however, is not a congruence since nothing
prevents a strategy from observing copy indices from Opponent. To
recover a well-behaved compositional setting, we need to ensure that
the strategy we consider behaves uniformly with respect to copy
indices from Opponent. It turns out that we need to add extra
structures on strategies, under the form of an isomorphism family.

\begin{definition}
  Consider a "tcg" $\A$.  A ""uniformity witness"" for an essential strategy
  $\sigma  : S \rightharpoonup  A$ is an isomorphism family $\tilde S$ on $S$
  such that:
  \begin{enumerate}
  \item $ \sigma $ becomes a "map of event structures with symmetry"
    $(S, \tilde S) \rightarrow  (A, \tilde A)$;
  \item if $ \theta  : x  \cong  y \in \tilde S$ and $ \sigma \,  \theta $ extends to
    $ \varphi  : x'  \cong  y' \in \tilde A$ with $x \subseteq^- x'$, then $ \theta $ extends
    to a $ \theta '$ such that $ \sigma \,  \theta ' =  \varphi $;
  \item if $ \theta  : x  \cong  y \in \tilde S$ is the identity on negative
    elements of $x$, then $ \theta $ is the identity on $x$.
  \end{enumerate}
  A ""uniform strategy"" on $\A$ is a "strategy" $ \sigma $ on $A$
  along with a "uniformity witness" for $ \sigma $.
\end{definition}

\noindent
We can lift the equivalence relation $\similarMaps$ from "maps@mapess" to strategies: two "uniform strategies" $ \sigma  : S  \rightarrow
A$ and $ \tau  : T  \rightarrow  A$ are ""weakly isomorphic"" (written $ \sigma
\weakIso  \tau $) when there exists an isomorphism of "event structures with
symmetry" $ \varphi  : \mathcal S  \cong  \mathcal T$  such that $ \tau  \circ
\varphi  \similarMaps  \sigma $.

\begin{theorem}[\cite{these}]
  Uniformity witnesses compose and "weak isomorphism" is a congruence
  on uniform strategies. As result, "tcgs" and "uniform strategies" up
  to "weak isomorphism" form a compact-closed category
  $\intro*{\UStrat}$.
\end{theorem}
The weak products of $\Strat$ induce weak products in $\UStrat$ in a
straighforward manner. We also get the desired In this settting, we
get the desired (co)monadic structure:
\begin{restatable}{lemma}{comonad}
  The construction $\oc(-)$ extends to an exponential comonad
  $\UStrat  \rightarrow  \UStrat$, that is it comes equipped a functorial action plus the following strategies:
$$\textbf{return:}\ \intro*{\dereliction \A} : \oc \A \profto \A\qquad
\textbf{digging:}\ \intro*{\digging \A} : \oc \A \profto \oc \oc \A
\qquad \textbf{contraction:}\ \intro*{\contraction \A} : \oc A \profto
\oc A \parallel \oc A.$$ satisfying standard laws \cite{mellies2008categorical}.
\end{restatable}

\mysubsubsection{Recursion} We finish by detailing the structure in
$\UStrat$ to interpret infinite processes. An ""embedding"" of
$ \sigma $ into $ \tau $ (both strategies on $\A$) is a map
$f : \mathcal S \rightarrow \mathcal T$ such that $ \tau \circ f \sim \sigma $ and such
that $f$ restricted to its image defines an isomorphism. We write
$f: \sigma \embedding {} \tau $. This notion is useful to define the
meaning of infinite objects by taking a limit of increasing
strategies.

\begin{restatable}{lemma}{Recursion}\label{lemma:rec}
  Any countable chain of embeddings
  $(f_i : \sigma _i \embedding{} \sigma _{i+1})_{i \in \mathbb{N}}$
  has a colimit, ie. there exists a unique a strategy $ \sigma $ and
  embeddings $g_i: \sigma _i \embedding{} \sigma $ such that
  $g_{i+1} \circ f_i = g_i$.
\end{restatable}


%% file: interpretation.tex
We are ready to interpret $\MetalanguageName$ into $\UStrat$. In
\S~\ref{sec:interpinterp}, we show how to interpret  session types and
processes. In \S~\ref{sec:full abstraction}, we prove that the
interpretation is fully abstract with respect to a 
weak bisimulation on strategies. Finally, in \S~\ref{sec:cml
  semantics} we discuss the resulting causal semantics of $\CML$.

\subsection{Semantics interpretation}\label{sec:interpinterp}
\subsubsection*{\bf Interpretation of types}
The interpretation of types is straightforward, since we have seen the
semantic constructions corresponding to type formers in
\S~\ref{sec:games}:
$\llbracket \oc T \rrbracket = \oc \llbracket T \rrbracket$,  
$\llbracket \wn T \rrbracket = \wn \llbracket T \rrbracket$, 
\begin{gather*}
\llbracket T_1  \parallel   \ldots   \parallel  T_n \rrbracket =  \llbracket T_1 \rrbracket  \parallel   \ldots   \parallel   \llbracket T_n \rrbracket,  
\quad
\llbracket \with_{i \in I} \ell^-_i \cdot T_i \rrbracket = \with_{i \in I}
\ell_i^- \cdot \llbracket T_i \rrbracket 
\ \text{and}\
\llbracket \oplus_{i \in I} \ell_i^+ \cdot T_i \rrbracket = \oplus_{i \in I} \ell_i^+ \cdot \llbracket T_i \rrbracket.
\end{gather*}
Minimal events of $ \llbracket S \rrbracket$ are either of the form
$\ell \in \mathbb L$ (a label) or $\Request$ for starting new
requests. A session type $S$ is "rooted" if and only if all
minimal events of $ \llbracket S \rrbracket$ are in conflict. The interpretation of
contexts is a simple parallel composition:
$ \llbracket a_1: S_1,  \ldots , a_n: S_n \rrbracket =  \llbracket S_1 \rrbracket  \parallel   \ldots   \parallel   \llbracket S_n \rrbracket.$
Minimal events of $ \llbracket  \Delta  \rrbracket$ are of the form $(a, v)$ where
$a  \in  \dom{\Delta}$ and $v$ is the minimal event of $ \llbracket  \Delta (a) \rrbracket$.
\subsubsection*{\bf Interpretation of terms} To interpret terms, we
make use of the categorical structure presented in
\S~\ref{sec:games}. In particular, if $\ContextElaboration {\vec x} S$
is defined, then there is an isomorphism
$ \varphi _{\ContextElaboration {\vec x} S} : \llbracket
\ContextElaboration {\vec x} S \rrbracket \cong \llbracket S
\rrbracket$ obtained from the monoidal structure. Given an isomorphism
$ \varphi : \A \cong \B$ between tcgs, and $ \sigma : \A$ a "uniform
strategy", we write $ \varphi ^*( \sigma ) : \B$ for the reindexed
strategy on $\B$.

What remains to be constructed in the semantic world is how to
interpret codereliction, ie. how to introduce nondeterminism and
races. Such a construct is interpreted in $\UStrat$ through a morphism
$\codermap_A : \oc A \parallel A \profto \oc A$.

\begin{wrapfigure}[11]{r}{0.48\linewidth}
  \vspace{-1em}
  \scalebox{0.8} {
  \begin{tikzpicture}
    \matrix[diag, column sep=0] {
      \& \oc A \&  \parallel  \& A \& \rightarrow \& \& \oc A \\[-1mm]
      \& \& \&  \& \& |(bi)| \Request^-_i \& \& |(bj)| \Request^-_j \\
       \& \&  \& \& \& |(e1)|  \ast_i  \& \& |(e2)| \ast_j \\
       |(bj')| \Request^+_{j, i} \& |(bi')| \Request^+_{i, j} \& \&|(d)| \texttt{a}^-\& \& |(d1)|   \langle \texttt{a},i \rangle ^+\& \& |(d2)|  \langle \texttt{a}, j \rangle ^+ \\
       |(d2')|  \langle \texttt{a},j, i \rangle ^-\& |(d1')|  \langle \texttt{a}, i, j \rangle ^-\\[-1em]
 \& \& \& \& \&       |(d2'')|  \langle \texttt{a},j, i \rangle ^+ \&  \&|(d1'')|  \langle \texttt{a},i, j \rangle ^+\\
    };
    \path (e1) edge[conflict] (e2);
    \path[cause]
    (bi) edge (e1) (bj) edge (e2) (e1) edge (d1) (e2) edge (d2)
    (bi) edge[bend right=20] (bi') (bj) edge[bend right=35] (bj')
    (e1) edge [bend right=20] (bj') (e2) edge[bend right=20] (bi')
    (bi') edge (d1') (bj') edge (d2')
    (d1') edge[bend left=10] (d1'') (d2') edge[bend right=20] (d2'')
    (d) edge (d1) (d) edge[bend right] (d2);
  \end{tikzpicture}
  }
    \vspace{-1em}
  \caption{Strategy $\codermap_A$ for $A = \mathtt{a}^+$}
  \label{fig:cA}
\end{wrapfigure}
The intuition is that requests on the right side race to get forwarded
to $A$; requests that did not win the race are then forwarded to the
left $\oc A$. Since this strategy has a complex operational behaviour,
let us start by an informal description, based on the example where
the game has a single positive move, ie. $A = a^+$. The desired
strategy $\codermap_A$ for a single move game $A = \mathtt{a}^+$ is (partially)
depicted in Figure \ref{fig:cA}.  The strategy starts by waiting for a
request on the rightmost $\oc A$. The different $\Request^-_i$ are all
racing together to be considered the \textbf{first} message
acknowledged by $\codermap_A$. Hence, one neutral event per
$\Request^-_i$  is triggered when the $i$th message wins the race. If
the $i$th message has won, then the copy of $A$ on the left is put in
contact with the successor of $\Request^-_i$, expressed by the
links $a^- \rightarrowtriangle  \langle a, i \rangle ^+$ and
$ \ast _i \rightarrowtriangle  \langle a,i \rangle ^+$. Then, if $i$ has won the race,
it means that any other $j \neq i$ lost. In that case, messages
on the $j$ component are forwarded to the leftmost $\oc A$:
$\Request^-_j  \rightarrowtriangle  \Request^+_{j, i}$ and
$ \ast _i  \rightarrowtriangle  \Request^+_{j, i}$. From there, we have a copycat strategy
between the $i$th copy of $A$ on the left and on the right.
Formally, the event structure $C_A$ is defined as follows.
\textbf{Events} are of one of the following form:
\begin{itemize}
\item an event $\Request_i^-$ for $i  \in   \mathbb{N} $ mapped to the initial move of the right $\oc A$;
  \item an internal event $ \ast _i$, representing the fact that
    the $i$th request has won the race
  \item an event $\Request_{i,j}^{+}$ for $i \neq j$,
    corresponding to the forwarding of $\Request_i^-$ when $j$ wins the
    race;
  \item an event $ \langle e, i \rangle $ for every $e \in \CC_A$ and $i \in  \mathbb{N} $
    representing the forwarding to $A$ when the race is won by $i$;
  \item an event $ \langle e, i, j \rangle $ for every distinct numbers
    $i, j \neq \mathbb{N} $ and $e  \in  \CC_A$, for forwarding the $i$th
    copy to $\oc A$ when $i$ loses the race to $j$.
  \end{itemize}
\textbf{Causality} includes the usual causal order on the copies of $\CC_A$, plus the following links:
  \begin{itemize}
  \item $\Request^-_i  \rightarrowtriangle   \ast _i$ and $ \ast _i  \leq   \langle e, i \rangle $ for all $e  \in  \CC_A$;
  \item $\Request^-_i  \rightarrowtriangle
    \Request_{i,j}^{+}$ and 
$\ast _j  \rightarrowtriangle  \Request_{i,j}^{+}$ for any
    $i \neq j$: $\Request_{i,j}^+$ is played when Opponent made the
    $i$-th request and $j$ won the race; and 
  \item $\Request_{i, j}^+  \leq   \langle e, i, j \rangle $ for $e  \in  \CC_A$.
\end{itemize}
{\textbf{Conflict}} is generated by asking that the $ \ast _i$ are all in
    mutual conflict.\\[1mm]
    \noindent $C_A$ might not be receptive if $A$ is positive (as in
    the example), so that the strategy $\codermap_A$ is obtained by
    precomposing with copycat to ensure receptivity. The
    interpretation of finite processes is given in
    Figure~\ref{fig:interp-metalanguage}. Infinite processes are
    obtained as limits of chain of interpretation of their approximants, since the
    interpretation is easily seen to be monotonic.

\begin{figure}
\small
  \begin{align*}
 \left \llbracket
      \infer{\Typing \CZero  \Delta }{ }
      \right \rrbracket \ = \ \weakening  \Delta 
& \quad \quad \quad 
  \left\llbracket\vcenter{
  \infer{\Typing {P \CFork Q}{ \wn \Gamma, \Gamma_1 ,  \Gamma_2}}
  {\Typing {P} { \wn \Gamma,\Gamma_1} \qquad \Typing Q { \wn \Gamma,\Gamma_2}}}
  \right\rrbracket \ = \ \contraction { \llbracket T \rrbracket}  \odot (\llbracket
  P \rrbracket  \parallel   \llbracket Q \rrbracket)
\\
\left\llbracket\vcenter{\infer
  {\Typing {\CSelect a {\ell_k} {\vec x} P}{ \Gamma , a: \oplus_{i \in I} \ell_i \cdot T_i}}
  {k \in I \quad \Typing {P} { \Gamma , \ContextElaboration {\vec x}{T_k}}}}
\right\rrbracket
                        &= ( \iota _k :  \llbracket T_k \rrbracket \profto  \llbracket \oplus_{i \in I} \ell_i \cdot T_i \rrbracket) \odot (  \varphi _{\ContextElaboration {\vec x} {T_k}}^
                         *(\llbracket P \rrbracket) :  \llbracket  \Gamma  \rrbracket^\perp \profto  \llbracket T_k \rrbracket)
  \\
  \left\llbracket\vcenter {
  \infer{\Typing {\CBranching a {i \in I} {\CBranch {\ell_i} {\vec x_i}{P_i}}}{ \Gamma , a: \with_{i \in I} \ell_i \cdot T_i}}
  { \forall i: I, \Typing {P_i} { \Gamma , \ContextElaboration{\vec x} {T_i}}}}
  \right\rrbracket &= \Pairing { \llbracket \varphi _{\ContextElaboration {\vec {x_i}} {T_i}}^
                         *(P_i) \rrbracket :  \llbracket  \Gamma  \rrbracket^\perp \profto  \llbracket T_i \rrbracket}_{i \in I}
  \\
  \left\llbracket\vcenter{
  \infer{\Typing{\cut a b T P} { \Gamma}}
  {\Typing {P} {a: T, b: T^\perp, \Gamma}}}
  \right\rrbracket &= ( \llbracket P \rrbracket :  (\llbracket T \rrbracket^\perp  \parallel   \llbracket T \rrbracket) \profto  \llbracket  \Gamma  \rrbracket) \odot ("\cc_{ \llbracket T \rrbracket}@copycat strategy" : \emptyset \profto  \llbracket T \rrbracket  \parallel   \llbracket T \rrbracket^\perp)
\\
\left\llbracket\vcenter{
  \infer
  {\Typing{\CProm a \PatternVar P} {\wn  \Gamma , a: \oc T}}
  {\Typing {P} {\wn  \Gamma , \ContextElaboration {\vec x}{T}}}}
  \right\rrbracket &=
(\ocG \varphi _{\ContextElaboration {\vec x} {T}}^
                         *(\llbracket P \rrbracket) : \ocG \ocG  \llbracket  \Gamma  \rrbracket^\perp \profto \ocG  \llbracket T \rrbracket)  \odot  \digging {\llbracket  \Gamma  \rrbracket^\perp}
\\
  \left\llbracket\vcenter{
  \infer
  {\Typing{\CDer a \PatternVar P} { \Gamma , a: \wn T}}
{\Typing{P}{ \Gamma , \ContextElaboration{\vec x}{T}}}}
\right\rrbracket &=
\dereliction {\llbracket T \rrbracket} \odot  ( \varphi _{\ContextElaboration {\vec x} {T}}^
                         *(\llbracket P \rrbracket) :  \llbracket  \Gamma  \rrbracket^\perp \profto  \llbracket T \rrbracket)
\\
\left\llbracket\vcenter{
  \infer
  {\Typing{\CCoder a \PatternVar P} {\Gamma , a: \oc T}}
  {\Typing {P} {\Gamma , a:\oc T, \ContextElaboration {\vec x}{T}}}}
  \right\rrbracket &= \codermap_{\llbracket T \rrbracket}  \odot \left(   \varphi _{\ContextElaboration {\vec x} {T}}^
                         *(\llbracket P
  \rrbracket) :  \llbracket  \Gamma  \rrbracket^\perp \profto  \llbracket \oc T \rrbracket  \parallel \llbracket T \rrbracket \right)
  \end{align*}
  \vspace{-3mm}
  \caption[Interpretation]{Interpretation of $\MetalanguageName$ into $\Strat$}
  \label{fig:interp-metalanguage}
   \vspace{-5mm}
\end{figure}

We start by showing the adequacy of our translation. To state it, we need a notion
of transition on strategies. Event structures come with a natural
notion of transition system, given by configurations. Given an event
structure $E$ and $x \in \config (E)$, we define $E/x$ as featuring
those events $e \in E \setminus x$ that are not in conflict with any
events in $x$, with causality and conflict directly inherited from
$E$.  This can be lifted to the level of strategies. Given a "uniform
strategy" $ \sigma : S \rightharpoonup \A$, and a configuration
$x \in \config (S)$, we build the "uniform strategy"
$ \sigma /x : \mathcal S/x \rightharpoonup \A/ \sigma \, x$ (read $ \sigma $
after $x$). We say that $ \sigma : A $ can do a transition to $ \tau $
with visible actions $y \in \config (A)$, written
$ \sigma \mathrel{\intro*{\belowStrategies y}} \tau $, when there
exists a configuration $x \in \config ( \sigma )$ such that
$ \sigma /x \weakIso \tau $ and $ \sigma \, x \cong y$ in $\tilde
A$. The particular case where $y = \emptyset$ corresponds to an
internal transition and is simply written
$ \sigma \belowStrategies {} \tau $.

\begin{restatable}{lemma}{InterpSound}
  \label{lemma:interp:sound}
  Consider $\Typing P  \Delta $.
  \begin{enumerate}
  \item {\rm (a)} If $P  \equiv  Q$, then $ \llbracket P \rrbracket \weakIso  \llbracket Q \rrbracket$; and {\rm (b)}
  If $P  \rightarrow  Q$ then $ \llbracket P \rrbracket \belowStrategies {}  \llbracket Q \rrbracket$
  \item If $P  \equiv  \CSelect a \ell {\vec x} Q$, then there exists
    $s  \in  \min( \llbracket P \rrbracket)$ mapped to $(a, \ell)$ and
    $ \llbracket P \rrbracket/\{s\} \weakIso  \llbracket Q \rrbracket$.
  \item If $P  \equiv  \CDer a {\vec x} Q$, then there exists
    $s  \in  \min( \llbracket P \rrbracket)$ mapped to $(a, \Request)$ and
    $ \llbracket P \rrbracket/\{s\} \weakIso  \llbracket Q \rrbracket$.
  \end{enumerate}

\end{restatable}

\begin{restatable}[Adequacy]{lemma}{interpAdequacy}
  \label{lemma:interp:adequate}
  Consider a process $\Typing P  \Delta $. We have:
  \begin{enumerate}
  \item If $ \llbracket P \rrbracket \belowStrategies {}  \sigma $, then there exists
    $\Typing Q  \Delta $ with $P  \rightarrow^*  Q$ and $ \llbracket Q \rrbracket \weakIso  \sigma $.
  \item If $s  \in   \llbracket P \rrbracket$ is a minimal positive event mapped to
    $(a, v)  \in   \llbracket  \Delta  \rrbracket$ then
   {\rm (a)}
    if $v = \Request$, then $P  \equiv  \CDer a {\vec x} Q$ and
      $ \llbracket Q \rrbracket \weakIso  \sigma /\{s\}$; and 
    {\rm (b)} if $v = \ell$, then $P  \equiv  \CSelect a \ell {\vec x} Q$ and
      $ \llbracket Q \rrbracket \weakIso  \sigma /\{s\}$.
  \end{enumerate}
\end{restatable}
\subsection{Full Abstraction}\label{sec:full abstraction}
Using the notion of transition on strategies defined in the previous section, we define weak bisimulations between strategies:
\begin{definition}
  A \emph{weak bisimulation} is an equivalence relation $\mathcal R$
  between "uniform strategies" on the same "tcg" such that if
  $ \sigma \mathcal R \tau $, for all configuration
  $y \in \config ( A )$, if $ \sigma \belowStrategies y \sigma '$,
  then there exists $ \tau '$ such that
  $ \tau \belowStrategies y { \tau '}$ with
  $ \sigma '\mathcal R \tau '$.
We write $\intro*{\Stratwb}$ for the largest weak bisimulation.
\end{definition}
\begin{lemma}\label{lemma:wb congruence}
  Weak bisimilarity is a congruence (stable under composition),
  hence $\sigma \weakIso \tau$ implies $ \sigma \Stratwb \tau $.
\end{lemma}
From this result, and using the standard method of action testers
\cite{Hennessy07}, we derive:
\begin{restatable}[Second-order Full Abstraction]{theorem}{FullAbs}\label{thm:fa}
  Consider $\Typing {P, Q} {\Delta}$. Then we have
  $P \MLObs Q$ iff $ \llbracket P \rrbracket \Stratwb  \llbracket Q \rrbracket$.
\end{restatable}
\subsection{Causal Semantics of $\CML$}\label{sec:cml semantics}
By composing the syntactic translation and the semantics
interpretation, we get an interpretation of types and programs of
$\CML$ in terms of event structures: $ \llbracket  \sigma  \rrbracket =  \llbracket \TranslateTerm  \sigma  \rrbracket$ and
$ \llbracket M \rrbracket =  \llbracket \TranslateTerm M \rrbracket$. From the previous results, we immediately get:
\begin{lemma}[Adequacy]
  Consider two terms $ \Delta   \vdash  M, N :  \sigma $. If $ \llbracket M \rrbracket \Stratwb  \llbracket N \rrbracket$, then
  $M \CMLObs N$.
\end{lemma}
As we discussed in \S~\ref{subsec:translation:correctness}, the
converse does not hold (see Appendix~\ref{subsec:discussion}).
However, we have \emph{second-order full
  abstraction}. An interface $ \Delta   \vdash   \sigma $ is second-order when (1) $ \sigma $ is a
base type and (2) types in $ \Delta $ are either base types, or arrow types
between base types. We have:
\begin{restatable}[Full Abstraction]{theorem}{SecondOrderFa}\label{thm:second-order-fa}\label{thm:fullsecond}
  Consider two terms $M, N$ well-typed on a second-order interface
  $ \Delta   \vdash   \sigma $. Then $M \CMLObs N$ if and only if $ \llbracket M \rrbracket \Stratwb  \llbracket N \rrbracket$.
\end{restatable}
This theorem is useful when applied to first-order functions
calling other external first-order functions (from say, libraries).

%% file: implementation.tex
\section{Implementation}\label{sec:implementation}
To illustate the causal model, we have used the model presented in
this paper to implement a causal interpretation of a subset of OCaml
corresponding to $\CML$ into event structures, representing
strategies. The metalanguage process is built implicitly but not
explicitly represented as they are not very informative and extremely
verbose. The (anonymised) prototype is a web application, available at:
\begin{center}
  \url{http://programminggamesemantics.github.io/index.html} 
\end{center}
The prototype allows entering OCaml code restricted to functions,
product, record types and sum types. The standard library of OCaml is
replaced by a simple kernel implemented by specific strategies to
represent integer references (called \texttt{var}) and
parallelism. See the webpage for details. The event structure is not
displayed in its entirety as it would often be infinite (because of
infinite datatypes and recursion) but instead the user can explore
interactively branches of the computation they are interested in by
clicking on Program events to unlock the next step of computation.


%% file: related-work.tex
\subsubsection*{Metalanguage and Process Representation for Strategies
  and Games}
\citet{DBLP:conf/fpca/HylandO95} first studied a relationship between
game semantics and the $\pi$-calculus, where $\pi$-calculus processes
are used to denote plays of innocent strategies (for PCF). This idea
led to recast the traditional encoding of the call-by-value
$\lambda$-calculus into the $\pi$-calculus \cite{MilnerR:funp} into a
game semantics model for call-by-value PCF \cite{HondaKgamtheaocbyc:}.
In the sequential setting, the work by
\citet{DBLP:journals/entcs/Longley09} proposed a programming language
to describe sequential innocent strategies as a whole.  Later,
\citet{DBLP:conf/popl/Goyet13} proposed an abstract calculus for
sequential strategies, close to the $\pi$I-calculus.

\citet{DBLP:conf/icfem/DimovskiL04,DBLP:conf/tacas/GhicaM06} represent
strategies as CSP terms for use with model checkers, but CSP is only
used a way to represent strategies, rather than a target language for
syntactic translations - in particular all reasoning must be done at
the level of the model. In a  similar vein,
\citet{DBLP:conf/lics/DisneyF15} argue for a reading of strategies in
terms of processes in the sequential setting, for type soundness.

\subsubsection*{Game Semantics for Concurrency and Nondeterminism}
The first concurrent game semantics model, based on traces, is due to
\citet{DBLP:journals/entcs/Laird01} for a message-passing language,
extended to call-by-name shared-memory later by \citet{cia}, angelic models
which are fully abstract for may-equivalence. Causal models arrived
later with \cite{Sakayori:2017:TCG:3080372.3080402} (asynchronous
$ \pi $-calculus) and \cite{concur16} (IPA: call-by-name
shared-memory), which are both angelic (ie. only adequate for
may-testing).  \citet{DBLP:conf/lics/HarmerM99} provide the first
non-angelic model of game semantics in the sequential setting based on
\emph{stopping traces}. This approach is tailored to must-equivalence.

Our approach thus gives the first non-angelic game model of a
realistic higher-order concurrent programming language, capturing
faithfully the nondeterministic branching behaviours of shared-memory
concurrent programs.

Recently, \citet{pamdill} gave a games semantics model for $\DiLL$
based on templates games. It differs from ours in two ways: (1) his
model is synchronous, while our model is asynchronous (due to courtesy);
and (2) it ignores deadlocks, preventing from modelling $\CML$
adequately. \citet{pam-ls-templates} extended it later to
model Concurrent Separation Logic. 

\subsubsection*{Extensions of the 
  Linear-Logic and Session Types Correspondence}
In this paper, we use
a variation of Differential Linear Logic (\DiLL) to express 
nondeterminism built on deterministic computations.
\citet{DBLP:journals/entcs/Beffara06} presents
a model of Linear Logic in terms of processes of the
$\pi$I-calculus, used to interpret concurrent extensions of the
$ \lambda $-calculus.
In the context of proof-nets, \citet{EhrhardLaurent10a} investigate
encoding a finitary
$\pi$-calculus in a transition system of
labelled differential interaction nets.  

Following the work by 
\citet{ToninhoCP12} on an encoding of the
simply-typed $\lambda$-calculus,
\citet{TY2018} have proven
that there exist 
\emph{mutually inverse} and \emph{fully abstract}
encodings 
between a polymorphic session $\pi$-calculus in \cite{CairesPPT13} and a linear
formulation of System F.
To gain expressiveness beyond 
functional and strong normalising behaviours, 
several extensions have been proposed, eg.~the lock primitives 
for nondeterminism 
\cite{DBLP:journals/pacmpl/BalzerP17}; 
dynamic monitoring 
\cite{Jia16,GommerstadtJP18}; exceptional handling
\cite{CairesP17}; 
multiparty interactions \cite{CMSY2015,CLMSW16}; 
hyperenvironments 
to capture non-blocking I/O \cite{Kokke:2019:BLN:3302515.3290337};
and a bounded linear logic based-extension to model 
racy conditions \cite{KokkeMW19}.
\citet{BalzerPT18} show an encoding of untyped asynchronous communication in
\cite{DBLP:journals/pacmpl/BalzerP17}, and  
\citet{BTP19} successfully
enabled deadlock-freedom  
by introducing 
extra-logically imposed partial orders.

Our approach differs from them, aiming to \emph{describe} game semantics
translations more accurately by the metalanguage arisen from 
event structures grounded on \DiLL{}. 
None of the above calculi provides 
a fully abstract encoding from
nondeterministic and concurrent languages (ie.~$\CML$)
nor a causal semantic interpretation. 


%% file: ccl.tex
Our contribution in this paper is twofold: we
have presented a factorisation of the usual game semantics methodology
in two steps: (1) a syntactic translation and (2) a semantics
interpretation which allows for separating concerns when designing new
game semantics models. We have applied this methodology to build the
first non-angelic interactive model of a call-by-value language with
shared memory concurrency. We have used our framework to implement a causal
modelling of a subset of a concurrent extension of OCaml. We believe our methodology
proves promising to model mainstream programming languages and systems 
where semanticists will need inputs from language or system 
designers who do not have an immediate access to game semantics. 

Another avenue would be to use the model to reason
about open terms, and inserts itself naturally in some new research
programs such as \cite{shao-refinement-games,
  Gu:2018:CCA:3192366.3192381} which aim at building large certified
systems by assembling components proved correct in isolation.

Other future work include relaxed shared memory using the techniques
presented in \citep{jfla}, and trying to characterise $\UStrat$ (or a
subset of it) as an initial category.


%% file: appendix-calculus.tex
\section{Appendix of Section~\ref{sec:calculus}}
\label{app:calculus}

\subsection{Proofs}
\label{sec:calculus:proofs}

\SR*

\begin{proof}
{\bf Case (1: \rulename{race}))} Assume
$P_0=(\CDer a \PatternVar P \CFork \CCoder b \PatternVari Q \CFork R)$ and
\begin{equation}\label{eq:race_one}
\Typing{P_0}{\Gamma,a:?A,b:A^\bot}  
\end{equation}  
Then by the intervion of \rulename{par}, we have: 
\begin{equation}\label{eq:race_two}
 \Typing{\CDer a \PatternVar P}
 {\Gamma_1,?\Gamma_0,a:?A}
 \quad \text{and} \quad 
 \Typing{\CCoder b \PatternVari Q}{\Gamma_2,?\Gamma_0,b:A^\bot}
        \quad \text{and} \quad
        \Typing{R}{\Gamma_3,?\Gamma_0}
        \quad 
        \text{with}
        \quad
        \Gamma=\Gamma_1, \Gamma_2, \Gamma_3,?\Gamma_0
\end{equation}
From (\ref{eq:race_two}), and inversion of \rulename{req} and
\rulename{nd}, we have
\begin{equation}\label{eq:race_three}
 \Typing{P}
 {\Gamma_1,?\Gamma_0,a:?A,\ContextElaboration {\PatternVar} T}
\quad\text{and}\quad 
 \Typing{Q}
        {\Gamma_2,?\Gamma_0,b:A^\bot, \ContextElaboration {\PatternVari} T'}
        \ \mbox{with}\  A^\bot = !T', \  T' = T^\bot
\end{equation}
From (\ref{eq:race_three}), applying \rulename{par} and \rulename{res},
we have
\begin{equation}\label{eq:race_four}
  \Typing{\CCut {\PatternVar} {\PatternVari} {(P \parallel Q)}}{
\Gamma_1,a:A,\Gamma_2,?\Gamma_0,b:A^\bot
  }
\end{equation}
Applying
\rulename{par} to 
(\ref{eq:race_four}) and
(\ref{eq:race_two}),
we have:
\begin{equation}\label{eq:race_five}
  \Typing{R\CFork \CCut {\PatternVar} {\PatternVari} {(P \parallel Q)}}{
\Gamma_1,a:A,\Gamma_2,\Gamma_3,?\Gamma_0,b:A^\bot
  }
\end{equation}
Applying \rulename{res} to (\ref{eq:race_five}), we obtain the result. \\

\smallskip

\noindent{\bf Case (2: \rulename{cxt} and \rulename{str}))} Straightforward
by IH.
\end{proof}


%% file: appendix-translation.tex
\section{Proofs of \S~\ref{sec:interp ML}}
We prove here Theorem \ref{the:trans:correct}.

\subsection{Approximation and normal form}
Consider a finite
semiclosed term $M$. Because there are no fixpoint, we know that $M$
must reduce to a normal form $\Nf M$ for $ \rightarrow $ (since we are
simply in the simply-typed $ \lambda $-calculus with uninterpreted
constants). By Lemma \ref{lemma:soundness}, we know that
$\TranslateTerm {M}_{c, o} \equiv \TranslateTerm {\Nf M}_{c, o}$. A
term is ""stuck"" when it is in normal form but not a value. Any finite term $M$ has a normal
form written $\Nf M$. We say that a machine is finite when the term it
contains is finite. In that case, we extend the notation $\Nf \cdot$:
$\Nf {\machineDefault M} := \machineDefault {\Nf M}$. If $ \sigma $ is a base
type, we write $\Nf  \sigma $ for the set of normal forms of semiclosed terms
of type $ \sigma $.

\begin{lemma}\label{lemma:syntax for stuck}
The set of "semiclosed" "stuck" terms is contained in the following grammar:
$$S ::= ( \lambda \vec x.\, M)\, S\, N\,  \ldots \, N \mid \IfTerm S M M \mid \RefAlloc (\underline n) \mid \oc r \mid r := \underline n.$$
where $M$ means any term, and $N$ means normal form (not necessarily
stuck).
\end{lemma}
\begin{proof}
  Consider a stuck term $S$. We proceed by induction on its syntax.
  \begin{itemize}
  \item It cannot be a $ \lambda $-abstraction since "semiclosed" terms do not have arrow types.
  \item If it is a conditional $\IfTerm M N {N'}$, then $M$ must be
    normal as otherwise $S$ would not be normal. If $M$ was a value,
    then it would have to be an immediate boolean since "semiclosed"
    terms do not have free variables of type $\BoolType$. But then,
    $S$ would reduce to $N$ or $N'$, which is absurd. So $M$ must be
    stuck.
  \item If it is a variable: impossible, variable are values.
  \item If it is application: then $S$ has the form
    $M\, N_1\,  \ldots \, N_k$ where $M$ is not an application. Since $S$ is
    "semiclosed", $M$ can only be an abstraction or a constant.
    \begin{itemize}
    \item If it is an abstraction: then all the $N_i$ must be in
      normal form. Moreover, the first argument must not be a value
      otherwise, there would be a reduction.
    \item If it a constant, then we see that the only possible cases
      are $\RefAlloc {(\underline n)}$, $\oc r$, or $r := \underline n$.
    \end{itemize}
  \end{itemize}
\end{proof}

\subsection{Finite adequacy}
\begin{lemma}\label{lemma:cml values nf}
  For any semiclosed value $V$, $\machineDefault V$ cannot do any
  transition.
\end{lemma}
\begin{proof}
  Easy inspection.
\end{proof}
\begin{lemma}[Finite adequacy]\label{lemma:finite adequacy lemma}
  Consider a finite machine $\machineDefault M$. If
  $\TranslateTerm{\machineDefault M} \xrightarrow{ \alpha } P$, then $ \alpha  =  \tau $, and there are two cases:
  \begin{itemize}
  \item Either $\machineDefault M \xrightarrow{ \tau } \machineDefaultt N$
    with $\TranslateTerm {\machineDefaultt N} \equiv P$,
  \item Or $P  \equiv   \alpha .\, Q$ and
    $\machineDefault M \xrightarrow{ \alpha } \machineDefaultt N$ with
    $\TranslateTerm {\machineDefaultt N} \equiv Q$.
  \end{itemize}
\end{lemma}
\begin{proof}
  First, by Lemma \ref{lemma:soundness}, we can assume without loss of
  generality that $M$ is in normal form. Moreover, given the shape of
  $\TranslateTerm {\machineDefault M}$ it is clear that $ \alpha $ must be
  $ \tau $ as there are no visible actions at top level.  By Lemma
  \ref{lemma:cml values nf}, we know that $M$ must be stuck. We
  examine the cases given by Lemma \ref{lemma:syntax for stuck}:
  \begin{itemize}
  \item If $M = \RefAlloc (\underline n)$, then we have a
    contradiction as this term cannot reduce.
  \item The cases for $\oc r$, are $r := \underline n$ are
    straightforward calculation. For instance, assume that
    $M = \oc r$. In this case, there are only two possible reductions:
    indeed $\TranslateTerm M_{c, o}$ starts with a dereliction that
    has to communicate with the codereliction on $r$. That must have
    caused the $ \tau $-transition. Then there are two cases, whether
    $r  \in  \vec y$ or not.
    \begin{itemize}
    \item If $r \not\in \vec y$: then the read is performed, and the
      value is sent back to $M$, and we have
      $P  \equiv  \TranslateTerm {\machineDefault {\underline{ \mu (n)}}}$ as
      desired.
    \item Otherwise, $r$ is visible and there is a visible transition on $\mathfrak l$ in front of
      the continuation, and we end up on the same process after
      performing it.
    \end{itemize}
  \item If $M = \IfTerm {M_0} N N'$, then we know that $M_0$ must be
    "stuck". By construction of the translation, we must have
    $\machineDefault {M_0} \xrightarrow{ \alpha } P'$ and $P$ can be obtained
    from $P'$ by the interpretation of if. Then we can apply the
    induction hypothesis.
  \item If $M = ( \lambda \vec{x}.\, M_0)\, \vec N$: then once again by the
    definition of the interpretation the reduction must occur within
    one of the $N_i$, and we conclude by induction.
  \end{itemize}
\end{proof}

\subsection{Conclusion}
We can now conclude the adequacy result and the correctness result of
our translation of machines.

\Adequacy*
\begin{proof}
  Write $\mathfrak m$ for the input machine.  Clearly $ \alpha  =  \tau $ for the
  same reason as in Lemma~\ref{lemma:finite adequacy lemma}. Consider now a
  finite approximation $P_0$ of $P$: by definition of the LTS, there
  exists a finite approximation $Q_0$ of $\TranslateTerm m$ with
  $Q_0 \xrightarrow{ \tau } P_0$. By continuity, we can find a finite
  approximation $M_n$ of $M$ such that
  $Q_0  \leq  \TranslateTerm {\machineDefault {M_n}}  \leq 
  \TranslateTerm{\machineDefault M_n}$. Then, we can apply
  Lemma~\ref{lemma:finite adequacy lemma}, and do a case distinction --- the
  two cases are similar. Say that
  $\machineDefault {M_n} \xrightarrow { \tau } \machineDefaultt {R_n}$ with
  $P_0  \equiv   \leq  \TranslateTerm{\machineDefaultt {R_n}}  \leq   \equiv  P$. This implies
  that $\machineDefault M \xrightarrow{ \tau } \machineDefaultt R$. Then it
  is easy to see that $\TranslateTerm{\machineDefaultt R}$ is the
  limit of the $\TranslateTerm{\machineDefaultt R_n}$, hence
  $\TranslateTerm{\machineDefault R} \equiv P$ as desired.
\end{proof}
\TranslationCorrect*
\begin{proof}
  $(1) \Leftrightarrow (2)$: To show the result, it is enough to show that

  $$
  \begin{aligned}
    \mathcal R_0 &= \{ (\TranslateTerm{\mathfrak m}, \TranslateTerm{\mathfrak n}), ( \alpha^{\mathfrak l} .\, \TranslateTerm{\mathfrak m},  \alpha^{\mathfrak l} .\, \TranslateTerm{\mathfrak n}) \mid \mathfrak m \CMLbisim \mathfrak n \} \\
    \mathcal R_1 &= \{ (\mathfrak m, \mathfrak n) \mid \TranslateTerm {\mathfrak m} \MLbisim \TranslateTerm {\mathfrak n}\}
  \end{aligned}$$
  are weak bisimulations.

  \begin{enumerate}
  \item For $\mathcal R_0$:
    \begin{itemize}
    \item If
      $( \alpha^{\mathfrak l} .\, \TranslateTerm{\mathfrak m}, \alpha^{\mathfrak l} .\,
      \TranslateTerm{\mathfrak n}) \in \mathcal R_0$, then it is easy
      to see that they both can only do $ \alpha^{\mathfrak l} $ before getting to
      $(\TranslateTerm {\mathfrak m}, \TranslateTerm{\mathfrak n})$
      which is still in $\mathcal R_0$.
    \item If
      $(\TranslateTerm{\mathfrak m}, \TranslateTerm{\mathfrak n}) \in
      \mathcal R_0$: if
      $\TranslateTerm{\mathfrak m} \xrightarrow{ \alpha } P$, then we know
      that $ \alpha  =  \tau $ by Lemma \ref{lemma:adequacy lemma}, and there are two cases:
      \begin{itemize}
      \item If $\mathfrak m \xrightarrow{ \tau } \mathfrak m'$ with
        $\TranslateTerm {\mathfrak m'} \equiv P$, then
        $\mathfrak n \xrightarrow{ \tau } \mathfrak n'$ for some
        $\mathfrak n'$ with $\mathfrak m' \CMLbisim \mathfrak n'$ and
        we conclude via Lemma \ref{lemma:cml sim}.
      \item If $\mathfrak m \xrightarrow{ \alpha} \mathfrak m'$ and
        $\TranslateTerm{\mathfrak m'}  \equiv   \alpha^{\mathfrak l} .\, P$, then
        $\mathfrak n \xrightarrow{ \alpha } \mathfrak n'$ with
        $\mathfrak m' \CMLbisim \mathfrak n'$. By Lemma \ref{lemma:cml
          sim}, we know that
        $\TranslateTerm{\mathfrak n} \xrightarrow{\alpha} Q$ and
        $Q \equiv  \alpha^{\mathfrak l} .\, \TranslateTerm{\mathfrak n'}$ hence
        $(P, Q)  \in  \mathcal R_0$ as desired.
      \end{itemize}
    \end{itemize}
  \item For $\mathcal R_1$: Assume that
    $(\mathfrak m, \mathfrak n)  \in  \mathcal R_1$, and
    $\mathfrak m \xrightarrow{ \alpha } \mathfrak m'$. Then by Lemma
    \ref{lemma:cml sim}, we have
    $\TranslateTerm{\mathfrak m} \By{ \alpha }\TranslateTerm{\mathfrak
      m'}$. By construction, this implies that
    $\TranslateTerm {\mathfrak n} \By{ \alpha } \TranslateTerm{\mathfrak n'}$
    for some $\mathfrak n'$, and we conclude by Lemma
    \ref{lemma:adequacy lemma}.
  \end{enumerate}

  $(2) \Leftrightarrow (3)$ For this, we need the full abstraction result of Theorem
  \ref{thm:fa}. Indeed, it is easy to see that machines $\mathfrak m$, $\mathfrak m'$, we have
  $$ \llbracket \mathfrak m \rrbracket \CMLbisim  \llbracket \mathfrak m' \rrbracket \Leftrightarrow  \llbracket \mathfrak m \rrbracket \Stratwb  \llbracket \mathfrak m' \rrbracket.$$
\end{proof}
  

\subsection{Discussion on full abstraction}
\label{subsec:discussion}
In this subsection, 
we offer a detailed discussion on the
full abstraction. This is a well-known phenomenon in game semantics due to some
information that is lost during the translation: the difference
between calling and returning. At the level of the metalanguage, the
distinction is gone which allows us to write contexts that can for
instance, evaluate an argument \emph{and} return in parallel which is
not possible in $\CML$.

We now show a concrete examples, using the following two terms:

$$
\begin{array}{rll}
  M&=  &f ( \lambda x. ()); \bot\\
  N&=&\mathtt{let}\, r = \RefAlloc\, 0\, \texttt{in} \ 
  \mathtt{let}\, s = \RefAlloc\, 0 \, \texttt{in}\\
  &&f ( \lambda x.\, r := 1; s := 1);
  r := !r + 1;\\
   &&\mathtt{if} {(!r = 1\, \mathtt{and}\, \oc s = 0)}\, \mathtt{then}\, {()}\,\mathtt{else}\, {\bot}\\
\end{array}$$

$M$ and $N$ are higher-order terms, with one higher order parameter
$f: (\UnitType \rightarrow \UnitType) \rightarrow \UnitType$. $M$ calls its
parameter on a dummy function, and then diverges. $N$ does almost the
same, except that there is a possibility for it converge: if at the
end, $r$ equals one and $s$ zero. From the point of view of
$\Context$, the two arguments are equivalent, as they always return
after being called. For $N$ to converge, only $r$ must be set to one,
which is impossible to achieve using a context written in $\CML$ as
calls to $f$ will always finish completely before the control is
passed back to $N$. However, there is a process that can interact
differently with $M$ and $N$. Indeed, their interpretations at $o$ are
processes over the context
$$ \Gamma  = o: \Return {}^+, f: \wn (\Call {\lambda}^+ \cdot (\oc \Call {}^- \cdot \Return {}^+  \parallel  \Return{}^-))$$
On this interface, we can build a context that will trigger the events
$\Call{}^-$ and $\Return{}^-$ in parallel, which simulates a function
that calls its argument \textbf{while} returning. In $N$, this
triggers several races, including one between the write on $s$ and the
read on $s$: this race makes it possible for the condition to be
evaluated after $r$ is set to one, but before $s$ is. Such a context
can be written as follows:
$$
\begin{aligned}
  C[] = \CCut {o} {\bar o} {(&[] \mid \CBranchingOneS {\bar o}{\Return {}}  \alpha  \mid \CProm f x \CBranchingOne x {\lambda} {x, r} (\CDer x {x_0} \CPositiveActionS {x_0} {\Call {}} \CFork \CPositiveActionS r {\Return {}}))}.
\end{aligned}$$
However as noted in the main part of the paper, we can derive the full abstraction
result for the second-order (Theorem \ref{thm:fullsecond}).

%% file: appendix-games.tex
\section{Proofs of \S~\ref{sec:games}}
\Recursion*
\begin{proof}
  Write $S_i$ for the underlying event structure of $ \sigma _i$. We define

  $$S = \bigcup_{i  \in   \mathbb{N} } S_i \setminus f_{i-1}(S_{i-1}),$$
  with the convention that $f_0(f_0) = \emptyset$. Causality is
  defined as follows: $s  \leq _S s'$ when there exists $i  \in   \mathbb{N} $ where both
  $s, s'$ can be embeded and where $s  \leq _{S_i} s'$. Similarly for
  conflict. It is easy to see that $S_i$ embeds into $S$ and that $S$
  is the smallest such event structure. Moreover, we can project $S$
  to $A$ by simply using the $ \sigma _i$.
\end{proof}
\productLinDet*
\begin{proof}
  Without loss of generality, we can assume that each $S_i$ contains
  the minimal strategy on $A^\perp  \parallel  B_i$. We define $S$ to be the
  union of the $S_i$, plus events $\{\ell_i \mid i \in I \}$, with all
  courteous links from $\ell_i$ to events in $S_i$, and conflict that
  of the $S_i$ plus all the minimal conflict between the $\ell_i$. The
  mapping $ \sigma $ from $S$ to
  $A^\perp  \parallel  \with_{k \in K} \ell_k^- \cdot B_k$ is given by the
  $ \sigma _i$. It is easy to check that this strategy satisfies the desired
  axiom.
\end{proof}

\subsection{Proof of the linear exponential comonad}
\mysubsubsection{Lifting and polarised categories}
We define two operations on game:
$\intro*{\LiftPlus \A} = \Request^+ \cdot \A$ and
$\intro*{\LiftNeg \A} = \Request^- \cdot \A$. We define the polarised
subcategories of $\UStrat$: $\UStrat^-$ is the category of negative
games and negative strategies; while $\UStrat^+$ is the category of
positive games and negative strategies. Duality induces an isomorphism
$(\UStrat^-)^{\text{op}} \cong \UStrat^+$. This operation on "tcgs"
extends to a functor $\LiftNeg - : \UStrat \rightarrow \UStrat^-$:
given $ \sigma : \A \profto \B$, $\LiftNeg { \sigma }$ starts by
acknowledging the negative $\Request^-$ on the right, before playing
$\Request^+$ on the left, and continuing as $ \sigma $. As a result,
$\LiftNeg { \sigma }$ is a "negative strategy".

\begin{lemma}
  The functor $\LiftNeg {(-)} : \UStrat  \rightarrow  \UStrat^-$ is a right adjunt
  to the inclusion $\UStrat^- \subseteq \UStrat$.
\end{lemma}
\begin{proof}
  The adjunction is fairly simple to describe:
  \begin{itemize}
  \item From $ \sigma : \LiftPlus A \profto B$ in $\UStrat^+$, by removing
    the initial negative event on $\LiftPlus A$, we get a strategy
    $A \profto B$ in $\UStrat$ (not necessarily negative).
  \item From $ \sigma : A \profto B$ in $\UStrat$, we can build a strategy
    $\Request^- \cdot  \sigma  : \LiftPlus A \profto B$ which is
    well-defined because $B$ is positive.
  \end{itemize}
  It is clear that these operations are inverse of each other;
  naturality is a simple calculation.
\end{proof}

\mysubsubsection{Lifting the monad} We are now ready to prove the result:

\comonad*
\begin{proof}
From \cite{these}, we know that
$\sharpG {\cdot}$ is an exponential comonad on $\UStrat^-$. We notice that
$\oc A = \sharpG {\LiftNeg {\A}}$ and we conclude by the fact that
$\LiftNeg {\A}$ is a right adjunct to the inclusion: it thus transports
exponential comonads on $\UStrat^-$ to exponential comonads on
$\UStrat$ as desired.
\end{proof}


%% file: appendix-interpretation.tex
\section{Proofs of \S~\ref{sec:interp}}

\subsection{Adequacy}
\InterpSound*
\begin{proof}
  \begin{enumerate}
  \item To show the result on $ \equiv $, since $\weakIso$ is a
    congruence, it is enough to show that it satisfies the rules
    listed in Figure~\ref{fig:calc:cong}.

    The first block is a consequence of the monoidal structure and the
    compact-closure. The third block is a direct consequence of the
    weak product structure and the exponential comonad structure.

    For the second block:
    \begin{itemize}
    \item \rulename{nil}: In the composition induced by $\CCut a b$,
      the initial move is a negative move on $a$. That means that
      every positive move will be dependent on this negative move, in
      particular any positive move. This means that the composition
      can only contain negative events on channels distinct to $a, b$,
      so the process must be isomorphic to the empty strategy.
    \item \rulename{res}: Consequence of compact closure
    \item \rulename{id}: Consequence of the categorical setting: the forwarder behaves as an identity
    \item \rulename{swap}: If $c \neq \{a, b\}$, then we can safely
      push it outside the restriction: indeed, this will not create
      new causal links from the prefix on $c$ to positive actions in
      $P_i$: such actions are already waiting on $\mathfrak a$. Hence,
      they are waiting on an action on $b$, that if happens, must
      occur in $\mathfrak c^-[Q_i]$, hence after $\mathfrak c^-$.
    \end{itemize}

  \item If $P \equiv \CSelect a \ell {\vec x} Q$, then we know that
    $ \llbracket P \rrbracket \weakIso \CSelect a \ell {\vec x} Q$. The interpretation of
    the latter is $ \iota _k  \odot   \llbracket Q \rrbracket$: the strategy $ \iota _k$ has $(a, \ell)$ as a
    minimal event which remains after composition since it is not in
    the hidden part. Similar reasoning for
    $P \equiv \CDer a {\vec x} Q$.
  \item If $P  \rightarrow  Q$, then we proceed by induction on $ \rightarrow $. The rules
    \rulename{str} follows from the previous point and rule
    \rulename{ctx} follows from the fact that evaluation contexts can
    not postpone a neutral event: any minimal neutral event of $ \llbracket P \rrbracket$
    is a minimal neutral event of $ \llbracket E[P] \rrbracket$. The most interesting rule
    is \rulename{race}. This is a consequence of the construction of
    $C_A$: if
    $\CCut a b {(\CDer a {\vec x} P  \parallel  \Coder a {\vec y} Q  \parallel  R)}  \rightarrow  \CCut
      a b {(R  \parallel  \CCut {\vec x}{\vec y}{(P  \parallel  Q)})}$, then the
    $\Request_i^+$ emitted by the dereliction meets one initial
    $\Request_i^-$ of the codereliction. This synchronisation is
    internal and unlocks the desired neutral event $\ast_i$. Then,
    $C_A$ behaves as copycat, which are the counterpart of the
    restrictions appearing in the right-hand-side of the syntactic
    rule.
  \end{enumerate}
\end{proof}
\interpAdequacy*
\begin{proof}
  \begin{enumerate}
  \item This reduction can be already be done by a finite approximant
    $P_0$ of $P$: $ \llbracket P_0 \rrbracket  \rightarrow   \sigma _0$ and $ \tau $ finite approximation of
    $ \sigma $. We can proceed by induction on $P_0$.
    \begin{itemize}
    \item If $P_0 = P_1  \parallel  P_2$: then the neutral event must come from
      $ \llbracket P_1 \rrbracket$ or $ \llbracket P_2 \rrbracket$, and we can conclude by induction.
    \item If $P_0$ starts with a negative prefix, then it does not have a
      neutral minimal event.
    \item If
      $P_0$ starts with a positive prefix, we conclude by induction.
    \item If $P_0$ is of the form $\CCut {\vec a} {\vec b}
      {P_1}$: we consider the neutral event $s$ of
      $P_0$. It corresponds to a neutral event $s'$ of
      $ \llbracket P_1 \rrbracket$ which might not be minimal. Without loss of generality,
      we can assume that all deterministic reductions in
      $P_1$ are reduced. This means that
      $P_1$ must be of the form $\Coder a {\vec x} Q \mid \CDer {\bar
        a} {\vec y} R$ where $a$ and $\bar a$ are connected by a nu,
      and we can conclude.
  \end{itemize}
\item Same reasoning, by induction on a finite approximant.
\end{enumerate}

\end{proof}

\subsection{Proof of full abstraction}

\FullAbs*
\begin{proof}
  From Lemmata \ref{lemma:wb congruence}, \ref{lemma:interp:sound},
  and \ref{lemma:interp:adequate}, we get easily the right-to-left
  direction.

  For the left-to-right direction, we use action testers. We show that
  $\mathcal R = \{ ( \llbracket P \rrbracket,  \llbracket Q \rrbracket) \mid P \MLObs Q\}$ is a weak
  bisimulation.

  \begin{itemize}
  \item Assume that $ \llbracket P \rrbracket \belowStrategies {}  \sigma $, then we directly apply
    adequacy and the definition of $\MLObs$ to conclude.
  \item Assume that $ \llbracket P \rrbracket \belowStrategies {e^-}  \sigma $, then we can conclude by
    receptivity.
  \item The interesting case is when $ \llbracket P \rrbracket \belowStrategies {(a, v)^+}
     \sigma $. There are two cases:
    \begin{itemize}
    \item If $v = \ell$, then $P  \equiv  \CSelect a \ell {\vec x}
      P_0$. Then, by definition of $\MLObs$, we must have
      $Q  \equiv  \CSelect a \ell {\vec x} {Q_0}$. Then we consider the
      context:
      $$C[]:=\CCut a b {[] \mid \CBranchingOne b \ell {\vec y}. \Forwarder {\vec x} {\vec y}}.$$
        We have $C[P]  \equiv  P_0$ and $C[Q]  \equiv  Q_0$. Hence, because $\MLObs$
        is a congruence, we have that $P_0 \MLObs Q_0$ as desired.
      \item If $v = \Request$, then the reasoning is the same.
    \end{itemize}
\end{itemize}
\end{proof}
\subsection{Second-order full abstraction}
We now show Theorem \ref{thm:second-order-fa}.
\begin{lemma}\label{lemma:sub}
  Consider a base type $ \sigma $ and $x:  \sigma ,  \Delta   \vdash  M, N :  \tau $. Then
  $M \CMLObs N$ if and only if for every value $v$ of type $ \sigma $, $M[x := v] \CMLObs N[x := v]$
\end{lemma}
\begin{proof}
  Left-to-right is by definition of $\CMLObs$. Right-to-left, it is
  easy to see that a context can only use $M$ and $N$ after it has fed
  a value for $x$.
\end{proof}

\SecondOrderFa*
\begin{proof}
  We only need to show the left-to-right inclusion. By Lemma
  \ref{lemma:sub}, we can assume that $ \Delta $ only contains
  function types, ie.
  $ \Delta = f_1 : \sigma _1 \rightarrow \tau _1, \ldots , \sigma _n
  \rightarrow \tau _n$. We provide a context on
  $ \Delta \vdash \sigma $ that translate function calls into a
  reference write followed by a reference read.  We can without loss
  of generality suppose that all base types involved are $\NatType$.
  
  $$
  \begin{aligned}
    C[] ::= &\LetIn {\vec{a_i}} {\overrightarrow{\mathtt{var}\, 0}} {}\\
    &\LetIn {\vec{r_i}} {\overrightarrow{\mathtt{var}\, 0}} {}\\
    &[] [f_i :=  \lambda i. a_i := i; !o_i]
  \end{aligned}$$ This context creates two references per external
  function (one to pass the argument value, and one to receive the
  return) and instanciates $f_i$ with a function that writes the
  argument, and tries to read back the value.

  By assumption, we have that $C[M] \CMLbisim C[N]$, which in turn
  implies that
  $ \llbracket \TranslateTerm {C[M]} \rrbracket \Stratwb \llbracket
  \TranslateTerm {C[N]} \rrbracket$. But it is easy to see that this
  implies that
  $ \llbracket M \rrbracket \Stratwb \llbracket N \rrbracket$: each
  write to $a_i$ corresponds to a call to $f_i$ with the corresponding
  argument, and each read to $r_i$ correspond to the return value. We
  conclude by full abstraction.
 \end{proof}
